\documentclass[a4paper,11pt]{article}
\pdfoutput=1 

\usepackage{jheppub} 
\usepackage{amssymb}
\usepackage{mathtools}
\usepackage{slashed}
\usepackage{cleveref}
\usepackage{dsfont}
\usepackage{amssymb}
\usepackage{mathtools}
\usepackage{slashed}
\usepackage{cleveref}
\usepackage[mathscr]{eucal}
\usepackage{enumitem}
\usepackage[table]{xcolor}

\newcommand{\be}{\begin {strip} \begin{equation} }
\newcommand{\ee}{\end{equation}}
\newcommand{\bea}{\begin{eqnarray}}
\newcommand{\eea}{\end{eqnarray}}

\newcommand\ddfrac[2]{\frac{\displaystyle #1}{\displaystyle #2}}
\newcommand{\epm}{e^+e^-}

\newcommand{\soft}[1]{\mathbb{S}_{\mbox{\small {#1}-h}}}
\newcommand{\ftsoft}[1]{\widetilde{\mathbb{S}}_{\mbox{\small {#1}-h}}}
\newcommand{\coll}{\mathbb{C}}

\newcommand{\lips}[1]{{d^3 \vec{#1}}/{2 E_{#1}}}
\newcommand{\kmeas}[2]{\frac{d^D #1_{#2}}{(2 \pi)^D}}

\newcommand{\eps}{\varepsilon}
\newcommand{\tplus}[1]{\left( \frac{#1}{\tau} \right)_+}
\newcommand{\zplus}[1]{\left( \frac{#1}{1-z} \right)_+}
\newcommand{\meas}[2]{\frac{d^{#2} #1}{(2\pi)^{#2}}}
\newcommand{\pFq}[2]{{}_{#1}F_{#2}}
\newcommand{\aS}{\frac{\alpha_S}{4\pi}}
\newcommand{\FT}{\int d^{2-2\eps}\vec{k}_T \, e^{i \vec{k}_T \cdot \vec{b}_T}}
\newcommand{\logbT}[1]{\log^{#1}{\left( \frac{b_T \, \mu}{c_1} \right)}}
\newcommand{\logb}[1]{\log^{#1}{\left( \frac{b}{c_1} \right)}}
\newcommand{\logmu}[1]{\log^{#1}{\left( \frac{\mu}{Q} \right)}}

\newcommand{\loggen}[3]{\log^{#3}{\left( \frac{#1}{#2} \right)}}
\newcommand{\TrC}{\frac{\text{Tr}_C}{N_C}}
\newcommand{\TrD}{\frac{\text{Tr}_D}{4}}

\newcommand{\hyp}[1]{\hypersetup{linkcolor=black}\ref{hyp:#1}}

\title{\boldmath 
Kinematic regions in the $e^+e^- \to h \, X$ factorized cross section 
with thrust}

\author{M. Boglione}
\author{and A. Simonelli}
 
\affiliation{Dipartimento di Fisica, Universit\`a di Torino,\\
                Via P.~Giuria 1, I-10125 Torino, Italy}
\affiliation{INFN, Sezione di Torino, Via P.~Giuria 1, I-10125 Torino, Italy}

\emailAdd{boglione@to.infn.it}
\emailAdd{andrea.simonelli@unito.it}

 \abstract{
 Factorization theorems allow to separate out the universal, non-perturbative content of the hadronic cross section from its perturbative part, which can be computed in perturbative QCD, up to the desired order.
In this paper, we derive a rigorous proof of factorization of the $\epm \to h\,X$ cross section, sensitive to the transverse momentum of the detected hadron with respect to the thrust axis, in a completely general framework, based on the Collins-Soper-Sterman approach. The results are explicitly computed to NLO-NLL accuracy and subsequently generalized to all orders in perturbation theory. 
This procedure naturally leads to a partition of the $\epm \to h\,X$ kinematics into three different regions, each associated to a different factorization theorem.
In one of these regions, which covers the central and widest range, the factorization theorem has a new structure, which shares the features of both TMD and collinear factorization schemes. In the corresponding cross section, the role of the rapidity cut-off is investigated, as its physical meaning becomes increasingly evident.
An algorithm to identify these three kinematic regions, based on ratios of observable quantities, is provided. 
}

\begin{document} 

\maketitle
 

\section{Introduction\label{sec:intro}}


\bigskip

Modern studies of high energy QCD processes are based on factorization theorems, that play a pivotal role in the study of strong interactions, as they allow to write the cross sections of hadronic processes in a form suitable for phenomenological analyses. Perturbative QCD alone is not sufficient to exploit the whole theory's predictive power since, even at lowest orders, several physical observables are affected by uncanceled infrared divergences. 
Through the factorization procedure, the divergent contributions are separated from the finite, computable parts and are collected into universal factors, as they can be extracted from a small set of experimental data and then used to predict any other observable that requires their contribution. Crucially, if universality is preserved, then the theory can be predictive.
Therefore, research in this field has a double purpose: in addition to the pure theoretical investigation, aimed to provide a solid proof of factorization of the processes that still lack of a proper factorization theorem, the phenomenological applications of the theory are fundamental in the extraction of universal non-perturbative factors.

\bigskip

If $Q$ is some typical hard energy scale for a certain process (c.m. energy, momentum transfer, etc...), then, at the cost of an error suppressed by powers of $m/Q$, where $m$ is a typical low energy hadronic mass scale (typically around $\sim 1$ GeV) , a factorization theorem recasts the cross section in terms of a convolution of hard, collinear and soft contributions.

As long as the physical observables are not sensitive to the transverse motion of the partons inside their parent hadron, the factorization procedure can be carried out rather simply, giving solid factorization theorems for a large set of processes.
Since the information on the transverse motion of partons is neglected, such theorems are often labeled as ``collinear". 
Many well known observables obey a collinear factorization theorem, like the cross section of Deep Inelastic Scattering (DIS) and $e^+e^-$ annihilation \cite{Bjorken:1968dy, Bjorken:1969ja, Feynman:1969ej, Drell:1969jm, Drell:1969wd, Drell:1969wb}.
In all of these cases the contribution of the soft part is trivial. In particular, any time in addition to the collinear partons there are real emissions with large transverse momentum (compared to $Q$), the soft factor fully factorizes and its value reduces to unity. This is due to the fact that the soft gluons are kinematically overpowered, and do not correlate the collinear parts anymore: then each collinear cluster of partons is totally independent from any other. 
As a consequence, only hard and collinear contributions appear explicitly in the final cross section.

\bigskip

When the 3D-motion of partons is considered, finding a way to properly separate the various contributions becomes a very tough task. 
On the other hand, Transverse Momentum Dependent (TMD) observables show a much richer structure, that may disclose some of the inner properties of hadronization and confinement. 
When the information on transverse motion survives in the final result, the corresponding factorization theorem is usually labeled as ``TMD". 
In such cases, the soft factor does not reduce to unity, and soft gluons have a non-trivial impact on the cross section, as they correlate the collinear parts.
This correlation originates from momentum conservation laws in the transverse direction. In fact, with no real emission carrying large transverse momentum entering into the game, the low transverse momentum components of soft and collinear particles cannot be neglected anymore. 
As a consequence, it is not possible to associate a PDF or a FF to the collinear contributions: parton densities are now related to different and more general objects, known as Transverse Momentum Dependent parton functions, either TMD PDFs or TMD FFs,  depending on whether they refer to an initial or a final state hadron.

\bigskip

The factorization procedure proposed by Collins, Soper and Sterman in the '80s \cite{Collins:1984kg,Collins:1989bt,Collins:1989gx} has become a benchmark for successive approaches designed to provide factorization theorems for hadronic processes.  
The most recent and complete form of such factorization procedure was devised by John Collins in Ref.~\cite{Collins:2011zzd} and in the following we will refer to this scheme as the ``Collins factorization formalism".  
It correctly reproduces the well-known collinear factorization theorems; in addition, and most importantly, it can successfully be applied to develop the TMD factorization theorems for three different processes: $e^+e^-$ annihilation into two back-to-back hadrons, SIDIS and Drell-Yan, in the proper kinematic regions. In Ref. \cite{Boglione:2020cwn} we showed how these processes belong to the same hadron-class, called the ``$2$-h class" as it includes all reactions involving two observed hadron  states.  
An urgent question is whether this factorization scheme can be extended to other processes, which do not belong to the same hadron-class \cite{Boglione:2020cwn,Boglione:2021vug}.
Until the end of 2018, one of the processes that were still lacking of a solid factorization theorem was $e^+e^-$ annihilation into a single hadron, sensitive to the transverse momentum of the detected hadron with respect to the axis of the jet of particles to which it belongs. At the beginning of 2019, the BELLE collaboration at KEK published the results of the measurements of this cross section 
\cite{Seidl:2019jei}, with the transverse momentum of the observed hadron measured with respect to the thrust axis. 
This is one of the measurements which go closer to being a direct observation of a partonic variable, the transverse momentum of the hadron with respect to its parent fragmenting parton. As such, it has indeed triggered a great interest of the high energy physics community, especially among the experts in the phenomenological study of TMD phenomena and factorization~\cite{DAlesio:2020wjq,DAlesio:2021dcx,Boglione:2020cwn,Boglione:2020auc,Boglione:2021vug,Kang:2020yqw,Makris:2020ltr,Gamberg:2021iat}. 
In this paper, we apply the Collins factorization formalism to such a QCD process and we show how to properly factorize it in each of the kinematic region in which it can occur.
In particular, we show that, if the size of the transverse momentum of the detected hadron is neither too large to affect significantly the topological configuration of the final state nor too small to be sensitive to the deflection due to soft radiation, then the cross section of $\epm \to h \, X$ factorizes in the convolution of a partonic cross section, fully computable in perturbation theory, and a TMD Fragmentation Function.
This is a new kind of structure, never encountered before in any known factorization theorem. It is a sort of hybrid of TMD and collinear factorization, and for this reason it will be dented as ``collinear-TMD factorization theorem''.
One of its most relevant features is related to the treatment of rapidity divergences. 
These arise as unregulated infinities into soft and collinear parts and they are due to the approximations introduced by the factorization procedure. Despite the various long-distance contributions are rapidity divergent, the full factorized cross-section is finite. Therefore, the rapidity divergences must cancel in the convolutions of the final result. 
There are many ways to regulate them. In the Collins factorization formalism, they are regulated by tilting the Wilson lines associated to the soft contributions off the lightcone. This operation is totally analogous to the insertion of a sharp rapidity cut-off that prevents the integral on rapitidities to diverge.
Clearly, regardless of the chosen regulator, the final cross section should not depend on it. 
However, for the collinear-TMD factorization theorem devised for $e^+e^- \to h X$ this does not happen. One could claim that this is a symptom of non-consistency of the corresponding factorization theorem (see for instance Ref.~\cite{Makris:2020ltr}). We do not interpret this as a failure of the factorization itself, but rather as a limit of the Collins factorization formalism in this particular process. In fact, as we have shown in Ref. \cite{Boglione:2020auc} and we will discuss in more detail in the following, the rapidity cut-off is intimately related to thrust, $T$, which in the case of $e^+e^- \to hX$ is measured. Hence, as we will show, it acquires a physical meaning, beyond the role of a mere divergence regulator. 

\bigskip

Like other hadronic processes, also $\epm \to h\,X$ can occur in different kinematic regions. In particular, we will show that  there are three of them, each  associated to a different factorization theorem. In particular, when the transverse momentum of the detected hadron is extremely small and sensitive to the deflection caused by the soft radiation, the resulting factorization theorem has a structure very similar to the TMD factorization \cite{Collins:2011zzd}. Instead, when the transverse momentum of the detected hadron is large enough to significantly affect the topology of the final state (and, ultimately, the measured value of thrust) the factorization theorem does not involve a TMD FF anymore, and it shares many of the features of collinear factorization. We will call these two kinematical ranges Region 1 and Region 3, respectively. The intermediate region  discussed in the previous paragraph, called Region 2, is the widest in terms of phase space and also the most interesting from the point of view of the factorization theorems, as it embeds the properties of both TMD and collinear factorization schemes. This matches with the nomenclature introduced in Ref.~\cite{Makris:2020ltr}.

The framework presented in this paper not only includes the well-known TMD factorization theorems developed for $e^+e^-$ annihilation into two back-to-back hadrons, SIDIS and Drell-Yan into a more general context, but it also  extends the investigation of TMD physics beyond these benchmark processes,  by developing proper factorization theorems for $\epm\to h \, X$.  Furthermore, it can potentially be applied to a much wider set of processes, also involving  more than two TMD functions. For these reasons, this approach looks very promising and it might be one of the future keys to fundamental QCD issues.

\bigskip


\section{Kinematic requirements and Region decomposition \label{sec:kin_Reg}}


\bigskip

One of the features that make the $\epm \to h\,X$ cross section, as measured by BELLE~\cite{Seidl:2019jei}, particularly relevant for TMD physics is the determination of the thrust axis, with respect to which the transverse momentum $P_T$ of the detected hadron is measured. The thrust axis is the direction $\vec{n}$ that maximizes the thrust $T$, defined as:
\begin{equation} \label{eq:thrust_def}
T = \dfrac{\sum_i |\vec{P}_{(\text{c.m.}),\,i} \cdot \widehat{n}|}{\sum_i |\vec{P}_{(\text{c.m.}),\,i}|},
\end{equation}
where the sum runs over all the detected particles in the c.m. frame (e.g. the LAB frame).
The variable $T$ describes the topology of the final state and it ranges from $0.5$ to $1$, where the lower limit corresponds to a spherical distribution of final state particles, while the upper limit realizes a pencil-like event.
In the following, the thrust axis will be identified with the axis of the jet to which the hadron $h$ belongs, coinciding with the direction of the fragmenting parton, eventually tilted by soft radiation recoil. 
Notice that the choice of including or leaving out the effects of soft recoiling affects the choice of the proper jet algorithm and, ultimately, the final factorization theorem~\cite{Neill:2016vbi}. Anyway, the identification of the thrust axis with the jet axis
is crucial to relate $\vec{P}_T$ with the transverse momentum $\vec{k}_T$ of the fragmenting parton with respect to the direction of $h$, which is the variable appearing in explicit perturbative computations. In fact, the two vectors are related as:
\begin{align}
\vec{P}_T \left[ 1 + \mathcal{O}\left( \frac{P_T^2}{Q^2} \right) \right] = - z \, \vec{k}_T
\label{eq:PT_vs_kT}
\end{align}
where $z = {P^+}/{k^+}$ is the collinear momentum fraction of the detected hadron with respect to the fragmenting parton.
For a detailed description of the kinematics of this process we refer to Ref.~\cite{Boglione:2020auc}.

Moreover, the cross section is differential also in the fractional energy $z_h$ of the detected hadron, defined as:
\begin{equation} 
\label{eq:zh_def}
z_h = 2 \, \frac{P \cdot q}{q^2} = \frac{P^+}{q^+} = 2 \frac{E_{\text{c.m.}}}{Q}
\end{equation}
where $E_{\text{c.m.}}$ is the energy of the detected hadron in the center of mass frame. 

\bigskip

%
\begin{figure}[t]
  \centering 
\begin{tabular}{c@{\hspace*{25mm}}c}  
      \includegraphics[width=5.5cm]{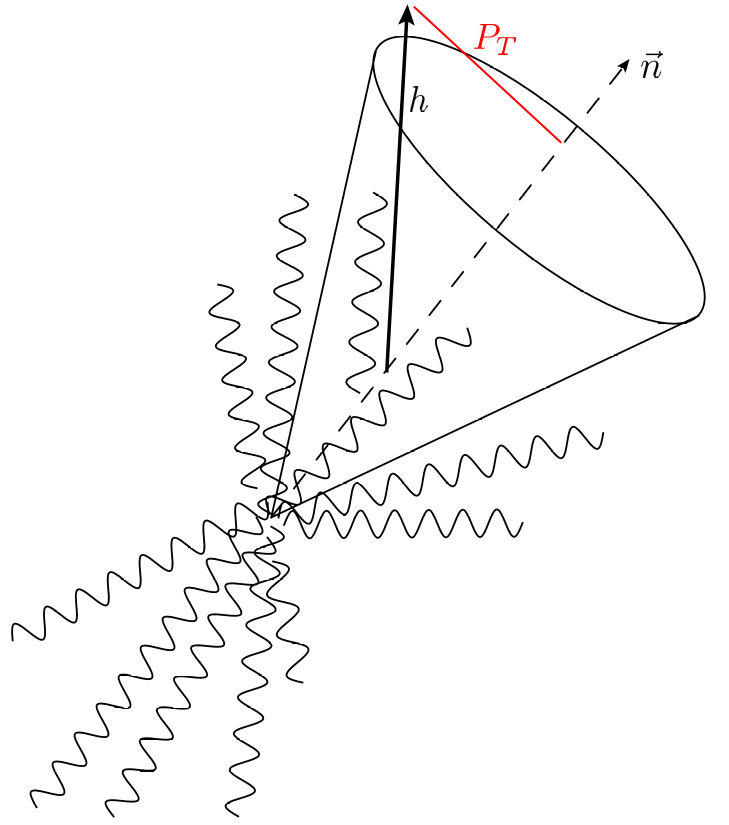}
  &
      \includegraphics[width=5.5cm]{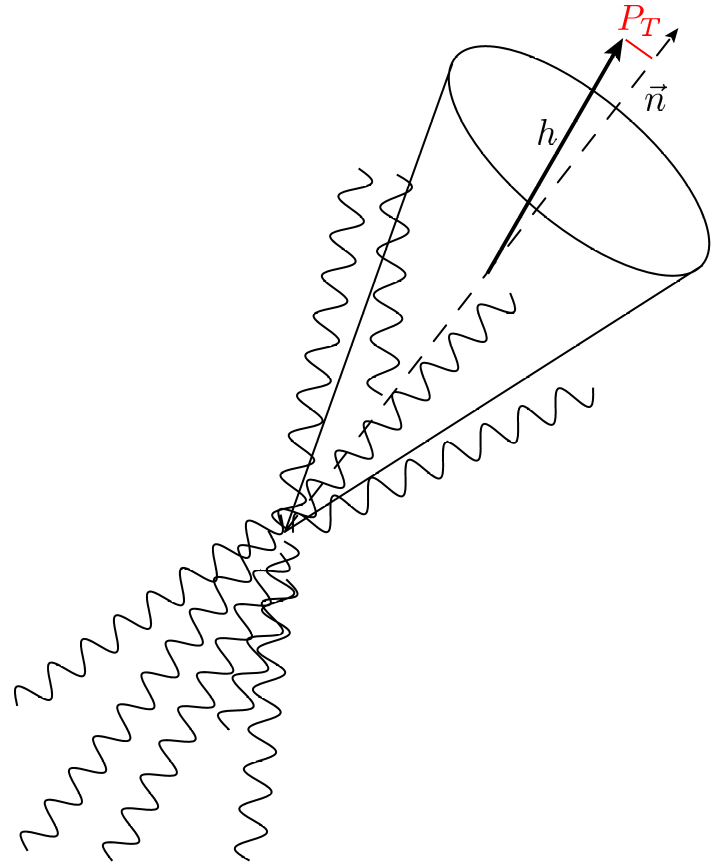}
  \\
  (a) & (b)
\end{tabular}
 \caption{Pictorial representation of $2$-jet configurations in which the kinematic requirements \ref{hyp:1} in (a) and \ref{hyp:2} in (b) do not hold true. In (a) the detected hadron is rather close to the jet boundary, causing a significant spread of the jet which affects the final state configuration, and ultimately the  measured value of $T$. In (b) the detected hadron momentum is very close the thrust axis as its transverse momentum is extremely small. In this case, the soft radiation affects significantly the final measured $P_T$.}
 \label{fig:hyps}
\end{figure}
%
All particles associated to the same jet of the detected hadron, including $h$ itself, are characterized by a small transverse momentum (relative to the thrust axis) and a large and positive rapidity. 
However, this general features are not enough to determine uniquely the kinematics and, ultimately, the proper factorization theorem, associated to the process.
In fact, depending on where $h$ is located within the jet, the underlying kinematic configuration can be 
remarkably different, resulting in distinct factorization theorems.
The most probable scenario occurs when the detected hadron is neither too close to the jet boundary, nor too close to the thrust axis, which are somehow the extreme regions where it can be found. Therefore, these two kinematic requirements can be used to identify such intermediate configuration, to which the majority of experimental data is expected to belong to:
\begin{enumerate}[label=\textbf{H.\arabic*}]
\item \label{hyp:1} 
The detected hadron $h$ is not extremely close to the jet boundary.

If this hypothesis does not hold true, then the detected hadron actively concurs to 
the jet transverse size, affecting the final state configuration, and ultimately the measured value of $T$, Fig.~\ref{fig:hyps}(a). In this case, its transverse momentum $P_T$ is moderately small: enough to consider $h$ as part of the jet but large if compared to the average transverse momenta of the other particles of the jet.
\item \label{hyp:2}
The hadron detected hadron $h$ is not extremely close to the thrust axis.

If this hypothesis does not hold true, Fig.~\ref{fig:hyps}(b), the size of $P_T$ is so small that the soft radiation concurs actively to the transverse deviation of the detected hadron and hence to the measured value of its transverse momentum. In this case, $h$ belongs to the jet mostly because it has a very small transverse momentum rather than because of its large rapidity.
\end{enumerate}
In this paper, we are primarily interested in investigating the kinematic configuration where both \hyp{1} and \hyp{2} hold true, which will be labeled Region 2. The other two regions, corresponding to the hadron $h$ detected very close to the thrust axis or to the jet boundary will be labeled Region 1 and Region 3, respectively. They correspond to the configurations associated to the cases where one of the kinematic requirements above is false. In fact, in Region 1 \hyp{1} is still true but \hyp{2} is false. On the other hand, in Region 3 \hyp{2} keeps its validity but \hyp{1} does not hold true anymore. Notice that the two hypotheses cannot be false together at the same time. The table~\ref{tab:kinRef_hyp} summarizes the relations between the kinematic regions and the validity of the kinematic requirements defined above.
%
\begin{figure}[t]
  \centering 
   \includegraphics[width=5.5cm]{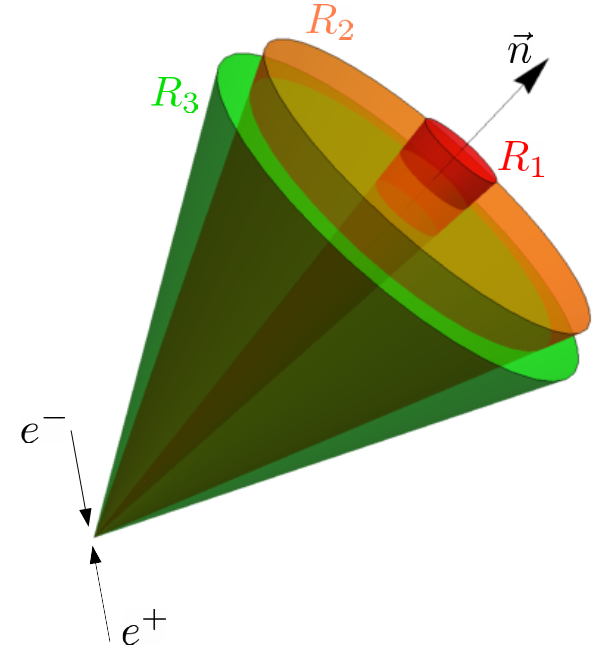}
 \caption{Pictorial representation of the three kinematic regions in the jet in which the hadron $h$ is detected. In Region 1 (red) $h$ is detected very close to the thrust axis $\vec{n}$ and its transverse momentum is so small that it is significantly affected by the deflections caused by soft radiation. In contrast, in Region 3 (green) $h$ is detected close to the boundary of the jet, hence it actively participates to the jet spreading, affecting the final state configuration and therefore the measured value of thrust. The most probable configuration, however, is Region 2 (orange), where $h$ is detected neither extremely close to the jet's axis, nor etremely close to the jet's boundary.}
 \label{fig:jetcone}
\end{figure}
%

The decomposition into kinematic regions discussed above not only allows for an organized and consistent development of proper factorization theorems, but also helps to determine the correct function to use in each case to describe the hadronization process, even without any knowledge of the details of the factorization procedure. 
\begin{table}[ht]
\centering
\begin{tabular}{c || c || c ||}
    {}   & \hyp{1}  & \hyp{2}   \\
 \hline\hline
    $R_1$   &       true        &        false	\\
 \hline\hline
    $R_2$   &       true        &        true 	\\
 \hline\hline
    $R_3$   &       false        &       true 	\\
 \hline
\end{tabular}
\caption{Kinematic regions and initial assumptions.} 
\label{tab:kinRef_hyp}
\end{table}

Consider Region 1. Since 
here the soft radiation must play an important role in the fragmentation of the parton that leads to the detected hadron, the TMD FF alone, being associated to strictly collinear radiation, is not sufficient to describe the physics of this region. A proper description of the hadronization process must necessarily take into account also the non-perturbative content associated to soft gluons.

On the other hand, in Region 3 
the function describing the hadronization process must be explicitly sensitive to the topology of the final state, i.e. to the value of thrust. Again, a TMD FF cannot provide a correct description of the physics of this region, as its non-perturbative content\footnote{Encoded in the functions $M_D$ and $g_K$, see Appendix~\ref{app:review_collins}.} must depend only on variables such as the collinear momentum fraction and transverse momentum, that are associated only to the fragmenting parton and the detected hadron, but oblivious of the rest of the process.

Therefore, the non-perturbative effects of the hadronization process can be described by a TMD FF only in Region 2, where both the kinematic requirements \hyp{1} and \hyp{2} are true. This means that the cross section in this region provides an extremely clean way to access TMD FFs, as in this case they are the only non-perturbative objects that need to be extracted in a phenomenological analysis. This is indeed one of most important reasons for investigating the factorization properties of $\epm \to h\,X$, as the TMD factorization theorems associated to the benchmark processes, besides the issues related to soft radiation~\cite{Boglione:2020cwn,Boglione:2020auc}, always present a convolution between two TMDs, which can be hardly disentangled.
Hence, the study of Region 2 is not only important for single hadro-production in $\epm$ annihilation, but it  also plays a crucial role in future phenomenological studies, where data from different processes will be consistently combined within global analyzes.

\bigskip


\section{General structure of the cross section \label{sec:struct_xs}}


\bigskip

The kinematic requirements discussed in the previous section provide an easy classification criterion for the three kinematic regions characterizing $\epm \to h\,X$. Their implementation into the proof of the corresponding factorization theorems is strictly related to the structure of the cross section, and in particular to the role of its different parts in generating TMD effects. Therefore, in this section we briefly review the main features of the cross section of $\epm \to h\,X$. More details can be found in Ref.~\cite{Boglione:2020cwn}.

\bigskip

The general structure of the cross section of $e^+e^- \to h\,X$ is given by the Lorentz contraction of the leptonic tensor $L_{\mu\,\nu}$,  corresponding to the initial state configuration, with the hadronic tensor $W^{\mu\,\nu}_h$, 
which describes the strong-interaction contribution to the process. Labeling $P$ the momentum of the detected hadron, we have:
\begin{equation} \label{eq:epm_crosssec}
\frac{d \sigma}{\left(\lips{P}\right)_{\text{LAB}} \, d^2 \vec{P}_T \, dT} = \frac{4 \alpha^2}{Q^6} L_{\mu \, \nu} 
W^{\mu \, \nu}_h.
\end{equation}
%
%
%
The leptonic tensor is defined as the lowest order of the electromagnetic vertex $\epm \to \gamma^\star$ with unpolarized leptons, and it is given by:
\begin{align}
L^{\mu\nu} = 
l_1^\mu \, l_2^\nu +
l_2^\mu \, l_1^\nu -
g^{\mu \nu} \, l_1 \cdot l_2,
\label{eq:leptonic_tensor_def}
\end{align}
where $l_1$ and $l_2$ are the momenta of the electron and the positron, respectively. 

\bigskip

The hadronic tensor $W^{\mu \, \nu}_h$ depends on the momentum $P$ of the outgoing hadron and on the momentum $q$ of the boson connecting the initial with the final state.
Furthermore, it encodes the whole dependence on the thrust $T$, as it describes the final state contribution to the process.
Its formal definition is:
\begin{align}
&W^{\mu \, \nu}_h(P,\,q,\,T) = 
\notag \\
&=
4 \pi^3 \, \sum_X \delta \left( p_X +P - q \right) \, \delta\left(  T - T_{\text{def.}}(p_X,\,P)\right)
\langle 0 | \, j^\mu(0) | \,P,\,X,\,\mbox{out} \, \rangle \,
 \langle P,\,X,\,\mbox{out} |j^\nu(0) | \,0 \rangle = 
\notag \\
&= 
\frac{1}{4\pi} \, \sum_X \, \int d^4 z\,
e^{i q \cdot z} \delta\left(  T - T_{\text{def.}}(p_X,\,P)\right)
\langle 0 | \,j^\mu\left(z/2\right) | \,P,\,X,\,\mbox{out} \,\rangle \,
\langle P,\,X,\,\mbox{out} | j^\nu\left(-z/2\right) | \,0 \rangle,
\label{eq:current_def_W}
\end{align}
where $j^\mu$ are the electromagnetic currents for the hadronic fields and 
$T_{\text{def.}}$ corresponds to the thrust as defined in Eq.~\eqref{eq:thrust_def}.
The final state is represented as $|P,\,X,\,\mbox{out} \, \rangle$ and corresponds to the topology associated with the thrust value and to the measured transverse momentum of the hadron with respect to the thrust axis.
The factor $1/(4\pi)$ in the last line coincides with the normalization 
used in Ref.~\cite{Collins:2011zzd}. The definition of Eq.~\eqref{eq:current_def_W} is hardly usable for explicit computational purposes. 
For this reason, it is useful to decompose the hadronic tensor in terms of Lorentz-invariant structure functions:
\begin{align} 
W^{\mu \, \nu}_h &= 
\left(-g^{\mu \, \nu} + \frac{q^\mu q^\nu}{q^2} \right) F_{1,\,h} \, +
\frac{\left( P^\mu - q^\mu \frac{P \cdot q}{q^2} \right) \left( P^\nu - q^\nu \frac{P \cdot q}{q^2} \right)}{P \cdot q} \,F_{2,\,h}.
\label{eq:had_tens_structfun}
\end{align}
Then, the projections of $W^{\mu \, \nu}_h $ onto its relevant Lorentz tensors are:
\begin{subequations}
\label{eq:W_projections}
\begin{align}
&-g_{\mu\,\nu}W^{\mu \, \nu}_h= 
3 F_{1,\,h} + \frac{z_h}{2} \, F_{2,\,h} + 
\mathcal{O} \left( \frac{M^2}{Q^2} \right); 
\label{eq:W_proj1_1} \\
&\frac{P_\mu P_\nu}{Q^2} W^{\mu \, \nu}_h = 
\left( \frac{z_h}{2} \right)^2 \, \left[ 
F_{1,\,h} + 
\frac{z_h}{2} \, F_{2,\,h}\right] + 
\mathcal{O} \left( \frac{M^2}{Q^2} \right),
\label{eq:W_proj2_1}
\end{align}
\end{subequations}
much easier to compute explicitly in perturbation theory.

\bigskip

Given the Lorentz structures of the leptonic tensor, Eq.~\eqref{eq:leptonic_tensor_def}, and of the hadronic tensor, Eq.~\eqref{eq:had_tens_structfun}, the full differential cross section can easily be decomposed into a transverse (T) and a longitudinal (L) contribution
%
\begin{align}
&\frac{d \sigma}{dz_h \, d^2\vec{P}_T \, dT} = 
\frac{d \sigma_T}{dz_h \, d^2\vec{P}_T \, dT} +
\frac{d \sigma_L}{dz_h \, d^2\vec{P}_T \, dT} .
\label{eq:xs_general_structfun}
\end{align}
Moreover, by exploiting Eqs.~\eqref{eq:W_projections}, the transverse and the longitudinal components of the cross section can be related to the structure functions of the hadronic tensor:
\begin{subequations}
\label{eq:xs_LT_struct}
\begin{align}
&\frac{1}{\sigma_B} \, 
\frac{d \sigma_T}{dz_h \, d^2\vec{P}_T \, dT} = 
z_h \, F_{1,\,h}(z_h, \, \vec{P}_T, \, T); 
\label{eq:xs_T}\\
&\frac{1}{\sigma_B} \, 
\frac{d \sigma_L}{dz_h \, d^2\vec{P}_T \, dT} = 
\frac{z_h}{2} \, \left(
F_{1,\,h}(z_h, \, \vec{P}_T, \, T) +
 \frac{z_h}{2} \, F_{2,\,h}(z_h, \, \vec{P}_T, \, T)
 \right),
\label{eq:xs_L}
\end{align}
\end{subequations}
where $\sigma_B$ is the Born cross section:
\begin{align}
\sigma_B = \frac{4 \pi \alpha^2}{3 Q^2}.
\label{eq:born_xs}
\end{align}
An interesting case occurs when the projection of the hadronic tensor with respect to $P_\mu \, P_\nu$, 
Eq.~\eqref{eq:W_proj2_1},
is zero (or it is so small that can be neglected). 
In this case, the two structure functions are not independent anymore and $F_{2,\,h} = -\frac{2}{z_h} \, F_{1,\,h} $.
As a consequence, the hadronic tensor can be written as:
\begin{align}
W^{\mu \, \nu}_{h;\,(T)}=
H_T^{\mu\nu} \,
F_{1,\,h},
\label{eq:k2prog_null_W}
\end{align}
where the transverse tensor is defined as:
\begin{align}
H_T^{\mu\nu} = 
-g^{\mu \nu} + 
\frac{P^\mu q^\nu + P^\nu q^\mu}
{P \cdot q} -
q^2 \, \frac{P^\mu P^\nu}
{\left(P \cdot q\right)^2}
\;.
\label{eq:HT_tensor}
\end{align}
Furthermore, the longitudinal cross section vanishes. Hence, in this case the detection of a hadron perpendicular to the beam axis is suppressed.

\bigskip

%
\begin{figure}[t]
\centering 
\includegraphics[width=7.5cm]{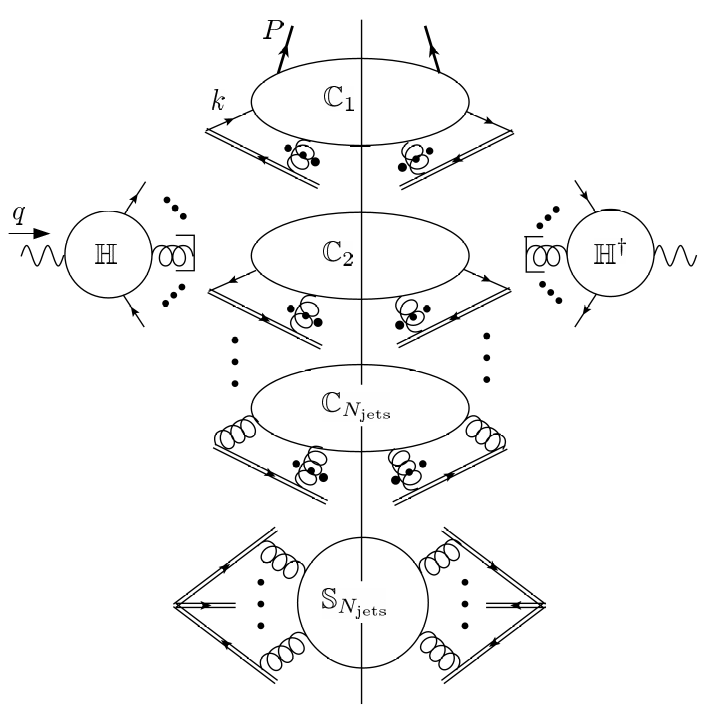}
\caption{Leading momentum regions for the hadronic tensor $W^{\mu \, \nu}_h$. The hard subgraphs, one on each side of the final state cut, are labeled by $\mathbb{H}$. The soft subgraph, labeled by $\mathbb{S}_{N_\text{jets}}$, represent the contribution of the soft gluons. It has as many Wilson lines as the number of jets in the final state. The collinear subgraphs are labeled by $\mathbb{C}_i$, for $i=1,\dots N_\text{jets}$.}
\label{fig:epm_regions}
\end{figure}
%
The determination of the hadronic structure functions $F_{1,\,h}$ and $F_{2,\,h}$ is obtained by applying the factorization procedure to the hadronic tensor.
The topology of the final state is determined by the value of thrust $T$: the closer $T$ to $1$, the lower the number $N_{\text{jets}}$ of observed jets. The minimum $N_{\text{jets}} = 2$ corresponds to a $2$-jet configuration while the limit $N_{\text{jets}} \to \infty$ is associated with an homogeneous spherical distribution of particles.
The general structure of $W^{\mu \, \nu}_h$ in terms of its hard, soft and collinear contributions is represented in Fig.~\ref{fig:epm_regions}, where each blob corresponds to a leading momentum region. 
The delta on thrust in Eq.~\eqref{eq:current_def_W} introduces a correlation among the total collinear and soft momenta flowing into the corresponding subgraph. As a consequence, each blob in Fig.~\ref{fig:epm_regions} acquires a dependence on $T$. 
The hard subgraph $\mathbb{H}$ represents the production of $N_{\text{jets}}$ partons dressed with all the required far off-shell virtual corrections. Its dependence on thrust sets the value of $T$ for the topology considered i.e. $T=1$ for $N_{\text{jets}}=2$, $T\sim2/3$ for $N_{\text{jets}}=3$ and so on.
Each parton exiting from the hard subgraph generates a collinear factor which results in a jet of particles. Therefore there are, in total,  $N_{\text{jets}}$ collinear factors $\coll_i$, all equipped with the proper subtraction of the soft-collinear overlapping terms as in Eq.~\eqref{eq:sub_coll}. Furthermore, there is a soft subgraph $\mathbb{S}_{N_{\text{jets}}}$ that correlates the collinear contributions. It has as many Wilson lines as $N_{\text{jets}}$\footnote{Notice that  $\mathbb{S}_{N_{\text{jets}}}$ is a matrix in color space. All its color indices are contracted with the hard subgraphs and the whole hadronic tensor is colorless, as required.}. 
The thrust dependence encoded into the soft and collinear contributions gives the deviation from the value associated to the hard scattering that reproduces the observed topology of the final state. For instance, in the $2$-jet case $T$ is close, but not exactly equal, to $1$; this value is obtained only after considering the proper contributions of soft and collinear emissions.
The pictorial representation of  Fig.~\ref{fig:epm_regions} corresponds to the following equation:
\begin{align}
&W^{\mu \, \nu}_h(z_h,\vec{P}_T,\,T) = 
\sum_{N_{\text{jets}} \geq 2} \, 
\sum_{j_1} \int \kmeas{k}{1} \, 
\sum_{j_2} \int \kmeas{k}{2} 
\prod_{\alpha = 3}^{N_{\text{jets}}} \,  
\int \kmeas{k}{\alpha} \coll_{\alpha}(k_\alpha)_{j_\alpha} 
\notag \\
&\times
\text{Tr}_D \left \{ 
\mathrm{P}_1 \coll_{1}(k_1,\,P)_{j_1} \overline{\mathrm{P}}_1 
\mathbb{H}_{j_1,\dots j_{N_{\text{jets}}}}^\mu(\widehat{k}_1,\dots\widehat{k}_{N_{\text{jets}}}) \,
\mathrm{P}_2 \coll_{2}(k_2)_{j_2} \overline{\mathrm{P}}_2 \,
(\mathbb{H}^\dagger)_{j_1\, \dots j_{N_{\text{jets}}}}^\nu
(\widehat{k}_1,\dots \widehat{k}_{N_{\text{jets}}})
\right\} 
\notag \\
&\times
\int \kmeas{k}{S} \mathbb{S}_{N_{\text{jets}};\,j_1\, \dots j_{N_{\text{jets}}}}(k_S) \, 
\delta(q - k_1 - k_2 - \sum_{\alpha} k_\alpha - k_S)\,
\notag \\
&\times
\delta\left( \vec{P}_T \left[ 1 + \mathcal{O}\left( \frac{P_T^2}{Q^2} \right) \right] + \frac{P^+}{k_1^+} 
\vec{k}_{T}( \vec{k}_{1,\,T}, \,\vec{k}_{S,\,T} ) \right) \,
\delta\left(  T - T_{\text{def.}}(k_1,\,k_2,\dots,\,k_{N_{\text{jets}}},\,k_S)\right) ,
\label{eq:W_prelim}
\end{align}
In the last line of the previous equation, the first delta function sets the relation between the measured transverse momentum $\vec{P}_T$  of the detected hadron and the transverse momentum $\vec{k}_T$ of the fragmenting parton, according to Eq.~\eqref{eq:PT_vs_kT}. Both of them are considered with respect to the thrust axis, although in different frames. Notice that $\vec{k}_T$ does not necessarily coincide with the total transverse momentum $\vec{k}_{1,T}$ entering into the collinear factor $\coll_1$ associated to the detected hadron. In fact, it can also depend on the total soft transverse momentum $\vec{k}_{S,T}$, when the direction of the thrust axis is modified by soft recoiling.

\bigskip


\subsection{TMD-relevance \label{sec:tmd_rel}}


\bigskip

Each individual term contributing to the hadronic tensor can be expanded in series of $\alpha_S$ and approached within perturbative QCD. In fact, the decomposition in momentum regions presented in Eq.~\eqref{eq:W_prelim} can be performed explicitly, order by order, by applying the kinematic approximators defined in Ref.~\cite{Collins:2011zzd} and reviewed in Appendix~\ref{app:review_collins}. 
This kind of ``bottom-up" approach to factorization, based on perturbative QCD alone, makes the steps that lead to a factorization theorem extremely clear, as the various contributions can be readily disentangled and made more transparent, order by order in pQCD. 
The downside of this methodology is that the final result, deduced by fix order computations and then generalized to all orders, must be properly supplemented with the non-perturbative content of soft and collinear contributions, associated to TMD effects.
Moreover, perturbative computations may be remarkably cumbersome, especially when the formalism has to be adapted to include one additional (observable) variable, in this case the thrust $T$. 
Finally, the generalization to all orders is a potentially dangerous operation, as new effects may arise at higher orders in perturbation theory. 
Despite this disadvantages, explicit perturbative computations are essential for understanding the structure of factorized cross sections, and in this paper we will extensively use them to derive consistent factorization theorems.   
Most importantly, not only the kinematic requirements \hyp{1} and \hyp{2} can easily be implemented in perturbative computations, they also allow to classify the relevance of the contributions singled out by the factorization procedure to the study of TMD effects.

\bigskip

We will refer to the partonic counterpart of the hadronic tensor as the {\emph{partonic tensor}} $\widehat{W}_j^{\mu \nu}$. It describes the process $\gamma^\star \to j\,X$ where $j$ represents the fragmenting parton. 
In general, it could be a fermion of flavor $f$ or a gluon $g$, but in a $2$-jet topology it can only be a fermion. In fact, of the two jets observed in the final state, one is initiated by the quark, the other by the antiquark.
Any further jet would be associated to an extra power of $\alpha_S$. In this paper we are mainly interested in the $2$-jet topology, which is the most probable configuration. 
It is rather simple to derive the partonic analogue of the hadronic tensor from its expression in Eq.~\eqref{eq:W_prelim}. This can be obtained by making the following replacements:
\begin{itemize}
\item The fragmenting parton plays the role of the detected hadron, hence $h \mapsto j$, 
$P \mapsto k_1$ 
and $z_h \mapsto z = {k_1^+}/{q^+}$. 
\item The radiation collinear to $j$ is generated by a parton of momentum $k_1'$, i.e. $k_1 \mapsto k_1'$. Then the collinear momentum fraction becomes $\widehat{z} = {P^+}/{k_1^+} \mapsto \rho = {k_1^+}/{k_1'^+}$.
\item The role of $\vec{P}_T$ is played by the transverse momentum $\vec{k}_T$ of the fragmenting parton with respect to the thrust axis.
Hence we have the replacement $\delta\left(\vec{P}_T + \widehat{z} \; \vec{k}_T\right) \mapsto \delta\left(\vec{k}_T - \vec{k}'_T\right)$, where, in general, $\vec{k}'_T$ is a function not only of $\vec{k}_{1,T}'$, but also on $\vec{k}_{S,T}$, depending on whether the soft-recoiling is relevant or not for the kinematics considered.
Notice that in this case both $\vec{k}_T$ and $\vec{k}'_T$ are considered in the same frame.
\end{itemize}
The implementation of the kinematic requirements \hyp{1} and \hyp{2} into the perturbative computation is strictly connected to the determination of the transverse momentum $\vec{k}_T'$ in each of the leading momentum regions: hard, soft and collinear.
Since each of such terms returns a picture of the whole process in a specific kinematic approximation, they all somehow depend on  the transverse momentum of the fragmenting parton. Sometimes such dependence is trivial, as for the case of backward radiation in a $2$-jet final state topology, and the corresponding momentum region is irrelevant for TMD effects. In other cases, the relevance for TMD physics depends on whether the considered term contributes significantly to the transverse deflection of the detected hadron from the thrust axis.
In practice, when a contribution is ``TMD-relevant" the transverse momentum $\vec{k}_T'$ is non-zero and hence it will depend explicitly on $\vec{k}_T$. Then it must be considered in the Fourier conjugate space of $\vec{k}_T$, as this is the natural framework in which TMDs and soft factors are defined in terms of field operators.
Notice that the contributions of the momentum region corresponding to particles moving collinearly to the fragmenting parton are always TMD-relevant, as they embody the core and essence of the TMD effects. 
On the contrary, if a term is ``TMD-irrelevant" then the transverse momentum $\vec{k}_T'$ vanishes and the Fourier transform is reduced to an integration over the entire spectrum of $\vec{k}_T$, washing out the whole information on transverse momentum. This indeed holds for virtual contributions and also for all terms associated to radiation collinear to other direction than the thrust axis. All these configurations do not play any role in the transverse deflection of the fragmenting parton from the thrust axis.
Therefore, the question of being TMD-relevant or TMD-irrelevant is only pertinent to soft and soft-collinear 
radiations. 

In particular, in Region 1 the soft radiation plays an active role in generating TMD effects, as the deflection of the detected hadron from the thrust axis is significantly affected by soft gluons. Hence in this case the soft approximation will contribute to the final result as a TMD-relevant term.
Clearly, this consideration only refers to soft gluons radiated in the same portion of space occupied by the jet in which the hadron $h$ is detected.
In the $2$-jet case this is one of the two-hemispheres defined by the thrust axis. 
Moreover, if soft radiation is TMD-relevant, then also soft-collinear gluons must be so.

The same cannot be stated for Region 2, where soft radiation does not contribute to the transverse momentum of the detected hadron. However, this does not exclude the possibility that soft-collinear gluons can generate significant TMD effect. Therefore, in this kinematic region the soft approximation produces a TMD-irrelevant term but soft-collinear contributions are instead TMD-relevant.

Finally, in Region 3 the transverse momentum of the detected hadron is too large to be affected either by soft and by soft-collinear radiations. Therefore, in this case both these contributions are associated to TMD-irrelevant terms.
The scheme presented in Table~\ref{tab:kinRef_tmdrel} summarizes this discussion.
\begin{table}[h]
\centering
\begin{tabular}{c || c || c || c ||}
    {}   & soft*  & soft-collinear & collinear   \\
 \hline\hline
    $R_1$   &   \cellcolor{blue!25}TMD-relevant   &  \cellcolor{blue!25}TMD-relevant & \cellcolor{blue!25}TMD-relevant	\\
 \hline\hline
    $R_2$   &   TMD-irrelevant   &  \cellcolor{blue!25}TMD-relevant & \cellcolor{blue!25}TMD-relevant	\\
 \hline\hline
    $R_3$   &   TMD-irrelevant   &  TMD-irrelevant & \cellcolor{blue!25}TMD-relevant	\\
 \hline
\end{tabular}
\caption{Kinematic regions and TMD-relevance. The symbol * reminds that the soft approximation in this case only refers to gluons radiated in the same portion of space occupied by the jet in which is detected the hadron $h$.} 
\label{tab:kinRef_tmdrel}
\end{table}
In this paper, we will refer to TMD-relevant (Fourier transformed) quantities as ``factors" indicating them with capital Greek letters, unless specified differently. TMD-irrelevant  quantities will be referred to as ``functions" and, if not stated differently, we will indicate them with capital italics Latin letters. The following scheme summarizes the notation:
\begin{subequations}
\label{eq:notation}
\begin{align}
&\text{TMD-relevant} \longleftrightarrow \FT \longleftrightarrow \substack{\mbox{factor}\\ \mbox{(capital Greek letters)}} \notag \\
&\text{TMD-irrelevant} \longleftrightarrow \int d^{2-2\eps}\vec{k}_T \longleftrightarrow  \substack{\mbox{function}\\ \mbox{(capital italics Latin letters)}}\notag
\end{align}
\end{subequations}

We will use the variable $\tau = 1-T$ instead of $T$, as it appears naturally in a $2$-jet limit.
We will denote the two hemispheres defined by the thrust axis as $\mathrm{S}_A$ and $\mathrm{S}_B$, where the first is the hemisphere in which hadron $h$ is detected. In the case of soft radiation, the particles can be emitted either in $\mathrm{S}_A$ or in $\mathrm{S}_B$ with the same probability. Therefore, the soft approximators are equipped with a further label ``$+$" or ``$-$" as an indication of the hemisphere in which the particle is emitted. Finally, the subgraphs singled out by the action of the approximators will be labeled by ``A", ``B", ``S" or ``H", depending on whether the involved particles are collinear to the fragmenting parton (A), collinear to the backward direction (B), soft (S) or hard, i.e. far off- shell (H). 

\bigskip


\subsection{Partonic tensor structure to Next to Leading Order \label{sec:What_NLO}}


\bigskip

In this section, we consider the structure of the partonic tensor up to next to leading order (NLO) for a $2$-jet final state topology. We will consider the fragmentation of a quark of flavor $f$.

\bigskip

The lowest order (LO) results was explicitly computed in Ref.~\cite{Boglione:2020auc}. Let us just recall that it corresponds to an exact, pencil-like, final state configuration. Therefore the dependence on is simply trivially given by $\delta(\tau)$.
Moreover, to LO the total transverse momentum of the radiation collinear to the fragmenting quark is trivially zero, as there is no radiation at all. This results in a delta function $\delta(\vec{k}_T)$ in transverse momentum space.
We have:
\begin{align}
&\widehat{W}_f^{\mu\nu\,\;[0]}(z,\,\tau,\,\vec{k}_T) =
\widehat{H}_T^{\mu \nu}\,
\widehat{F}_{1,\,f}^{[0]}(z,\,\tau,\vec{k}_T),
\label{eq:What_lo}
\end{align}
where the transverse tensor $\widehat{H}_T^{\mu \nu}$ is defined as in Eq.~\eqref{eq:HT_tensor}, but with $P \to k_1$, while the LO partonic structure function $\widehat{F}_{1,\,f}^{[0]}$ in transverse momentum space is given by
\begin{align}
\label{eq:F1_LO}
\widehat{F}_{1,\,f}^{[0]}(z,\,\tau,\,\vec{k}_T) = 
e_f^2 \,  N_C (1-\epsilon) \,
\delta \left( 1 - z \right)\,\delta(\tau) \, \delta(\vec{k}_T),
\end{align}
with $z = {k_1^+}/{q^+}$ the partonic version of $z_h$. 

%
\begin{figure}[t]
  \centering 
\begin{tabular}{c@{\hspace*{8mm}}c@{\hspace*{6mm}}c}  
      \includegraphics[width=3cm]{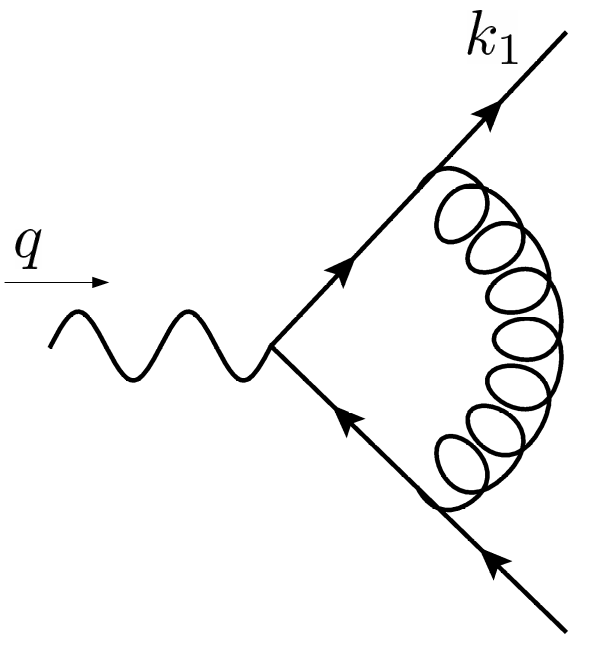}
  &
      \includegraphics[width=3.5cm]{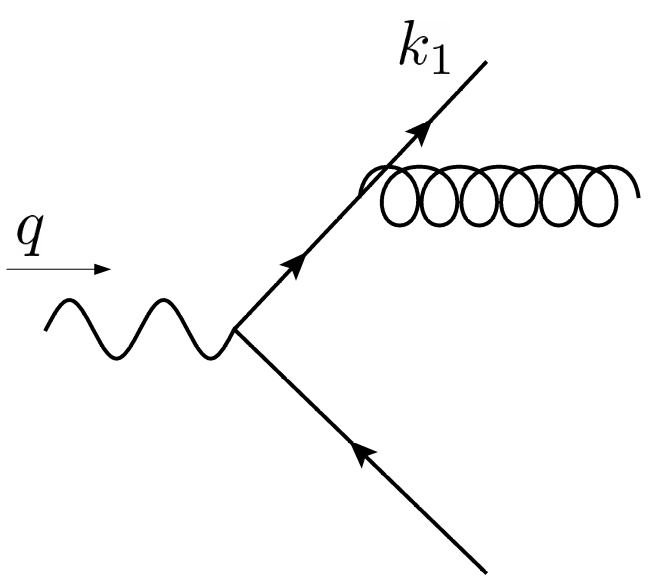}
  &
      \includegraphics[width=3.5cm]{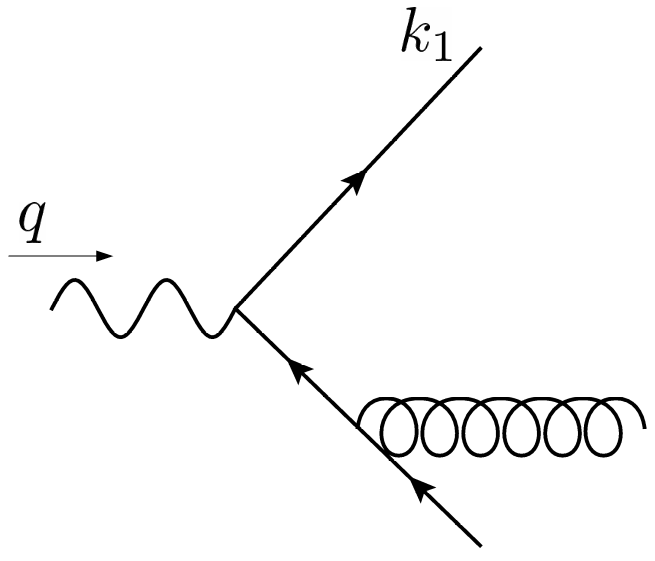}
  \\
  (a) & (b) & (c)
\end{tabular}
 \caption{The next lowest order Feynman graphs contributing to the partonic tensor for a $2$-jet topology.}
 \label{fig:nlo}
\end{figure}
%

To NLO there is a single gluon radiated, which can be either virtual or real. The corresponding Feynman diagrams are represented in Fig.~\ref{fig:nlo}.
We will start by considering the contribution of the virtual radiation.
The squared matrix element associated to this configuration is:
\begin{align}
&M_{f,\,V}^{\mu\,\nu}{}^{\;[1]}(\epsilon;\,\mu) = 
\left(\begin{gathered}
\includegraphics[width=4.3cm]{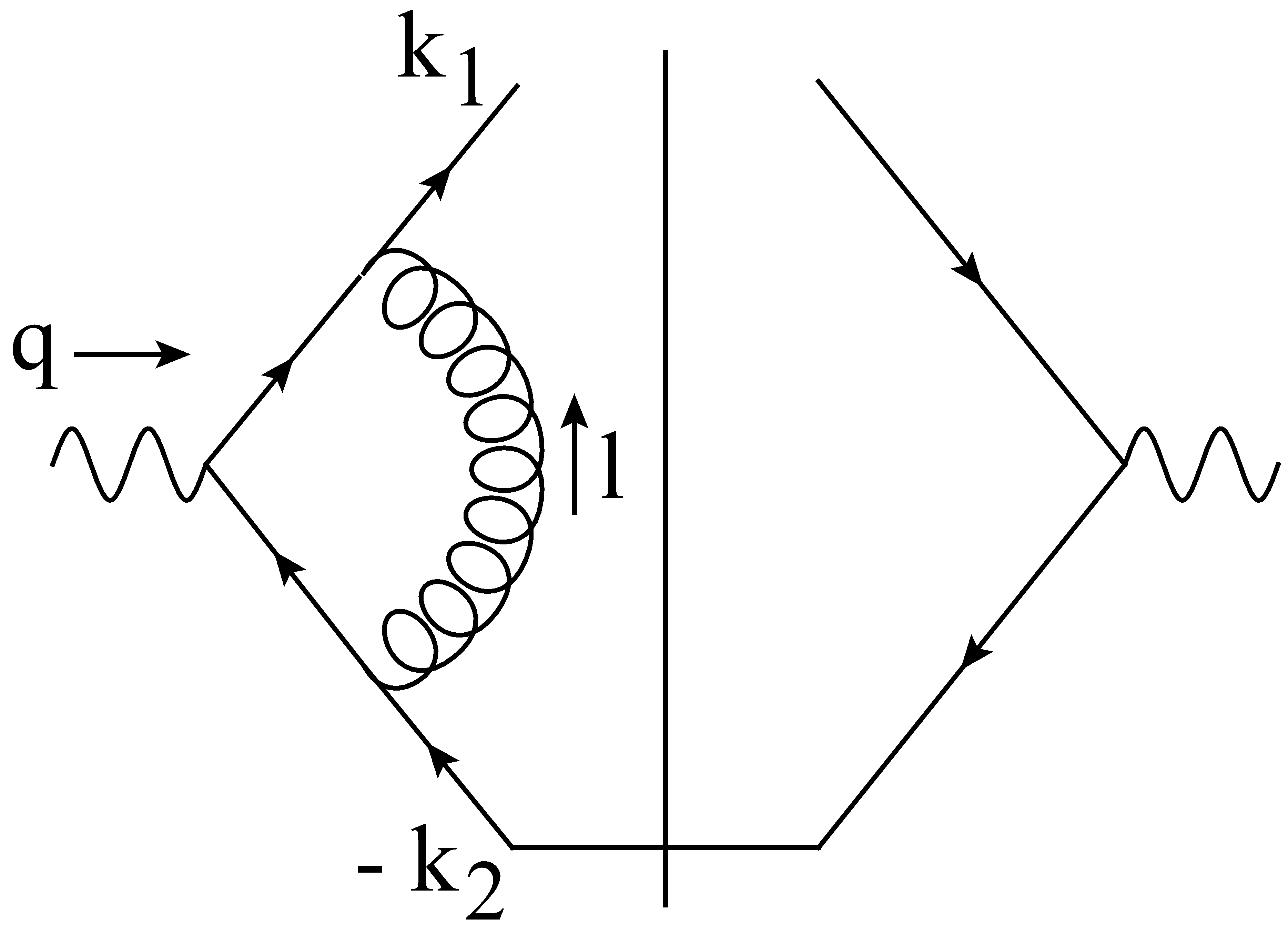}
\end{gathered}
+ h.c.\right)^{\mu \nu}.
\label{eq:M_virtual}
\end{align}
Regarding thrust and transverse momentum all the previous considerations for the LO case calculation hold unchanged for NLO. In fact, for the virtual radiation the NLO final state configuration is pencil-like and there are no TMD effects associated to virtual radiation. In other words, the virtual contribution is always TMD-irrelevant. 
The starting point for applying the factorization procedure is the following non-approximate partonic tensor:
\begin{align}
&\widehat{W}_{f,\,\text{V}}^{\mu \nu, \,[1]}\left(\eps;\,z , \tau ,  \vec{k}_T\right) =
\frac{1}{4\pi} \,
\int \frac{d^D k_2}{(2\pi)^{D}} \, 
M_{f,\,V}^{\mu\,\nu}{}^{\;[1]}(\epsilon;\,\mu)\,
(2\pi)^{D} \, 
\delta^{(D)}\left( q - k_1 - k_2 \right) \,
\delta(\tau)\,\delta(\vec{k}_T).
\label{eq:What_nlo_V}
\end{align}
The virtual gluon can be hard (far off-shell), soft or collinear to the fermionic lines. In full-dimensional regularization the only non-vanishing contribution is associated to the hard momentum region\footnote{The other momentum regions give vanishing contributions but they can be used to fix the necessary UV counterterms for canceling the UV divergences associated to the real emission terms.}, obtained through the action of the approximator $T_H$.
Therefore:
\begin{align}
\label{eq:1loop_xsfV_structure}
&\widehat{W}_{f,\,\text{virtual}}^{\mu \nu, \,[1]}\left(\eps;\,z , \tau ,  \vec{k}_T\right) = 
T_H \left[ \widehat{W}_{f,\,\text{V}}^{\mu \nu, \,[1]}\left(\eps;\,z , \tau ,  \vec{k}_T\right) \right] +  
\substack{\mbox{power suppressed}\\\mbox{corrections}}.
\end{align}
The approximator $T_H$, applied to Eq.~\eqref{eq:What_nlo_V}, only acts on the squared matrix element. A standard result is:
\begin{align}
&T_H \left[
M_{f,\,V}^{\mu\,\nu}{}^{\;[1]}(\epsilon;\,\mu)
\right] 
= 
\notag \\
&\quad= 
i \, e_f^2 \, g^2 \, \mu^{2 \epsilon} \, C_F \, N_C \,
\int 
\frac{d^{4-2\epsilon} \, l}
{(2 \pi)^{4-2\epsilon}} \,
\frac{\mbox{Tr} \left \{ 
\slashed{k}_1 \, \gamma^\alpha \, (\slashed{k}_1 - \slashed{l}) \,
\gamma^\mu \, (\slashed{k}_2 + \slashed{l}) \, \gamma_\alpha \,
\slashed{k}_2 \, \gamma^\nu
\right \}}
{\left[ (k_1 - l)^2 + i \, 0 \right] \, 
\left[ (k_2 + l)^2 + i \, 0 \right] \,
\left[ l^2 + i \, 0 \right]} 
+ h.c.=
\notag \\
&\quad=
M_f^{\mu\,\nu}{}^{\;[0]} \,
V^{[1]}(\epsilon;\,{\mu}/{Q}).
\label{eq:M_V_1}
\end{align}
The last step is obtained by decomposing the Dirac structure in its scalar, vector and tensor parts, by using momentum conservation and the Passarino-Veltman reduction formula~\cite{Passarino:1978jh}.
The 1-loop vertex function $V^{[1]}$ is given by:
\begin{align}
&\frac{\alpha_S}{4 \pi}
V^{[1]}(\epsilon;\,{\mu}/{Q}) = 
\notag\\
&=
-\frac{\alpha_S}{4\pi} \,2\, C_F \, 
S_\epsilon \,
\left[ 
\frac{2}{\epsilon^2} +
\frac{2}{\epsilon} \,
\left( 
\frac{3}{2}+\log{\frac{\mu^2}{Q^2}}
\right)+
8 - \pi^2 + 3\log{\frac{\mu^2}{Q^2}} +
\left(\log{\frac{\mu^2}{Q^2}}\right)^2\,
\right] + 
\mathcal{O}(\eps).
\label{eq:vertex_1loop_2}
\end{align}
Finally the virtual gluon contribution at 1-loop can be represented as:
\begin{align}
\label{eq:1loop_TH_structure}
&\widetilde{\widehat{W}}_{f,\,\text{virtual}}^{\mu \nu, \,[1]}\left(\eps;\,z , \tau ,  b_T\right) = 
\int_z^1 \frac{d \rho}{\rho} 
{}^{\star}\widehat{W}^{\mu \nu,\,[0]}_f ({z}/{\rho})  
\delta (1-\rho ) \delta(\tau) \, V^{[1]}(\eps,\,{\mu}/{Q}) +  
\substack{\mbox{power suppressed}\\\mbox{corrections}},
\end{align}
where${}^{\star}W^{\mu \nu,\,[0]}_f$ is the partonic tensor at LO, computed in Eq.~\eqref{eq:What_lo}, but considered without its (trivial) dependence on thrust and transverse momentum. The previous result holds in
$b_T$-space. In fact, the Fourier transform is trivial, as TMD-irrelevant contributions are proportional to $\delta(\vec{k}_T)$. 

\bigskip

When the emitted gluon is real, the squared matrix element is given by:
\begin{align}
&M_{f, R}^{\mu\nu;\,[1]}(\epsilon; \,\mu, \,\{ k_i \}) = 
\notag \\
&=
\left(
\begin{gathered}
\includegraphics[width=3.8cm]{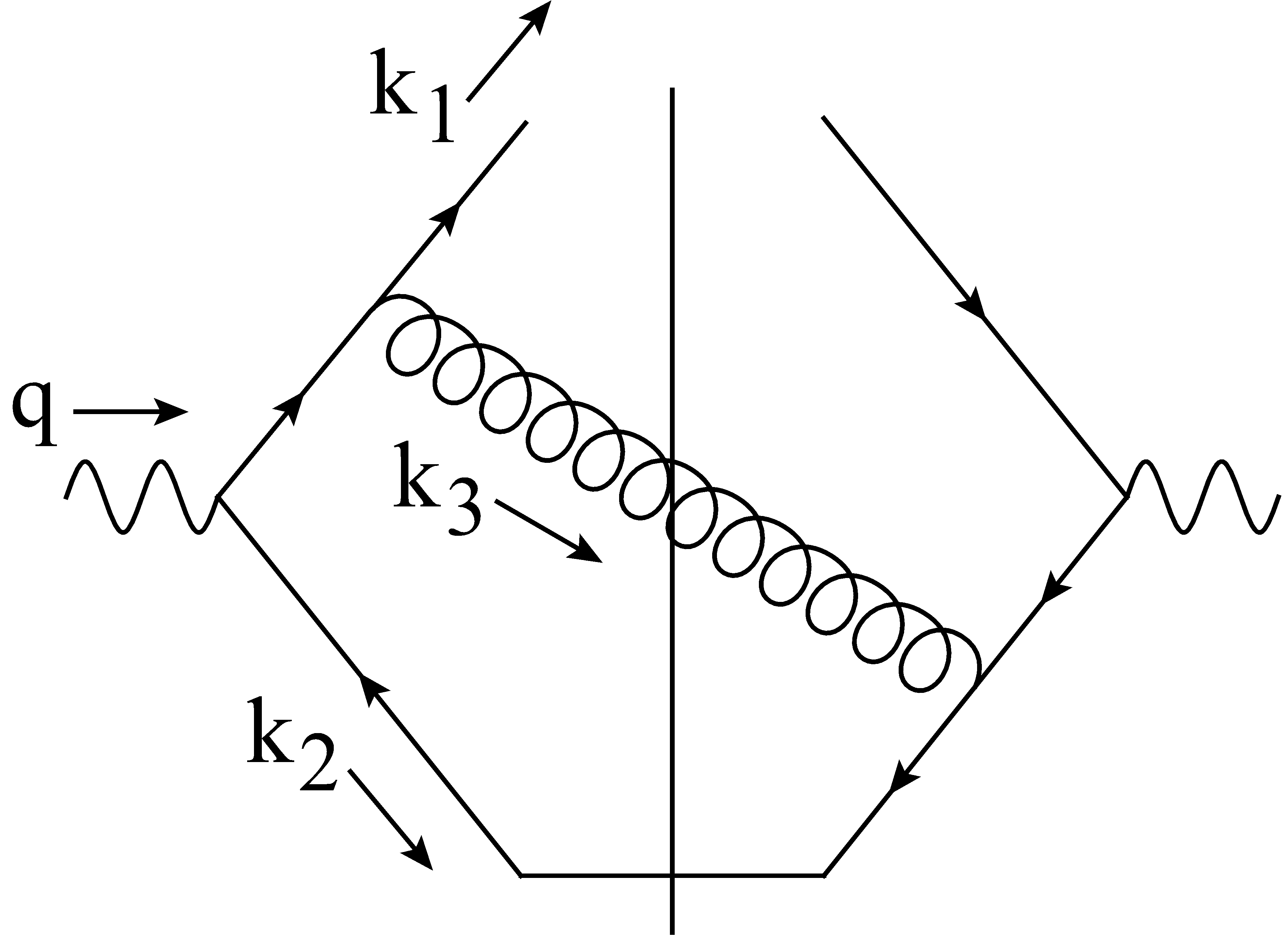}
\end{gathered} + h.c. +
\begin{gathered}
\includegraphics[width=3.8cm]{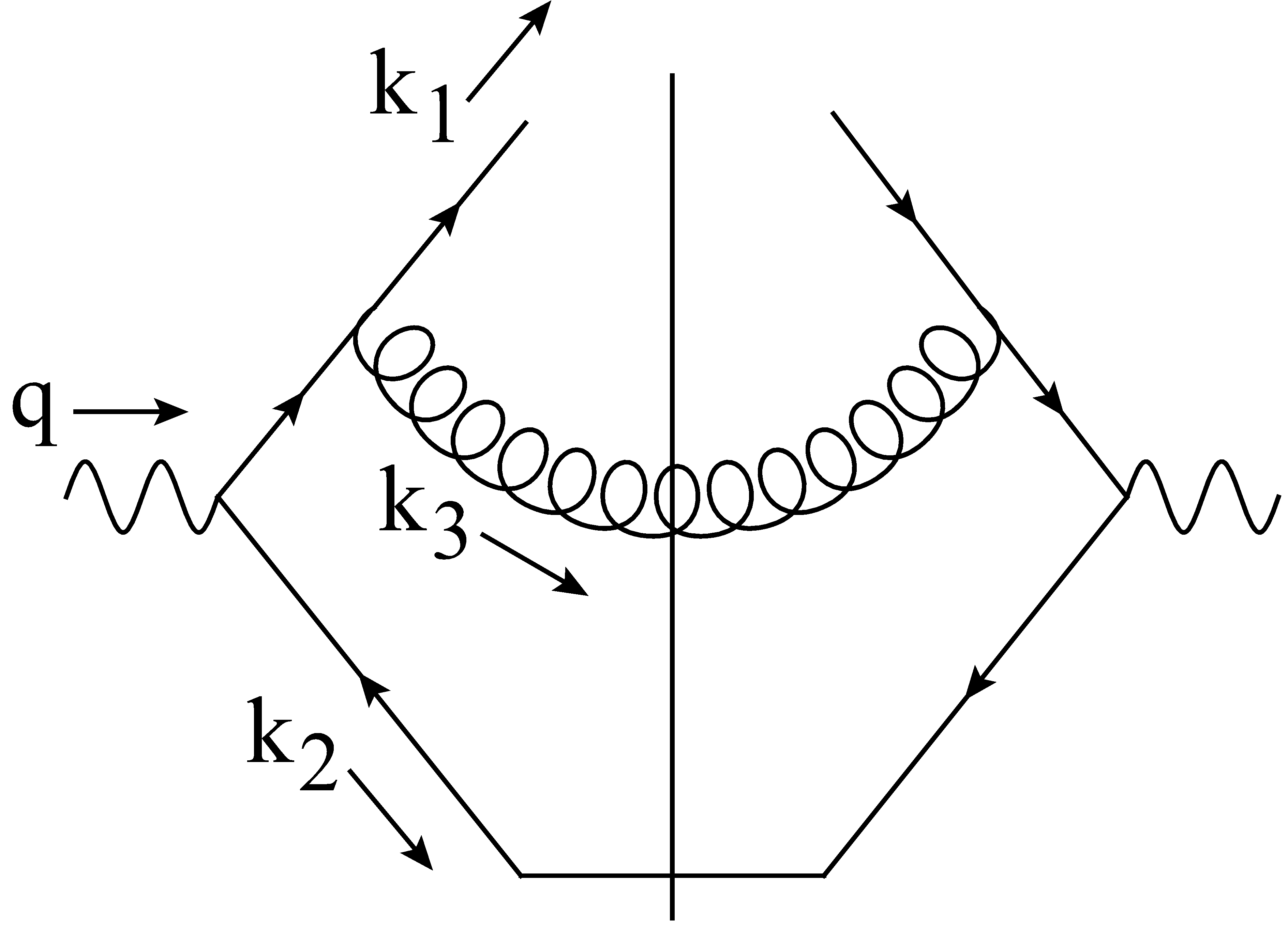}
\end{gathered} +
\begin{gathered}
\includegraphics[width=3.8cm]{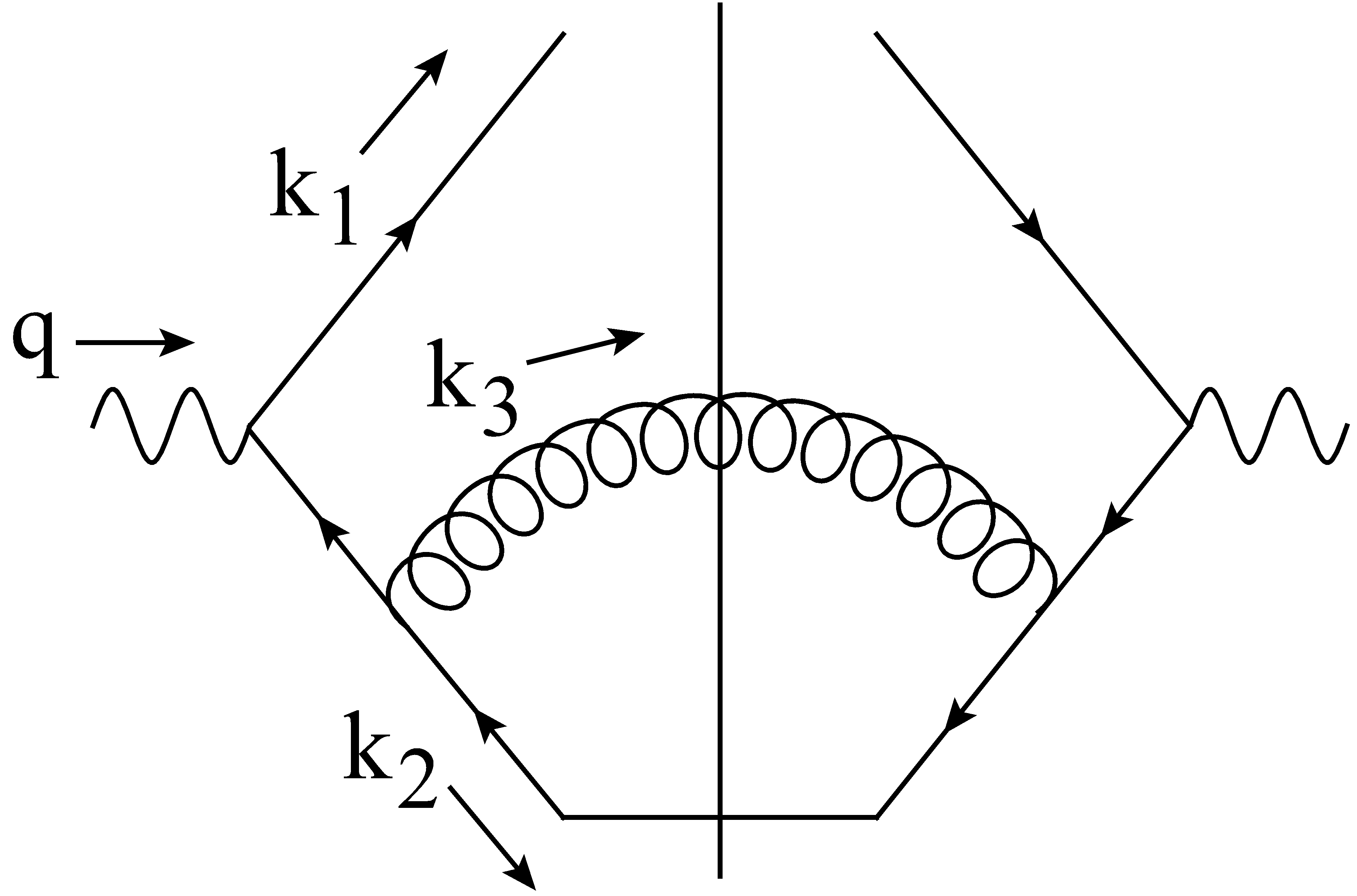}
\end{gathered}
\right)^{\mu \nu}
\label{eq:Mf_real}
\end{align}
In this case, the computations are more difficult and require non-standard mathematical techniques. Not only because of the larger number of Feynman diagrams and a wider available phase space for the particles in the final state, but mostly because of the further non-trivial dependence on the thrust and on the transverse momentum of the fragmenting parton. 
The starting point for the application of the factorization procedure is the following unapproximated partonic tensor:
\begin{align}
&\widehat{W}_{f,\,\text{R}}^{\mu \nu, \,[1]}\left(\eps;\,z , \tau ,  \vec{k}_T\right) =
\frac{1}{4\pi} \,
\int \frac{d^D k_2}{(2\pi)^{D}} \, 
\frac{d^D k_3}{(2\pi)^{D}}
M_{f,\,R}^{\mu\,\nu}{}^{\;[1]}(\epsilon;\,\mu,\,{k_i})\,
(2\pi)^{D} \, 
\delta^{(D)}\left( q - k_1 - k_2 - k_3 \right) \,
\notag \\
&\quad\times
\delta(\tau - \tau_{\text{def.}}({k_i}))\,
\delta(\vec{k}_T - \vec{k}_T'({k_i})),
\label{eq:What_nlo_R}
\end{align}
where $\tau_{\text{def.}}({k_i})$ refers to the definition of $\tau = 1-T$ as it is given in Eq.~\eqref{eq:thrust_def}, by using the momenta $k_1$, $k_2$ and $k_3$ of the particles involved\footnote{It is possible to show that the phase space is divided into three regions $U_j$. In each of these regions, the value of the trhust $\tau$ corresponds to the variable $y_j = 2{k_i \cdot k_l}/{Q^2}$, with $i,l \neq j$. See Ref.~\cite{Boglione:2020auc} for more details.}.
The action of the kinematic approximators does not only involves the squared amplitudes in Eq.~\eqref{eq:Mf_real}, it also modifies the values of $\tau_{\text{def.}}({k_i})$ and $\vec{k}_T'({k_i})$ according to the considered momentum region.
In this regard, the value of $\vec{k}_T'({k_i})$ is crucial to determine the TMD-relevance of the corresponding contribution and ultimately the selection of the kinematic region.
The final result will be obtained by summing together the contributions of backward and forward radiation, corresponding to the case where the gluon is emitted in the $\mathrm{S}_B$ and in the $\mathrm{S}_A$ hemisphere, respectively. In each of these configurations, the gluon can be collinear to the fermion or soft. It certainly cannot be hard, as being a real particle it is considered on-shell.
The implementation of the kinematic requirements will be crucial in the study of the forward radiation

The rest of the paper will be devoted to the detailed analyses of the 1-loop contributions of each of the leading momentum regions. 
This will lead to the determination of the factorization theorem in each one of them.

\bigskip


\section{Backward Radiation \label{sec:backward_rad}}


\bigskip

In this section we will consider the contribution of the radiation emitted in the direction opposite to the fragmenting quark, which will be denoted as backward radiation. 
If the gluon is emitted in the $\mathrm{S}_B$-hemisphere, then there are two leading momentum regions.  
The first is associated to the configuration in which the gluon is collinear to the antiquark and is obtained through the action of the approximator $T_B$. The other regards the emission of a soft gluon, and it is given by applying the approximator $T_S^-$. These two momentum regions overlap, hence we have to remove the double counting of the same contributions. The overlapping region is associated to a gluon which can be considered soft-collinear: it has a very small energy but also a large (and, in this case, negative) rapidity. Such contribution is obtained by the combination of approximators $T_B T_S \equiv T_S T_B$. Notice that the label ``$-$" is redundant in such combination, as the action of $T_B$  already encodes the information about the selection of the hemisphere.

All these approximators act on the squared matrix elements as well as on the integration of the phase space, as defined in the previous Section. In particular, in a backward approximation, there is no transverse momentum contributing to the transverse deflection of the fragmenting parton (which is the other hemisphere), and hence $\vec{k}'_T = 0$. This confirms that all the contributions associated to the backward radiation are TMD-irrelevant.

In conclusion, the 1-loop contribution of the backward radiation to the partonic tensor can be written as:
\begin{align}
\label{eq:1loop_xsf-_structure}
&\widehat{W}_{f,\,\text{backward}}^{\mu \nu, \,[1]}\left(\eps;\,z , \tau ,  k_T\right) = 
\left(T_S^- - T_S T_B + T_B \right) \left[ \widehat{W}_{f,\text{R}}^{\mu \nu, \,[1]}\left(\eps;\,z , \tau ,  k_T\right) \right] +  
\substack{\mbox{power suppressed}\\\mbox{corrections}}.
\end{align}
In the following, the three contributions involved in the previous expression will be considered separately.

\paragraph{Soft approximation} \hfill

\noindent The action of $T_S^-$ leads to the following approximation:
\begin{align}
\label{eq:1loop_TS-}
&T_S^-\left[\widehat{W}_{f,\text{R}}^{\mu \nu, \,[1]}\left(\eps;\,z , \tau ,  k_T\right) \right] =
\int_z^1 \frac{d \rho}{\rho} 
{}^{\star}\widehat{W}^{\mu \nu,\,[0]}_f ({z}/{\rho},\,Q) \,
\delta (1-\rho ) \mathscr{S}_{-}^{[1]}\left(\eps;\tau\right) \, \delta \left(\vec{k}_T\right),
\end{align}
where we used the same conventions adopted in the previous section. In the previous expression, we have introduced the \textbf{generalized soft thrust function} $\mathscr{S}$. Its definition is obtained by integrating (instead of Fourier transforming) the $2$-h soft factor, defined as in Eq.~\eqref{eq:S2h_bTspace}, with the further explicit dependence on thrust, implemented as for the usual soft thrust function. In practice, it is defined modifying the usual soft thrust function by introducing the rapidity divergence regulator. In the case of the Collins factorization formalism, this is achieved by tilting the two Wilson lines off the light-cone, see Appendix~\ref{app:review_collins}. Therefore, besides the dependence on  thrust, $\mathscr{S}$ depends also on the rapidity cut-offs.

The  label ``$-$" associated to the generalized soft thrust function in Eq.~\eqref{eq:1loop_TS-} reminds that only the contribution associated to the hemisphere $\mathrm{S}_B$ has to be taken into account. 
This is realized by imposing that the rapidity of the backward radiations cannot be positive. At 1-loop order, it is sufficient to set $l^- > l^+$, where $l$ is the momentum of the radiated soft gluon. Therefore, we have:
\begin{align}
\label{eq:cssSthrustfunction_minus}
&\mathscr{S}_{-}^{[1]}\left(\eps;\,\tau,\,y_1,\,y_2\right) =
\int \meas{l}{D} \theta\left( l^- - l^+\right)
\delta\left(\tau - \frac{l^+}{q^+}\right)
\left(
\begin{gathered}
\includegraphics[width=4.cm]{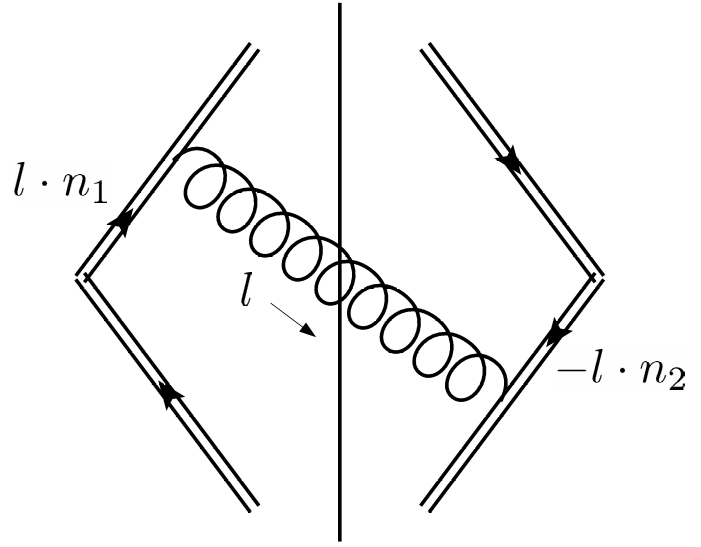}
\end{gathered}
+ h.c.
\right) = 
\notag \\
&\quad=
\aS 4 \, C_F \, S_\eps \left(\frac{\mu}{Q}\right)^{2\eps} \tau^{-1-2\eps} 
\int_{-\infty}^0 dy \; e^{2\eps y}
\frac{1+e^{-2(y_1-y_2)}}
{\left(1-e^{-2(y_1-y)} \right)
\left(1-e^{2(y_2-y)} \right)} + h.c.,
\end{align}
where in the last step we made a change of variables in order to expose the range of the rapidity of the gluon $y = \frac{1}{2}\log\left(\frac{l^+}{l^-}\right)$, which is negative as required for backward radiation.
The integration gives:
\begin{align}
\label{eq:rapint_Sminus}
&\int_{-\infty}^0 dy \; e^{2\eps y}
\frac{1+e^{-2(y_1-y_2)}}
{\left(1-e^{-2(y_1-y)} \right)
\left(1-e^{2(y_2-y)} \right)} =
\frac{1}{2} 
\frac{1+e^{-2(y_1-y_2)}}
{1-e^{-2(y_1-y_2)}}
\frac{1}{1-2\eps}
\notag \\
&\quad\times
\Big\{
-e^{-y_2} \left[
\pFq{2}{1}\left( 1,\,1+2\eps;\,2+2\eps;\,e^{-y_2}\right)
-
\pFq{2}{1}\left( 1,\,1+2\eps;\,2+2\eps;\,-e^{-y_2}\right)
\right] +
\notag \\
&\quad+
e^{-y_1} \left[
\pFq{2}{1}\left( 1,\,1+2\eps;\,2+2\eps;\,e^{-y_1}\right)
-
\pFq{2}{1}\left( 1,\,1+2\eps;\,2+2\eps;\,-e^{-y_1}\right)
\right]
\Big\} =
\notag \\
&=
\frac{1}{2} \left(
\frac{1}{\eps} +
\left( -e^{2 y_2}\right)^\eps 
\Gamma(-\eps)\Gamma(1+\eps)+
\mathcal{O}\left( e^{-2 y_1},\, 
e^{2 y_2},\,e^{-2(y_1-y_2)}\right)
\right).
\end{align}
Inserting this result in Eq.~\eqref{eq:cssSthrustfunction_minus} and neglecting the errors due to the vanishing rapidity cut-offs, we obtain:
\begin{align}
\label{eq:cssSthrustfunction_minus_2}
&\mathscr{S}_{-}^{[1]}\left(\eps;\,\tau,\,y_2\right) =
\aS 2 \, C_F \, S_\eps \left(\frac{\mu}{Q}\right)^{2\eps} \tau^{-1-2\eps} 
\left(
\frac{1}{\eps} +
\left( -e^{2 y_2}\right)^\eps 
\Gamma(-\eps)\Gamma(1+\eps) + h.c.
\right) .
\end{align}
Notice that whole dependence on the rapidity cut-off $y_1$ is suppressed in the final result. Therefore, we dropped it from the l.h.s of the previous equation. This is a general feature: only the dependence on the rapidity cut-off relevant for the considered hemisphere survives. In fact, the analogous contribution of the hemisphere $\mathrm{S}_A$ to the generalized soft thrust function is obtained from Eq.~\eqref{eq:cssSthrustfunction_minus_2} by replacing\footnote{Notice that this is the same replacement derived in Ref.~\cite{Boglione:2020cwn} in studying the behavior of the TMDs under a $Z$-axis reflection.} $y_2$ with $-y_1$:
\begin{align}
\label{eq:genS_+-}
\mathscr{S}_{+}\left(\eps;\,\tau,\,y_1\right)  =  \mathscr{S}_{-}\left(\eps;\,\tau,\,-y_1\right) 
\end{align}
There is a straightforward factorization theorem that relates the contributions of the two hemispheres to the total generalized soft thrust function:
\begin{align}
\label{eq:generalized_S_fact_theo}
\mathscr{S}\left(\eps;\,\tau,\,y_1 - y_2\right) = \mathscr{S}_{+}\left(\eps;\,\tau,\,y_1\right) \mathscr{S}_{-}\left(\eps;\,\tau,\,y_2\right).
\end{align}
Finally, it is interesting to point out that the result of Eq.~\eqref{eq:cssSthrustfunction_minus_2} can be equivalently written as:
\begin{align}
\label{eq:genSthr_2}
&\mathscr{S}_{-}^{[1]}\left(\eps;\,\tau,\,y_2\right) =
S_{-}^{[1]}\left(\eps;\,\tau\right) + 
\aS 2 \, C_F \, S_\eps \left(\frac{\mu}{Q}\right)^{2\eps} \tau^{-1-2\eps} 
\Gamma(-\eps)\Gamma(1+\eps)
\left(\left( -e^{2 y_2}\right)^\eps 
 + h.c.\right),
\end{align}
where $S_{-} \equiv \frac{1}{2} \, S$ is the backward radiation contribution to the usual 1-loop soft thrust function. If we had removed the rapidity cut-offs from the very beginning in the definition of the generalized soft thrust function, Eq.~\eqref{eq:cssSthrustfunction_minus}, this would have been the whole final result.
However, retaining $y_1$ and $y_2$ in the calculation leads to an extra term depending on the leading cut-off of the hemisphere, in this case $y_2$. Such extra term will have to cancel out when we will subtract the overlapping with the collinear contribution.

\paragraph{Soft-collinear approximation (overlapping)} \hfill

\noindent The action of $T_S T_B$ produces the following approximation:
\begin{align}
\label{eq:1loop_TSTB}
&T_S T_B
\left[\widehat{W}_{f,\text{R}}^{\mu \nu, \,[1]}\left(\eps;\,z , \tau ,  k_T\right) \right] =
\int_z^1 \frac{d \rho}{\rho} 
{}^{\star}\widehat{W}^{\mu \nu,\,[0]}_f ({z}/{\rho},\,Q) \,
\delta (1-\rho ) \mathscr{Y}_{-}^{[1]}\left(\eps;\tau\right) \, \delta \left(\vec{k}_T\right).
\end{align}
In this expression we have introduced the \textbf{soft-collinear thrust functions} $\mathscr{Y}_{\pm}$. 
In particular, the contribution associated to the hemisphere $\mathrm{S}_B$ is involved in Eq.~\eqref{eq:1loop_TSTB}.
Differently from $ \mathscr{S}$, this functions does not have a counterpart among the usual thrust functions reviewed for instance in Ref.~\cite{Boglione:2020auc}. It is defined as the subtraction term of the TMDs\footnote{Here the TMDs are defined according to the factorization definition, Eq.~\eqref{eq:sub_coll}.}, integrated (instead of Fourier transformed) over $\vec{k}_T$, and modified to include the explicit dependence on the thrust. In practice, $\mathscr{Y}_{\pm}$ are defined similarly to the usual soft thrust function $\mathscr{S}$, but tilting only the Wilson line pointing in the reference direction indicated by the collinear approximator. Therefore, the soft-collinear thrust functions acquire a dependence on the rapidity cut-off associated to the tilted Wilson line. At 1-loop order and for the backward radiation contribution we have:
\begin{align}
\label{eq:cssSCBthrustfunction}
&\mathscr{Y}_{-}^{[1]}\left(\eps;\,\tau,\,y_2\right) =
\int \meas{l}{D}
\delta\left(\tau - \frac{l^+}{q^+}\right)
\left(
\begin{gathered}
\includegraphics[width=4.cm]{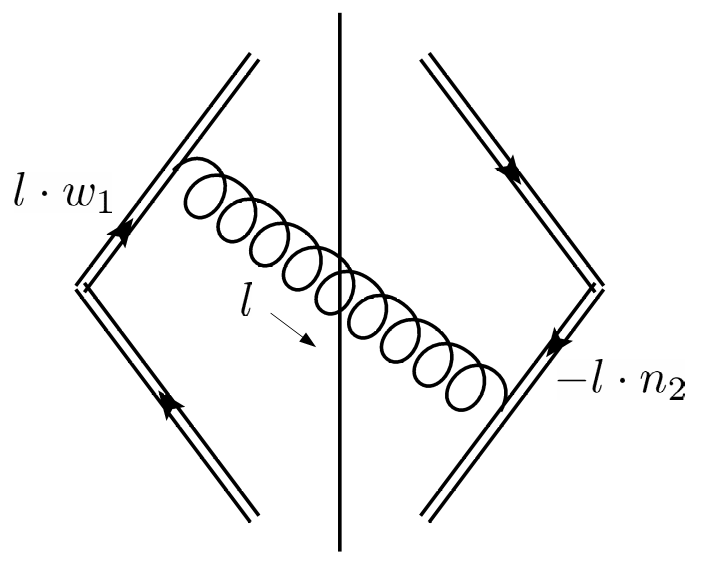}
\end{gathered}
+ h.c.
\right) = 
\notag \\
&\quad=
\aS 4 \, C_F \, S_\eps \left(\frac{\mu}{Q}\right)^{2\eps} \tau^{-1-2\eps} 
\int_{-\infty}^{+\infty} dy \; e^{2\eps y}
\frac{1}
{\left(1-e^{2(y_2-y)} \right)} + h.c.
\end{align}
Notice that the action of $T_B$ makes the Heaviside theta that selects the (-)-hemisphere redundant, as $T_B \theta(l^- - l^+) = \theta(l^-)$. However this condition is already encoded into the requirement that the emitted gluon is on-shell at the final state cut.
As a consequence, differently from Eq.~\eqref{eq:cssSthrustfunction_minus}, the integration on the rapidity of the gluon is unbounded from below.
The integral in Eq.~\eqref{eq:cssSCBthrustfunction} has the following solution:
\begin{align}
\label{eq:rapint_SB}
&\int_{-\infty}^{+\infty} dy \; e^{2\eps y}
\frac{1}
{\left(1-e^{2(y_2-y)} \right)} =
\frac{1}{2} 
\left(e^{2 y_2}\right)^\eps
\left[
B_{e^{2 y_2}}\left( -\eps,\,0 \right)
-
B_{e^{-2 y_2}}\left( 1+\eps,\,0 \right)
\right] = 
\notag \\
&\quad=
\frac{1}{2} \left(-e^{2 y_2}\right)^\eps
\,\Gamma(-\eps) \Gamma(1-\eps)
+\mathcal{O}\left( e^{2 y_2}\right),
\end{align}
where $B$ is the incomplete Beta function. Inserting this result into Eq.~\eqref{eq:cssSCBthrustfunction} we obtain:
\begin{align}
\label{eq:cssSCBthrustfunction_2}
&\mathscr{Y}_{-}^{[1]}\left(\eps;\,\tau,\,y_2\right) =
\aS 2 \, C_F \, S_\eps \left(\frac{\mu}{Q}\right)^{2\eps} \tau^{-1-2\eps} 
\Gamma(-\eps)\Gamma(1+\eps)
\left(\left( -e^{2 y_2}\right)^\eps 
 + h.c.\right).
\end{align}
As for the soft case, the analogous contribution in the opposite hemisphere, resulting from the action of $T_S T_A$ can be easily obtained from Eq.~\eqref{eq:cssSCBthrustfunction_2} by replacing $y_2$ by $-y_1$:
\begin{align}
\label{eq:SB_+-}
\mathscr{Y}_{+}\left(\eps;\,\tau,\,y_1\right)  =  \mathscr{Y}_{-}\left(\eps;\,\tau,\,-y_1\right) 
\end{align}
Notice that if we had removed the rapidity cut-off from the very beginning, already in the definition of $\mathscr{Y}$ in Eq.~\eqref{eq:cssSCBthrustfunction}, then the integration over the rapidity of the emitted gluon would have been scaleless and hence vanishing in full dimensional regularization. This is the reason for which the soft-collinear thrust functions do not have a counterpart among the thrust-dependent functions usually encountered in the factorization of $\epm$ annihilation processes.

In this case, the fact that r.h.s of Eq.~\eqref{eq:cssSCBthrustfunction_2} is non vanishing is crucial for the success of the subtraction mechanism. In fact, the result found for $\mathscr{Y}_{-}$ is exactly equal to the extra term, rapidity cut-off dependent, obtained in the calculation of $\mathscr{S}_{-}$, Eq.~\eqref{eq:genSthr_2}. Therefore:
\begin{align}
\label{soft-collB_subtraction}
\mathscr{S}_{-}^{[1]}\left(\eps;\,\tau,\,y_2\right) - \mathscr{Y}_{-}^{[1]}\left(\eps;\,\tau,\,y_2\right) =  S_{-}^{[1]}\left(\eps;\,\tau\right) .
\end{align}
In other words, the \emph{subtracted} soft contribution associated to the backward emission, obtained through the action of $\left(T_S^{-} - T_S T_B\right)$, is totally independent of the rapidity cut-off $y_2$.  Therefore, we can derive the following factorization theorems, which generalize  Eq.~\eqref{soft-collB_subtraction}, as well as the analogous equation holding for the $\mathrm{S}_A$-hemisphere, to all orders:
\begin{subequations}
\label{eq:Sthr_fact_theo_hemi}
\begin{align}
 S_{+}\left(\eps;\,\tau\right) = 
\frac{\mathscr{S}_{+}\left(\eps;\,\tau,\,y_1\right)}{ \mathscr{Y}_{+}\left(\eps;\,\tau,\,y_1\right)};
\label{eq:Sthr_fact_theo_plus}\\
 S_{-}\left(\eps;\,\tau\right) = 
\frac{\mathscr{S}_{-}\left(\eps;\,\tau,\,y_2\right)}{ \mathscr{Y}_{-}\left(\eps;\,\tau,\,y_2\right)}.
\label{eq:Sthr_fact_theo_minus}
\end{align}
\end{subequations}
In conclusion, this factorization theorems, together with Eq.~\eqref{eq:generalized_S_fact_theo}, leads to:
\begin{align}
 S\left(\eps;\,\tau\right) = 
\frac{\mathscr{S}_{+}\left(\eps;\,\tau,\,y_1\right)}{ \mathscr{Y}_{+}\left(\eps;\,\tau,\,y_1\right)} \, 
\frac{\mathscr{S}_{-}\left(\eps;\,\tau,\,y_2\right)}{ \mathscr{Y}_{-}\left(\eps;\,\tau,\,y_2\right)}.
\label{eq:Sthr_fact_theo}
\end{align}
which encodes the relations between the usual soft thrust function and the soft and soft-collinear thrust functions defined in this Section.

\paragraph{Collinear approximation} \hfill

\noindent Finally, the effect of the action of the $T_B$ approximator gives:

\begin{align}
\label{eq:1loop_TB}
&T_B
\left[\widehat{W}_{f,\text{R}}^{\mu \nu, \,[1]}\left(\eps;\,z , \tau ,  k_T\right) \right] =
\int_z^1 \frac{d \rho}{\rho} 
{}^{\star}\widehat{W}^{\mu \nu,\,[0]}_f ({z}/{\rho},\,Q) \,
\delta (1-\rho ) J^{[1]}\left(\eps;\tau\right) \, \delta \left(\vec{k}_T\right).
\end{align}
where $J$ is the usual jet thrust function at 1 loop\footnote{In principle, in the pure Collins factorization formalism, masses cannot be neglected in the collinear contributions. However, since in this case we are dealing with rather low scales ($Q \sim 10$ GeV for the BELLE experiment) that prevent the presence of heavy quarks, we will put all masses to zero.}. In this case, we do not have any rapidity cut-off, since in (unsubtracted) collinear parts the Wilson lines are defined along the light-cone.

\paragraph{Final result for backward radiation} \hfill

\noindent Combining the results above and inserting them into Eq.~\eqref{eq:1loop_xsf-_structure}, we can write the final expression for the contribution of the backward radiation to the partonic tensor. In transverse momentum space we have:
\begin{align}
\label{eq:1loop_xsf-_structure_after}
&\widehat{W}_{f,\,\text{backward}}^{\mu \nu, \,[1]}\left(\eps;\,z , \tau ,  k_T\right) = 
\notag \\
&\quad=
\int_z^1 \frac{d \rho}{\rho} 
{}^{\star}\widehat{W}^{\mu \nu,\,[0]}_f ({z}/{\rho},\,Q) \,
\delta (1-\rho ) \left[
 S_{-}^{[1]}\left(\eps;\,\tau\right) + J^{[1]}\left(\eps;\tau\right)
\right]\, \delta \left(\vec{k}_T\right) +  
\substack{\mbox{power suppressed}\\\mbox{corrections}},
\end{align}
where we used Eq.~\eqref{soft-collB_subtraction} to combine soft and soft-collinear contributions. The power suppressed terms contains both the errors due to the approximations introduced by the factorization procedure and also the terms neglected in the limit of large rapidity cut-off. Since the dependence on $\vec{k}_T$ is trivial, in $b_T$-space we have simply: 
\begin{align}
\label{eq:1loop_xsf-_structure_afterbT}
&\widetilde{\widehat{W}}_{f,\,\text{backward}}^{\mu \nu, \,[1]}\left(\eps;\,z , \tau ,  b_T\right) = 
\notag \\
&\quad=
\int \frac{d \rho}{\rho} 
{}^{\star}\widehat{W}^{\mu \nu,\,[0]}_f ({z}/{\rho},\,Q) \,
\delta (1-\rho ) \left[
 S_{-}^{[1]}\left(\eps;\,\tau\right) + J^{[1]}\left(\eps;\tau\right)\right] +  
\substack{\mbox{power suppressed}\\\mbox{corrections}}.
\end{align}

\bigskip


\section{Region 2: collinear-TMD factorization \label{sec:reg2}}


\bigskip

Contributions associated to the radiation emitted in the same hemisphere of the detected hadron are indeed the most interesting. They encode the whole information on the TMD effects, hence they are the keystone for exploring the rich kinematic structure underlying the process we are investigating. 
The leading momentum regions are the counterpart of those considered in the previous section, properly modified in order to describe the emission in the $\mathrm{S}_A$ hemisphere. Therefore, when the emitted gluon is soft, the partonic tensor is well approximated by the action of $T_S^+$, while for a gluon collinear to the fragmenting quark we rely on the $T_A$ approximator. Moreover, in the soft-collinear overlapping region, the gluon has a very low energy but also a large (and positive) rapidity and the corresponding contribution is well approximated by the action of $T_S T_A \equiv T_A T_S$. Therefore, the 1-loop contribution to the partonic tensor of the radiation emitted (forwardly) into the $\mathrm{S}_A$-hemisphere can be written as:
\begin{align}
\label{eq:1loop_xsf+_structure}
&\widehat{W}_{f,\,\text{forward}}^{\mu \nu, \,[1]}\left(\eps;\,z , \tau ,  k_T\right) = 
\left(T_S^+ - T_S T_A + T_A \right) \left[ \widehat{W}_{f,\text{R}}^{\mu \nu, \,[1]}\left(\eps;\,z , \tau ,  k_T\right) \right] +  
\substack{\mbox{power suppressed}\\\mbox{corrections}}.
\end{align}
The most important difference with respect to the cases discussed in Section~\ref{sec:backward_rad} is in the role that each of these contribution plays in the generation of significant TMD effects. In other words, according to the nomenclature introduced in Section~\ref{sec:tmd_rel}, not all of these kinematic configurations may be TMD relevant. Indeed, when the gluon is radiated collinearly to the fragmenting quark, it actively contributes to the deflection of the detected hadron with respect to the thrust axis. Therefore, as pointed out in Section~\ref{sec:tmd_rel}, the action of $T_A$ produces a TMD-relevant term.
For soft and soft-collinear gluons this is not as straightforward, first of all because, according to power counting, their transverse momentum has a much smaller size than that of the collinear emission. 

Region 2 gathers all the kinematic configurations in which the soft radiation is not TMD-relevant, while the soft-collinear contributions together with the collinear radiations remain TMD-relevant,
as in Table~\ref{tab:kinRef_tmdrel}.
%
\begin{figure}
\centering
\includegraphics[width=6cm]{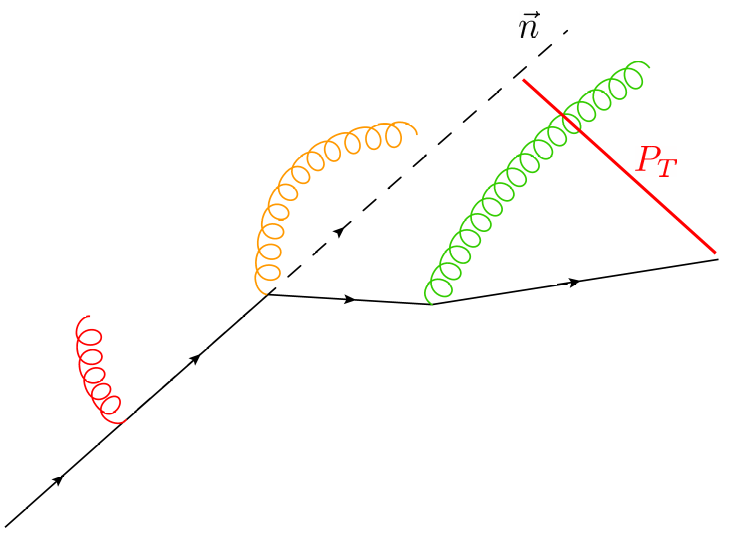}
\caption{Pictorial representation of the effect of the three kinds of radiation in Region 2. The final $P_T$, measured with respect to the thrust axis $\vec{n}$ , is not affected by the emission/absorption of soft radiation, as the soft gluons (red) are TMD-irrelevant. Instead, soft-collinear (orange) and collinear (green) radiations produce the deflections that lead to the size of the observed transverse momentum. }
\label{fig:Reg2_pictorial}
\end{figure}
%
In this case, the soft radiation does not produce any significant TMD effect, hence it does not affect the experimentally measured value of $P_T$. On the other hand, soft-collinear and collinear emissions play an active role in generating the transverse momentum of the detected hadron. 

\paragraph{Soft approximation} \hfill

\noindent Since in Region 2 soft radiation is TMD-irrelevant, the action of $T_S^+$ leads to a result which is perfectly analogous to its counterpart in the opposite hemisphere, found in Section~\ref{sec:backward_rad}.
In fact, we have: 
\begin{align}
\label{eq:1loop_TS+_reg2}
&T_S^+\left[\widehat{W}_{f,\text{R}}^{\mu \nu, \,[1]}\left(\eps;\,z , \tau ,  k_T\right) \right] =
\int \frac{d \rho}{\rho} 
{}^{\star}\widehat{W}^{\mu \nu,\,[0]}_f ({z}/{\rho},\,Q) \,
\delta (1-\rho ) \mathscr{S}_{+}^{[1]}\left(\eps;\tau\right) \, \delta \left(\vec{k}_T\right),
\end{align}
where $\mathscr{S}_{+}$ is the contribution of the $\mathrm{S}_A$ hemisphere to the generalized soft thrust function at 1-loop. Thanks to Eq.~\eqref{eq:genS_+-}, we can use the solution found for $\mathscr{S}_{-}$ in Eq.~\eqref{eq:cssSthrustfunction_minus_2}. 

\paragraph{Soft-collinear approximation (overlapping)} \hfill

\noindent In Region 2, soft-collinear radiation plays an active role in generating significant TMD effects. The result of the action of $T_S T_A$ gives a TMD-relevant contribution:
\begin{align}
\label{eq:1loop_TSTA_int}
&T_S T_A \left[\widehat{W}_{f,\text{R}}^{\mu \nu, \,[1]}\left(\eps;\,z , \tau ,  k_T\right) \right] =
\int \frac{d \rho}{\rho} 
{}^{\star}\widehat{W}^{\mu \nu,\,[0]}_f ({z}/{\rho},\,Q) \,
\delta (1-\rho ) \Upsilon_{+}^{[1]}\left(\eps;\tau, \, k_T\right),
\end{align}
where we have introduced the \textbf{soft-collinear thrust factor} $\Upsilon_{\pm}$, whose $\mathrm{S}_A$-hemisphere contribution is involved into Eq.~\eqref{eq:1loop_TS+_int}. This object is defined in the same way of the subtraction term of the TMDs,  Eq.~\eqref{eq:sub_coll}, but with an additional dependence on thrust. Therefore, $\Upsilon_{\pm}$ depends on the total soft-collinear transverse momentum, on the rapidity cut-off associated to its tilted Wilson line and also on thrust. To 1-loop order, the forward hemisphere contribution is then defined as:
\begin{align}
\label{eq:SCthrustfactor_kT}
&\Upsilon_{+}^{[1]}\left(\eps;\tau, \, k_T,\,y_1\right) =
\int \frac{d l^+ \, d l^-}{(2\pi)^D}
\delta\left(\tau - \frac{l^-}{q^-}\right)
\left(
\begin{gathered}
\includegraphics[width=4.cm]{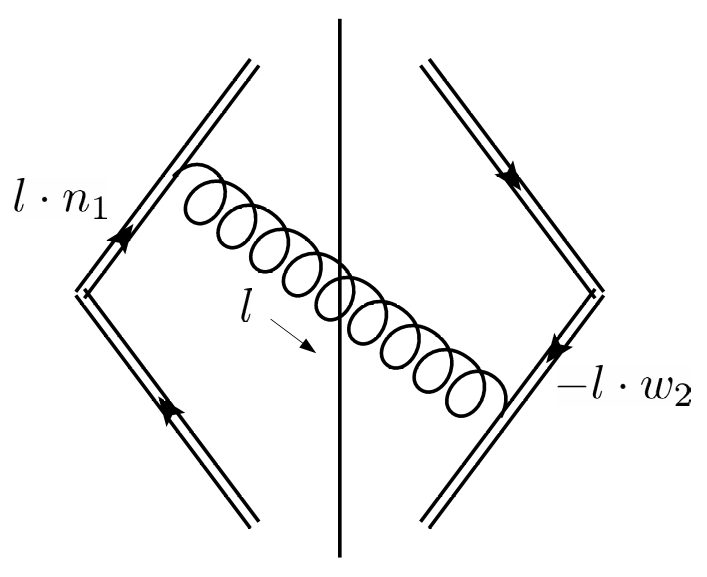}
\end{gathered}
+ h.c.
\right) = 
\notag \\
&\quad=
\aS 2 \, C_F \, S_\eps \,
\frac{\Gamma(1-\eps)}{\pi^{1-\eps}} \,
\mu^{2\eps}\,\frac{1}{k_T^2} \,
\int_{-\infty}^{+\infty} dy \,
\frac{1}
{1-e^{-2(y_1-y)}}
\delta \left( \tau - \frac{k_T}{Q} e^{-y}\right) + h.c.
\end{align}
Notice that, analogously to the case of backward radiation, in soft-collinear contributions the range of the radiated particle rapidity is unconstrained. In fact, 
the rapidity of the emitted gluon in Eq.~\eqref{eq:SCthrustfactor_kT} is unbounded from below.
Its Fourier transform leads to:
\begin{align}
\label{eq:SCthrustfactor_FT}
&\widetilde{\Upsilon}_{+}^{[1]}\left(\eps;\,\tau,b_T,\,y_1\right) = 
\FT \,\Upsilon_{+}^{[1]}\left(\eps;\,\tau,\vec{k}_T\,y_1\right) =
\notag \\
&\quad=
\aS 2 \, C_F \, S_\eps \,
\left( \frac{\mu}{Q} \right)^{2\eps}\,
\Gamma(1-\eps)\,
\left( \frac{b}{c_1} \right)^{\eps}\,
e^{-\eps \gamma_E} \,
\tau^{-1-\eps} \,
\int_0^\infty 
\frac{x^{\eps/2}}{x-e^{-2 y_1}}\,J_{-\eps}\left( \frac{\tau \, b}{\sqrt{x}} \right) + h.c.
\end{align}
The solution to the integral in the last line of the previous equation requires advanced mathematical tools and non-standard techniques, which are described in detail in Appendix~\ref{app:solutionsint}. The result can be found in Eq.~\eqref{eq:int_notsodifficult} and it is given by:
\begin{align}
\label{eq:int_SC}
\int_0^\infty 
\frac{x^{\eps/2}}{x-r_1}\,J_{-\eps}\left( \frac{\tau \, b}{\sqrt{x}} \right) = 
-2\left(-r_1\right)^{\eps/2}K_{-\eps} \left( \frac{\tau \, b}{\sqrt{-r_1}} \right) + 
\left( \frac{\tau \, b}{2} \right)^{\eps} \Gamma(-\eps),
\end{align}
where $r_1 =e^{-2 y_1}$. This is an exact result, as there are no terms suppressed in the limit $r_1 \to 0$. 
Inserting this result in Eq.~\eqref{eq:SCthrustfactor_FT} we obtain:
\begin{align}
\label{eq:SCthrustfactor_FT_2}
&\widetilde{\Upsilon}_{+}^{[1]}\left(\eps;\,\tau,b_T,\,y_1\right) = 
\aS 2 \, C_F \, S_\eps \,
\left( \frac{\mu}{Q} \right)^{2\eps}\,
\Bigg\{
\frac{1}{\tau}\,
\left( \frac{b}{c1} \right)^{2\eps} \,
e^{-2 \gamma_E\,\eps}\;\Gamma(1-\eps)\Gamma(-\eps)-
\notag \\
&\quad
-2 \tau^{-1-\eps} \Gamma(1-\eps)
\left(-e^{-2y_1}\right)^{\eps/2}
K_{-\eps}\left( \frac{\tau b}{\sqrt{- e^{-2y_1}}} \right)
\Bigg\}
 + h.c.
\end{align}
As widely discussed in Appendix~\ref{app:frag_gluon}, the $2$-jet limit is realized in $b_T$-space by the asymptotic behavior at large distances. This follows from the relation between the thrust and the transverse momentum in $k_T$-space, as the limit $\tau \to 0$ corresponds to the region of small transverse momenta, at least when the kinematic requirement \hyp{1} holds true.
The large-$b$ asymptotic of $\widetilde{\Upsilon}_{+}^{[1]}$ can be obtained exploiting the trick of Eq.~\eqref{eq:distr_expansion}.
Notice that the two contributions in Eq.~\eqref{eq:SCthrustfactor_FT_2} are separately divergent as $\tau \to 0$, but their sum is integrable. We have:
\begin{align}
\label{eq:SCthrust_largeb}
&\frac{1}{\tau}\,
\left( \frac{b}{c1} \right)^{2\eps} \,
e^{-2 \gamma_E\,\eps}\;\Gamma(1-\eps)\Gamma(-\eps)
-2 \tau^{-1-\eps} \Gamma(1-\eps)
\left(-r_1\right)^{\eps/2}
K_{-\eps}\left( \frac{\tau b}{\sqrt{- r_1}} \right) = 
\notag \\
&\quad=
\delta(\tau) \, \frac{1}{2} \Bigg\{
\left(-r_1\right)^{\eps} 
\Gamma(1 - \eps) \Gamma(1 + \eps) \Gamma(-\eps)
\Bigg[ 
\frac{\Gamma(-\eps)}{\Gamma(1-\eps)^2}  \pFq{1}{2}\left(-\eps;\,1-\eps, 1-\eps;\,
-\frac{b^2}{4 r_1}\right)+
\notag \\
&\quad+
\left( \frac{b}{c1} \right)^{2\eps} \,
e^{-2\eps \gamma_E} \left(-r_1\right)^{-\eps} 
G^{2,\,0}_{1,\,3}\left( \frac{b^2}{4 r_1} \middle\vert \begin{array}{c}
1\\
0,0,-\eps\\
\end{array}
\right)
\Bigg] - 
\notag \\
&\quad-
\left( \frac{b}{c1} \right)^{2\eps} \,
e^{-2\eps \gamma_E} 
\Gamma(1 - \eps) \Gamma(-\eps) \left( 
H_{-\eps} - 2 \log{\left(\frac{b}{c_1}\right)} + \log{r_1}
\right)
\Bigg\} + 
\notag \\
&\quad+
\left(
\frac{1}{\tau}\,
\left( \frac{b}{c1} \right)^{2\eps} \,
e^{-2 \gamma_E\,\eps}\;\Gamma(1-\eps)\Gamma(-\eps)
-2 \tau^{-1-\eps} \Gamma(1-\eps)
\left(-r_1\right)^{\eps/2}
K_{-\eps}\left( \frac{\tau b}{\sqrt{- r_1}} \right)
\right)_+ = 
\notag \\
&\quad=
\frac{1}{2 \eps^2}\delta(\tau) + \frac{1}{\eps}\left[ 
-\tplus{1} + \frac{1}{2}\log{(-r_1)\delta(\tau)}
\right] 
-2 \log{\left(\frac{b}{c_1}\right)} \,\tplus{1}-
\notag \\
&\quad
- \delta(\tau)  \left[
\log^2{\left(\frac{b}{c_1}\right)} -
\log{\left(\frac{b}{c_1}\right)}\log{(-r_1)}
\right] + 
\substack{{\mbox{terms suppressed}}\\{\mbox{in the limit }b\to \infty}} + \mathcal{O}(\eps)
\end{align}
Therefore, the contribution of soft-collinear gluons in a $2$-jet configuration is obtained by inserting this result into Eq.~\eqref{eq:SCthrustfactor_FT_2}. We have:
\begin{align}
\label{eq:SCthrustfactor_FT_3}
&\widetilde{\Upsilon}_{+}^{[1],\,\text{ASY}}\left(\eps;\,\tau,b_T,\,y_1\right) = 
\notag \\
&=
\aS 2 \, C_F \, S_\eps \,
\Bigg\{
\frac{1}{\eps^2}\delta(\tau) + \frac{1}{\eps}
\left[ 
-2\tplus{1} + 
\delta(\tau) \log{(e^{-2 y_1})} + 
2 \delta(\tau) \logmu{}
\right] +
\notag \\
&\quad+
2\logbT{}
\left[
-2\tplus{1} + 
\delta(\tau) 
\left( 
\log{(e^{-2 y_1})} - \logbT{} + 2 \logmu{}
\right)
\right]\Bigg\}.
\end{align}

\paragraph{Collinear approximation} \hfill

\noindent The last contribution to the partonic tensor is associated to the radiation collinear to the fragmenting quark, obtained through the action of $T_A$. This is a TMD-relevant quantity by default. The approximation gives:
\begin{align}
\label{eq:1loop_TA_int}
&T_A \left[\widehat{W}_{f,\text{R}}^{\mu \nu, \,[1]}\left(\eps;\,z , \tau ,  k_T\right) \right] =
\int \frac{d \rho}{\rho} 
{}^{\star}\widehat{W}^{\mu \nu,\,[0]}_f ({z}/{\rho},\,Q) \, 
\Gamma_{q/q}^{[1]}\left(\eps;\,\rho,\,k_T,\,\tau\right),
\end{align}
where $\Gamma_{q/q}$ is the quark-from-quark generalized Fragmenting Jet Function\footnote{In the literature~\cite{Jain:2011iu,Makris:2020ltr}, the GFJFs are usually indicated by $\mathcal{G}$. In this paper, we will be consistent with the nomenclature introduced in this Section. Furthermore, in analogy to TMD FFs the GFJFs are usually defined with a normalization factor of $1/z$, which here is not considered.} (GFJF), which is diagonal in quark's flavors. It is defined in momentum space as the unsubtracted collinear parts, although its explicit dependence on thrust is implemented as in the usual jet thrust function. At 1-loop order it is defined\footnote{The definition is obtained by neglecting all mass corrections, as in Eq.~\eqref{eq:1loop_TB}.} as:
\begin{align}
\label{eq:Cthrustfactor_kT}
&\Gamma_{q/q}^{[1]}\left(\eps;\,z,\,k_T,\,\tau\right) =
\int \frac{d l^-}{(2\pi)^D}
\delta\left(\tau - \frac{z}{1-z} \frac{k_T^2}{Q^2}\right)
\notag \\
&\quad \times
\TrC \TrD \gamma^+
\left(
\begin{gathered}
\includegraphics[width=3.1cm]{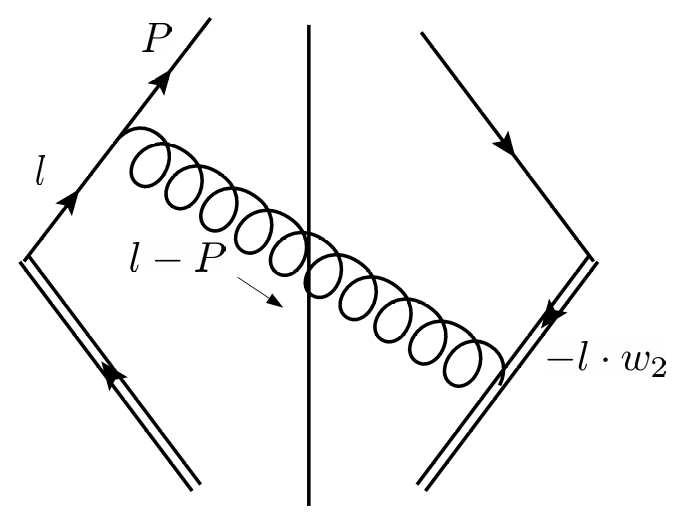}
\end{gathered} + h.c. +
\begin{gathered}
\includegraphics[width=2.8cm]{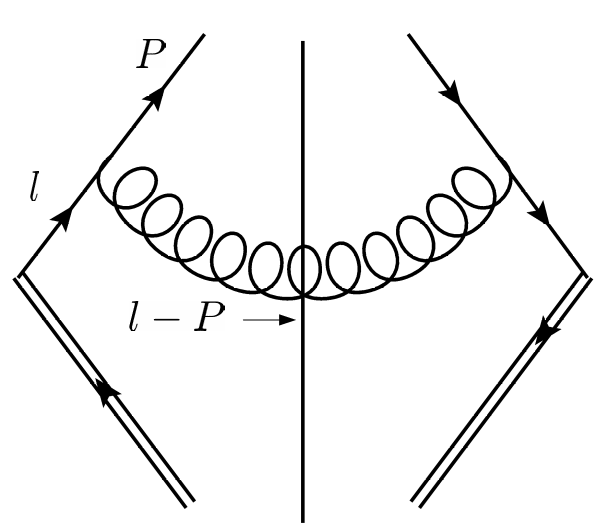}
\end{gathered}
\right) = 
\notag \\
&=
\aS 2 \, C_F \, S_\eps \,
\frac{\Gamma(1-\eps)}{\pi^{1-\eps}} \,
\mu^{2\eps}\,\frac{1}{k_T^2} \,
\left( \frac{2}{1-z} + (1-\eps)\,\frac{1-z}{z} \right)\,\theta(1-z)\,
\delta\left(\tau - \frac{z}{1-z} \frac{k_T^2}{Q^2}\right).
\end{align}
Notice that without the delta that fixes the thrust, the previous expression would have coincided with the unsubtracted quark-from-quark TMD FF in momentum space. Its Fourier transform gives:
\begin{align}
\label{eq:Cthrustfactor_FT}
&\widetilde{\Gamma}_{q/q}^{[1]}\left(\eps;\,z,\,b_T,\,\tau\right) =
\FT \Gamma_{q/q}^{[1]}\left(\eps;\,z,\,k_T,\,\tau\right)= 
\notag \\
&\quad=
\aS 2 \, C_F \, S_\eps \,
\left( \frac{\mu}{Q} \right)^{2\eps}
\,z^\eps 
\left(
2 (1-z)^{-1-\eps} + (1-\eps)\,\frac{(1-z)^{1-\eps}}{z}
\right)\,\times
\notag \\
&\quad \times
\tau^{-1-\eps} \,
\pFq{0}{1}\left( 1-\eps;\,-\tau \,\frac{1-z}{z}\frac{b^2}{4}\right).
\end{align}
The combination of $\tau^{-1-\eps}$ with the hypergeometric function $\pFq{0}{1}$ is computed in detail in Eq.~\eqref{eq:coll0F1_exp_2}
Such result holds for a function that is regular in $z \sim 1$. Therefore, we can use that solution only for computing the term proportional to $(1-z)^{1-\eps}$, but not for the term proportional to $(1-z)^{-1-\eps}$. 
The combination $(1-z)^{-1-\eps}\tau^{-1-\eps}\pFq{0}{1}$ can be treated by using a generalized version of the trick expressed in Eq.~\eqref{eq:distr_expansion}. However, the easiest way to obtain an expansion in distributions of $\tau$ and $z$ is to apply the usual trick of Eq.~\eqref{eq:distr_expansion} to $(1-z)^{-1-\eps}$ and $\tau^{-1-\eps}$ separately:
\begin{align}
\label{eq:ztau_0F1}
&z^\eps \, (1-z)^{-1-\eps} \, \tau^{-1-\eps} \pFq{0}{1}\left( 1-\eps;\,-\tau \,\frac{1-z}{z}\frac{b^2}{4}\right) = 
\notag \\
&\quad=
\frac{1}{\eps^2} \delta(\tau) \delta(1-z) -
\frac{1}{\eps} \left[
\delta(\tau)\zplus{1}+\delta(1-z)\tplus{1}
\right] + 
\notag \\
&\quad+
\delta(\tau) \zplus{\log{(1-z)}} -
\delta(\tau) \frac{\log{z}}{1-z} + 
\delta(1-z) \tplus{\log{\tau}} + 
\notag \\
&\quad+
\zplus{1} \, \tplus{1} \,
J_0\left( \sqrt{\tau \, \frac{1-z}{z}} \, b\right) + \mathcal{O}(\eps).
\end{align}
All terms containing either $\delta(1-z)$ or $\delta(\tau)$ are trivial, since the hypergeometric function evaluated in $\tau = 0$ and/or in $z=1$ gives just one. The only non-trivial term is the last line of the previous equation:
\begin{align}
\label{eq:ztau_combo_1}
\zplus{1} \, \tplus{1} \,
J_0\left( \sqrt{\tau \, \frac{1-z}{z}} \, b\right).
\end{align}
Next, we will make use of the following trick, valid for functions of a variable $x$ that varies in the range $0\leq x \leq1$, at most divergent as a simple pole at small values of $x$:
\begin{align}
\label{eq:trick_plusD}
\left(\frac{1}{x}\right)_+ f(x) = 
\delta(x) \, \int_0^1 d\alpha \left[f(\alpha)-f(0)\right] + 
\left(\frac{1}{x}\, f(x)\right)_+.
\end{align}
Proceeding in this way, we find:
\begin{align}
\label{eq:ztau_combo_2}
&\zplus{1} \, \tplus{1} \,
J_0\left( \sqrt{\tau \, \frac{1-z}{z}} \, b\right) = 
\notag \\
&=
\zplus{1} \Bigg\{
-\frac{1-z}{z}\frac{b^2}{4} \pFq{2}{3}\left( 
1,1;\,2,2,2;\,-\frac{1-z}{z}\frac{b^2}{4}
\right) \delta(\tau)
+ 
\left(\frac{1}{\tau} J_0\left( \sqrt{\tau \, \frac{1-z}{z}} \, b\right)\right)_+
\Bigg\}.
\end{align}
The contribution multiplying $\delta(\tau)$ can be treated similarly:
\begin{align}
\label{eq:ztau_combo_3}
&\zplus{1} \left[ 
-\frac{1-z}{z}\frac{b^2}{4} \pFq{2}{3}\left( 
1,1;\,2,2,2;\,-\frac{1-z}{z}\frac{b^2}{4}
\right)
\right] = 
\notag \\
&=
\delta(1-z) \left[
G^{3,\,0}_{1,\,3}\left( \frac{b^2}{4} \middle\vert \begin{array}{c}
1\\
0,0,0\\
\end{array}
\right)
-\frac{\pi^2}{6} - 2 \logb{2}
\right] +
\notag \\
&\quad+
\left( 
-\frac{1}{z}\frac{b^2}{4} \pFq{2}{3}\left( 
1,1;\,2,2,2;\,-\frac{1-z}{z}\frac{b^2}{4}
\right)
\right)_+ .
\end{align}
where $G^{3,\,0}_{1,\,3}$ is a Meijer G-function.
Next, we have to determine the large-$b$ asymptotic of $\widetilde{\Gamma}_{q/q}$ in order to investigate the $2$-jet limit of the collinear radiation term. At large $b_T$, the two terms in Eq.~\eqref{eq:ztau_combo_3} behaves as:
\begin{subequations}
\label{eq:ztau_combo_4}
\begin{align}
&G^{3,\,0}_{1,\,3}\left( \frac{b^2}{4} \middle\vert \begin{array}{c}
1\\
0,0,0\\
\end{array}
\right)
-\frac{\pi^2}{6} - 2 \logb{2}
=
-\frac{\pi^2}{6} - 2 \logb{2} + 
\substack{{\mbox{terms suppressed}}\\{\mbox{in the limit }b\to \infty}}\,;
\label{eq:ztau_combo_4_a}\\
&\left( 
-\frac{1}{z}\frac{b^2}{4} \pFq{2}{3}\left( 
1,1;\,2,2,2;\,-\frac{1-z}{z}\frac{b^2}{4}
\right)
\right)_+ = 
\notag \\
&=
-2 \logb{} \zplus{1} - 
\zplus{\log{(1-z)}} + 
\frac{\log{z}}{1-z} +  
\frac{\pi^2}{6} \delta(1-z) + 
\substack{{\mbox{terms suppressed}}\\{\mbox{in the limit }b\to \infty}}\,.
\label{eq:ztau_combo_4_b}
\end{align}
\end{subequations}
On the other hand, the contribution in the last line of Eq.~\eqref{eq:ztau_combo_2} has to be computed carefully, as it shows also a non-trivial dependence on $\tau$ besides that on $z$. 
A rather simple way to study such contribution is to investigate its action on two test functions $T(\tau)$ and $R(z)$. We have:
\begin{align}
\label{eq:ztau_combo_5}
&\int_0^1 dz R(z) \zplus{1} 
\int_0^1 d\tau T(\tau) \left(\frac{1}{\tau} J_0\left( \sqrt{\tau \, \frac{1-z}{z}} \, b\right)\right)_+ =
\notag \\
&\quad=
\int_0^1 d\tau \frac{T(\tau)-T(0)}{\tau} \Bigg\{
R(1) \int_0^1 \frac{dz}{1-z} \left[
J_0\left( \sqrt{\tau \, \frac{1-z}{z}} \, b\right)
\right]
+\notag \\
&\quad+
\int_0^1 dz \, \frac{R(z)-R(1)}{1-z}\,
J_0\left( \sqrt{\tau \, \frac{1-z}{z}} \, b\right)
\Bigg\} =
\notag \\
&\quad=
R(1) \int_0^1 d\tau \frac{T(\tau)-T(0)}{\tau}
\left[-2\logb{} - \log{\tau}\right]+ 
\substack{{\mbox{terms suppressed}}\\{\mbox{in the limit }b\to \infty}}.
\end{align}
Therefore, the difficult term in Eq.~\eqref{eq:ztau_combo_2} can be approximated as:
\begin{align}
\label{eq:ztau_combo_6}
&\zplus{1} 
\left(\frac{1}{\tau} J_0\left( \sqrt{\tau \, \frac{1-z}{z}} \, b\right)\right)_+ =
\notag \\
&\quad=
-\delta(1-z) \left[
2\logb{}\tplus{1} + \tplus{\log{\tau}}
\right] + 
\substack{{\mbox{terms suppressed}}\\{\mbox{in the limit }b\to \infty}}.
\end{align}
Combining together this result with Eqs.~\eqref{eq:ztau_combo_4}, we can finally write the large-$b$ behavior of the combination of distributions in Eq.~\eqref{eq:ztau_combo_1}:
\begin{align}
\label{eq:ztau_combo_final}
&\zplus{1} \, \tplus{1} \,
J_0\left( \sqrt{\tau \, \frac{1-z}{z}} \, b\right) = 
\notag \\
&\quad=
- 2 \logb{2} \delta(\tau) \delta(1-z) 
- 2 \logb{} \left[ \delta(\tau) \zplus{1} + \delta(1-z) \tplus{1} \right] -
\notag \\
&\quad-
\left[ \delta(\tau) \zplus{\log{(1-z)}} 
-\delta(\tau) \frac{\log{z}}{1-z}
+ \delta(1-z) \tplus{\log{\tau}} \right] + 
\substack{{\mbox{terms suppressed}}\\{\mbox{in the limit }b\to \infty}}.
\end{align}
%
\begin{figure}
\centering
\includegraphics[width=12cm]{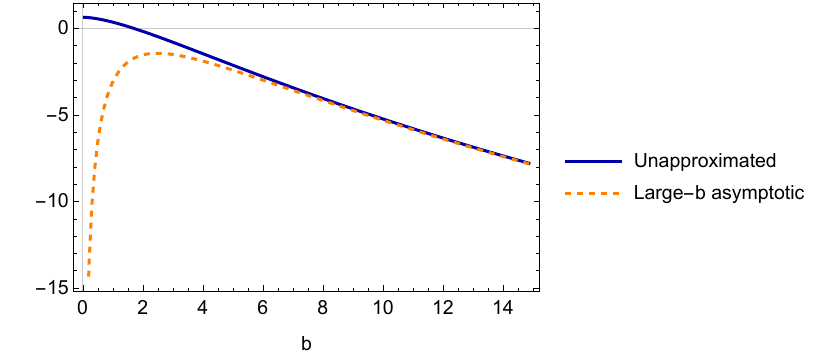}
\caption{The term $\zplus{1} \, \tplus{1} \,
J_0\left( \sqrt{\tau \, \frac{1-z}{z}} \, b\right)$ in Eq. \eqref{eq:ztau_combo_final} (solid, blue line) is compared to its large-$b$ behavior (orange, dashed line).  These lines are obtained by integrating 
with two test functions,  $T_{\tau}(\tau)=e^{-\tau}$ and $T_z(z)=e^{-(1-z)}$.}
\end{figure}
%
Notice how the last line of the previous equation cancels exactly the terms in the third line of Eq.~\eqref{eq:ztau_0F1}. In fact:
\begin{align}
\label{eq:results_for_Gamma_combodistr}
&z^\eps \, (1-z)^{-1-\eps} \, \tau^{-1-\eps} \pFq{0}{1}\left( 1-\eps;\,-\tau \,\frac{1-z}{z}\frac{b^2}{4}\right) = 
\notag \\
&\quad=
\frac{1}{\eps^2} \delta(\tau) \delta(1-z) -
\frac{1}{\eps} \left[
\delta(\tau)\zplus{1}+\delta(1-z)\tplus{1}
\right] -
\notag \\
&\quad
- 2 \logb{2} \delta(\tau) \delta(1-z) 
- 2 \logb{} \left[ \delta(\tau) \zplus{1} + \delta(1-z) \tplus{1} \right] +
\notag \\
&\quad+ 
\substack{{\mbox{terms suppressed}}\\{\mbox{in the limit }b\to \infty}} + \mathcal{O}(\eps).
\end{align}
This result, together with Eq.~\eqref{eq:coll0F1_exp_2}, allows to determine the large-$b$ asymptotic of $\widetilde{\Gamma}_{q/q}$:
\begin{align}
\label{eq:Cthrustfactor_FT_final}
&\widetilde{\Gamma}_{q/q}^{[1],\,\text{ASY}}\left(\eps;\,z,\,b_T,\,\tau\right)=
\notag \\
&\quad=
\frac{1}{z} Z_{q/q,\,\text{coll.}}^{[1]}\left(\eps;\,z\right) \delta(\tau) + 
\notag \\
&\quad+
\aS 2 \, C_F \, S_\eps \,\delta(1-z)
\left\{
\delta(\tau) \left[
\frac{2}{\eps^2} + \frac{1}{\eps} \left(
\frac{3}{2} + 4 \logmu{} \right)\right] + 
\frac{2}{\eps}\tplus{1}
\right\} + 
\notag \\
&\quad+
\aS 2 \, C_F \, S_\eps 
\Bigg\{
\delta(\tau)\Bigg[
2 \logbT{} \left( 
2 \zplus{1} + 1 -\frac{1}{z}
\right) -1 + \frac{1}{z} -
\notag \\
&\quad-
4 \delta(1-z) \logbT{} 
\left(\logbT{} -2 \logmu{} \right)
\Bigg] - 4 \delta(1-z)\tplus{1} \logbT{}
\Bigg\}.
\end{align}
Notice that all the non-trivial $z$-dependence associated with the poles is encoded into the function $Z_{q/q,\,\text{coll.}}$, which is the UV counterterm of the quark-from-quark collinear FF. This is not a coincidence, but rather the 1-loop expression of a crucial factorization theorem.

\bigskip

In fact, the ``pure collinear" radiation contribution is obtained after the subtraction of the overlapping terms in the soft-collinear momentum region. Therefore, combining the result of Eqs.~\eqref{eq:Cthrustfactor_FT_final} and~\eqref{eq:SCthrustfactor_FT_3}, we can write the following expression for the subtracted collinear part:
\begin{align}
\label{eq:subtract_collsoftA_1loop}
&\widetilde{\Gamma}_{q/q}^{[1],\,\text{ASY}}\left(\eps;\,\tau,\,z,\,b_T\right) - 
\delta(1-z) \, \widetilde{\Upsilon}_{+}^{[1],\,\text{ASY}}\left(\eps;\,\tau,b_T,\,y_1\right) = 
\delta(\tau) \, z \, \widetilde{D}_{q/q}^{[1],\,(0)}\left( \eps;\,z,\,b_T,\,y_1 \right),
\end{align}
where $\widetilde{D}_{q/q}^{(0)}$ is the \emph{bare} quark-from-quark TMD FF in $b_T$-space. It has both its characteristic collinear divergence, encoded into $Z_{q/q,\,\text{coll.}}$, and also the divergences that have to UV-renormalized by adding the counterterm $Z_{q/q,\,\text{TMD}}$.
By defining the \textbf{subtracted quark-from-quark GFJF} as:
\begin{align}
\label{eq:Gammaqq_subtr}
&\widetilde{\Gamma}_{q/q}^{\text{sub.}}\left(\eps;\,z,\,k_T,\,\tau,\,y_1\right) = 
\frac{\widetilde{\Gamma}_{q/q}\left(\eps;\,z,\,k_T,\,\tau\right)}{\widetilde{\Upsilon}_{+}\left(\eps;\,\tau,b_T,\,y_1\right)}
\end{align}
which is totally analogous to the factorization definition of the TMDs introduced in Eq.~\eqref{eq:sub_coll}, the result of Eq.~\eqref{eq:subtract_collsoftA_1loop} can be generalized to all orders, leading to:
\begin{align}
\label{eq:Gammaqq_subtr_ASY_fact_theo}
&\widetilde{\Gamma}_{q/q}^{\text{sub.}}\left(\eps;\,z,\,k_T,\,\tau,\,y_1\right) 
 \hspace{.1cm}\substack{2\text{-jet limit}\\ \sim}  \hspace{.15cm}
\delta(\tau) \, z \, \widetilde{D}_{q/q}^{(0)}\left( \eps;\,z,\,b_T,\,y_1 \right)
\end{align}
This result, together with Eq.~\eqref{eq:Sigma_ASY_fact_theo}, will be crucial in developing a suitable factorization theorem for Region 2.

\paragraph{Final result for forward radiation in Region 2}\hfill

\noindent Combining all the results of this section and inserting them into Eq.~\eqref{eq:1loop_xsf+_structure}, we obtain the final expression for the contribution of the radiation emitted in the $\mathrm{S}_A$-hemisphere to the partonic tensor, in Region 2. In $b_T$-space we have:
\begin{align}
\label{eq:1loop_xsf+_structure_after}
&\widetilde{\widehat{W}}_{f,\,\text{forward}}^{\mu \nu, \,[1]}\left(\eps;\,z , \tau ,  b_T\right) =
\notag \\
&= 
\int_z^1 \frac{d \rho}{\rho} 
{}^{\star}\widehat{W}^{\mu \nu,\,[0]}_f ({z}/{\rho},\,Q) 
\left[ 
\delta (1-\rho ) 
\hspace{-.1cm}
\left( \mathscr{S}_{+}^{[1]}\left(\eps;\tau, \, y_1\right) - 
\widetilde{\Upsilon}_{+}^{[1]}\left(\eps;\,\tau,b_T,\,y_1\right)
\right) +
\widetilde{\Gamma}_{q/q}^{[1]}\left(\eps;\,\tau,\,\rho,\,b_T\right)
\right]
\notag \\
&\hspace{.1cm}\substack{2\text{-jet limit}\\ \sim}  \hspace{.15cm}
\int_z^1 \frac{d \rho}{\rho} 
{}^{\star}\widehat{W}^{\mu \nu,\,[0]}_f ({z}/{\rho},\,Q) 
\delta(\tau) \, \left[ 
\delta (1-\rho ) \, \mathscr{S}_{+}^{[1]}\left(\eps;\tau, \, y_1\right)+
\rho \, \widetilde{D}_{q/q}^{[1],\,(0)}\left( \eps;\,\rho,\,b_T,\,y_1 \right)
\right],
\end{align}
where we have neglected all the power-suppressed terms.
Notice that we could easily have obtained the same result by neglecting from the very beginning, already in the transverse momentum space, any correlation between thrust and transverse momentum in the leading momentum regions associated to TMD-relevant contributions. Since in Region 2 the TMD-relevant terms constitute the \emph{subtracted} collinear radiation, such approximation is indeed the realization of assumption~\ref{hyp:1}. In fact, neglecting the relation between collinear transverse momentum and thrust corresponds to the kinematic configuration in which the detected hadron does not modify the topology of the final state and, ultimately, the measured value of $T$. This circumstance happens any time the detected hadron does not have a transverse momentum large enough to cause a significant spread of the jet to which it belongs to. 
The jet could well be wide, but not due to the direction of the detected hadron which, in Region 2, remains rather far from the jet external boundary.

\bigskip


\subsection{Factorization theorem for Region 2 \label{sec:reg2_fact_theo}}


\bigskip

The $b_T$-space expression for the partonic tensor, to 1-loop, in Region 2 follows from the results obtained in the previous sections for the various approximations. Summing the contributions of virtual Eq.~\eqref{eq:1loop_TH_structure}, backward Eq.~\eqref{eq:1loop_xsf-_structure_afterbT} and forward radiation Eq.~\eqref{eq:1loop_xsf+_structure_after} we obtain:
\begin{align}
\label{eq:1loop_xsR2_structure}
&\widetilde{\widehat{W}}_f^{\mu \nu, \,[1]}\left(\eps;\,z , \tau ,  b_T\right) =
\notag \\
&=
\widehat{H}_T^{\mu \nu} \, N_C \, e_f^2 \int \frac{d \rho}{\rho} 
\delta \left(1-{z}/{\rho}\right)  
\Bigg[
\delta (1-\rho ) \delta(\tau) \,  V^{[1]}(\eps) 
+ \delta(1-\rho) \left[  J^{[1]}\left(\eps;\tau\right) + S_{-}^{[1]}\left(\eps;\,\tau\right) \right] +
\notag \\
&
+ \delta(1-\rho) \,  \mathscr{S}_{+}^{[1]}\left(\eps;\tau,\,y_1\right) +
\delta(\tau) \, \rho \, \widetilde{D}_{q/q}^{[1],\,(0)}\left( \eps;\,\rho,\,b_T,\,y_1 \right) \Bigg] = 
\notag \\
&=
\widehat{H}_T^{\mu \nu} \, N_C \, e_f^2  
\Bigg[
\delta (1-z ) \Big( 
\delta(\tau) \, V^{[1]}(\eps) +J^{[1]}\left(\eps;\tau\right) + S_{-}^{[1]}\left(\eps;\,\tau\right) + \mathscr{S}_{+}^{[1]}\left(\eps;\tau,\,y_1\right) -
\notag \\
&-
\delta(\tau) \, Z_{q/q,\,\text{TMD}}(\eps;\,y_1)
\Big)
+ \delta(\tau) \, z \, \widetilde{D}_{q/q}^{[1]}\left( \eps;\,z,\,b_T,\,y_1 \right) \Bigg],
\end{align}
where we used Eq.~\eqref{eq:subtract_collsoftA_1loop} to rearrange  the combination of the large-$b_T$ asymptotic behavior of the quark-from-quark GFJF and that of the soft-collinear thrust factor into the bare quark-from-quark TMD FF.
Moreover, we added and subtracted the UV counterterm for the quark-from-quark TMD $Z_{q/q,\,\text{TMD}}$. In this way, we can drop the label ``$(0)$" from the TMD FF and consider it as a renormalized quantity. It still has a pole in $\eps$, but it is its expected collinear divergence, explicit in perturbative computations.

The combination of the other terms is a \emph{finite} quantity. This case be readily verified by substituting their explicit 1-loop expressions:
\begin{align}
\label{eq:partonicW_ch3}
&\delta(\tau) \,  V^{[1]}(\eps) 
 + J^{[1]}\left(\eps;\tau\right)  +
 S_{-}^{[1]}\left(\eps;\,\tau\right) + 
 \mathscr{S}_{+}^{[1]}\left(\eps;\tau,\,y_1\right) - \delta(\tau)  Z_{q/q,\,\text{TMD}}(\eps;\,y_1) =
\notag \\
&=
\aS \, C_F \, 
\Bigg[
\delta(\tau)\left( -9 + \frac{2\pi^2}{3} + \log^2{y_1} - 6\logmu{} + 4 \log{y_1} \logmu{}
 - 4\logmu{2}\right) -
 \notag \\
&\quad-
\tplus{1} \left( 3 + 4 \log{y_1} \right) - 4 \tplus{\log{\tau}}
\Bigg].
\end{align}
All the UV and rapidity divergences have been canceled. Moreover, we can readily verify the RG-invariance of the 1-loop cross section by deriving the previous expression with respect to $\log{\mu}$: the result is exactly \emph{minus} the anomalous dimension of the TMD to 1-loop. 
However, the dependence on the rapidity cut-off is problematic. The final result depends on $y_1$ as Eq.~\eqref{eq:partonicW_ch3} cannot compensate the variation with respect to the rapidity regulator encoded into the TMD. In fact, in the TMDs such dependence is ruled by CS-evolution, Eq. \eqref{eq:CSevo_TMD}, but clearly Eq.~\eqref{eq:partonicW_ch3} evolves differently, since it does not depend on $b_T$. This does not happen only at fix order in perturbation theory, but it will be a characteristic of the final, all-order result.

\bigskip

The final factorized cross section valid in Region 2 can be obtained by generalizing the 1-loop result of Eq.~\eqref{eq:1loop_xsR2_structure} to all orders. 
Although in general this is a potentially dangerous operation, in this case, the generalization of the 1-loop computation gives the correct final result. In fact, in Region 2 there is no asymmetry in the soft radiation, as it is considered TMD-irrelevant, regardless of whether it is emitted backward or forward with respect to the hemispheres identified by the thrust axis.
As all divergences cancel out among each other, we will drop the $\eps$-dependence in all  terms involved in the final result.

The generalization is achieved by reversing the operations that lead to the partonic tensor starting from its hadronic counterpart, as discussed in Section~\ref{sec:tmd_rel}. In particular, the relation between $\vec{P}_T$ and $\vec{k}_T$ involves the collinear momentum fraction, which is the Lorentz factor that relates the two frames in which the two transverse vectors are considered. As a consequence, the $z$ factor in front of the TMD FF found at fix order, Eqs.~\eqref{eq:1loop_xsf+_structure_after},~\eqref{eq:1loop_xsR2_structure}, is reabsorbed in the change of variables.
Therefore, the final result is:
\begin{align}
\label{eq:xs_R2_structure}
&\frac{d \sigma_{R_2}}{dz_h \, d^2 \vec{P}_T \, dT} = 
\sigma_B \, N_C \, V \, \hspace{-.1cm}
\int d \tau_{S_+} \, d \tau_{S_-} \, d\tau_B \; J(\tau_B) \, S_{-}(\tau_{S_-})  \mathscr{S}_{+}\left(\tau_{S_+},\,\zeta\right)
\delta(\tau - \tau_{S_+} - \tau_{S_-} - \tau_B) 
\notag \\
&\quad \times
\int \frac{d^2 \vec{b}_T}{(2\pi)^2} \, e^{i \, \frac{\vec{P}_T}{z_h} \cdot \vec{b}_T} \, 
\sum_f e_f^2 \;\widetilde{D}_{h/f}\left( z_h,\,b_T,\,\zeta \right) ,
\end{align}
where the rapidity cut-off has been recast into the variable $\zeta$, see Appendix \ref{app:review_collins}.

\bigskip


\subsubsection{Alternative proof \label{sec:reg2_topdown}}


\bigskip

It is important to notice that the same final result of Eq. \eqref{eq:xs_R2_structure}  could have been obtained directly 
by applying \hyp{1} and \hyp{2} in the hadronic tensor, as written in Eq.~\eqref{eq:W_prelim}. More precisely by implementing \hyp{1} and \hyp{2} straight into the delta functions which constrain the value of thrust 
$\delta\left(  T - T_{\text{def.}}(k_1,\,k_2,\dots,\,k_{N_{\text{jets}}},\,k_S)\right)$, and in the two delta functions constraining the momentum conservation and the measured transverse momentum of the final hadron, $\delta(q - k_1 - k_2 - \sum_{\alpha} k_\alpha - k_S)$ and 
$\delta\left( \vec{P}_T \left[ 1 + \mathcal{O}\left( \frac{P_T^2}{Q^2} \right) \right] + \frac{P^+}{k_1^+} 
\vec{k}_{T}( \vec{k}_{1,\,T}, \,\vec{k}_{S,\,T} ) \right)$, respectively.
This kind of ``top-down" approach to factorization, which was adopted in Ref.~\cite{Boglione:2020cwn},
is very powerful, as it allows us to avoid the traps and threats of the ``bottom-up" approach shown in the previous sections, entirely based on perturbation theory.
Here, we briefly sketch this methodology and contextualize it in the more general framework presented in this paper, especially highlighting the role of the kinematic requirements in the selection of the proper kinematic region.
Since the proof proceeds by exploiting very general arguments regarding momentum conservation and kinematics, it allows to derive the result for any considered topology of the final state, not only for $2$-jet configurations. 

Let's go back to Eq. \eqref{eq:W_prelim} and refer to the leading momentum regions of Fig.~\ref{fig:epm_regions}.
As discussed in Refs. \cite{Boglione:2020cwn,Boglione:2020auc}, among all collinear factors, only $\coll_1$ is actually relevant for studying the non-perturbative effects of hadronization, while all other $\coll_{i \neq 1}$ are effectively fully perturbative contributions, in the same sense of the hard factors.
Applying \hyp{1} to the delta on thrust (i.e. requiring the transverse momentum of the final hadron not to be extremely small, so that it does not affect the measured value of thrust) allows us to remove the whole $k_1$ dependence from $\delta(T - T_{\text{def.}})$. 
As $k_1$ is the only link that correlates $\coll_1$ to the thrust, this operation eliminates any $T$-dependent term from $\coll_1$. 
In other words, if \hyp{1} holds true, $\coll_1$ does not depend on thrust and it does not contribute to the thrust dependence of the cross section which, in turn, is insensitive to the topology of the final state. 

Much less trivial is the role of soft gluons in Eq.~\eqref{eq:W_prelim}, embedded in the soft factor $\mathbb{S}_{N_{\text{jets}}}$, defined similarly to Eq.~\eqref{eq:S2h_bTspace}, but with $N$ Wilson lines and modified to allow for the dependence on thrust, constrained to the value that $T$ assumes in the soft momentum region. 
Luckily, the momentum delta functions combined with the requirement \hyp{2} allows us to avoid its explicit, general calculation.
In fact, if soft transverse momenta cannot significantly deviate the detected hadron from the thrust axis, then any information about  $\vec{k}_{S,\,T}$ inside the delta that relates the measured $P_T$ to the transverse momentum of the fragmenting parton can be neglected,
so that in the last line of Eq.~\eqref{eq:W_prelim} we can set $\vec{k}_T \equiv \vec{k}_{1,\,T}$. 
Moreover, $\vec{k}_{S,\,T}$ can be deformed away from its power counting region in momentum conservation.
The consequence of this operation is that the residual dependence of the cross section on the weak components of $k_1$, $k_1^-$ and $k_{1T}$, 
is relagated to $\coll_1$, which is connected to the rest of the process only through a convolution on $k_1^+$.
Furthermore, at this point the soft factor has to be considered a fully perturbative object, on the same footage of the $\coll_\alpha$ factors, and 
the contribution of soft gluons, although non-trivial, turns out to be totally computable in pQCD as it involves only TMD-irrelevant terms. 

\bigskip

After applying \hyp{1} and \hyp{2}, we can follow the same strategy used for collinear factorization and gather all the perturbative contributions  (hard subgraphs, $\coll_\alpha$ factors, $\mathbb{S}_{N_\text{jets}}$) in a single thrust-dependent function playing the role of the partonic cross section. 
The structure of the final cross section will be the same of that obtained in a collinear factorization scheme but, due to the dependence on the transverse momentum of the detected hadron, the interpretation will be TMD.
The hadronic tensor can hence be written as:
\begin{align}
&W^{\mu \, \nu}_h(z_h,\,\vec{P}_T,\,T) = 
\sum_{j_1} \int_{P^+}^{{P^+}/{z_h}} d k_1^+ \,
\delta\left( \vec{P}_T \left[ 1 + \mathcal{O}\left( \frac{P_T^2}{Q^2} \right) \right] + \frac{P^+}{k_1^+} \vec{k}_{1,\,T} \right) 
\notag \\
&\quad \times
\int \frac{d k_1^- \, d^{D-2}\vec{k}_{1,\,T}}{(2 \pi)^D}
\, \mbox{Tr}_D \left\{ 
\mathrm{P}_1 \coll_{1}(k_1,\,P)_{j_1,\,H} \overline{\mathrm{P}}_1 
\mathcal{H}_{j_1}^{\mu \, \nu}(Q,\,k_1^+,\,T)
\right \}. \label{eq:W_prelim_2}
\end{align}
In the above equation, all the 
contributions that can be totally predicted by perturbative QCD have been collected in 
the factor $\mathcal{H}^{\mu \, \nu}_{j_1}$, which is clearly strictly related to the partonic tensor since $\widehat{W}_{j_1}^{\mu \nu} = \mbox{Tr}_D \{k_1^+ \gamma^- \mathcal{H}_{j_1}^{\mu \, \nu}\}$.
The TMD FFs are obtained by applying the Dirac projectors $\mathrm{P}_1$, $\overline{\mathrm{P}}_1$ to $\coll_1$.

Notice that since the partonic tensor is meant to represent the whole process at parton level, there is also a contribution associated to the radiation collinear to the fragmenting parton. Such term in the final cross section overlaps the collinear momentum region covered by the TMD FF and hence $\widehat{W}_{j_1}^{\mu \nu}$ must be equipped with a proper subtraction procedure~\cite{Boglione:2020cwn,Boglione:2020auc}. 
The final form of the factorization theorem for Region 2 can be written as:
\begin{align}
\label{eq:fact_theo_2}
&\frac{d \sigma}{dz_h \, d P_T^2 \,dT} = 
\pi \, \sum_{j} \int_{z_h}^1 \frac{d \widehat{z}}{\widehat{z}} \, 
\frac{d \widehat{\sigma}_j}
{d({z_h}/{\widehat{z}})\,dT} \;
D_{j,\,h}(\widehat{z},\, P_T) 
\, \left[1 + \mathcal{O}(\frac{P_T^2}{Q^2} ,\; \frac{M_h^2}{Q^2})\right].
\end{align}
where:
\begin{align}
&D_{j,\,h}(\widehat{z},\, P_T) = D_{1,\,h/j}(\widehat{z},\,P_T) 
\mp \frac{\widehat{z}}{M_h} S_T \, P_T \,
D_{1T,\,h/j}^{\perp}(\widehat{z},\,P_T).
\label{eq:TMDs_pT}
\end{align}

\bigskip

Since TMDs are properly defined in the Fourier conjugate space, it is more convenient to write the cross section using their $b_T$-space counterparts, given by:
\begin{equation} \label{eq:antiFT_PT}
D_{j,\,h}(\widehat{z}, P_T) \,
\left[1 + \mathcal{O}(\frac{P_T^2}{Q^2})\right]= 
\int \frac{d^{2} \vec{b}_T}{(2 \pi)^{2}} \, 
e^{i \, \frac{\vec{P}_T}{\widehat{z}} \cdot 
\vec{b}_T} \, \widetilde{D}_{j,\,h}(\widehat{z}, \,b_T),
\end{equation}
where the factor ${1}/{\widehat{z}}$ in the Fourier factor is due to the change of variables from $\vec{k}_T$ (h-frame) to $\vec{P}_T$ (p-frame), as $\vec{b}_T$ has been defined as the variable conjugate to $\vec{k}_T$.
The factorization theorem can also be written as:
\begin{align}
&\frac{d \sigma}{dz\, dP^2_T \,dT} = 
\pi \sum_{j} \int_{z_h}^1 \frac{d \widehat{z}}{\widehat{z}} \,
\frac{d \widehat{\sigma}_j}{d({z_h}/{\widehat{z}})\,dT} \,
\int \frac{d^{2} \vec{b}_T}{(2 \pi)^{2}} \, 
e^{i \, \frac{\vec{P}_T}{\widehat{z}} \cdot  \vec{b}_T} \, 
\widetilde{D}_{j,\,h}(\widehat{z}, \,b_T)
\, \left[1 + \mathcal{O}(\frac{M_h^2}{Q^2})\right].
\label{eq:fact_theo_bT}
\end{align}
Here the Fourier transform acts as an analytic continuation extending the TMD also beyond the original momentum region. This is an important effect to keep in mind when performing any phenomenological application based on formulae like Eq.~\eqref{eq:fact_theo_bT}. In fact, the TMD is originally  modeled in the $b_T$-space and afterward tested on data, in the transverse momentum space. Therefore, the cross section showed in Eq.~\eqref{eq:fact_theo_bT}, even if formally well defined for any value of $P_T$,
can only be trusted where $P_T \ll Q$ or, more precisely, where $P_T \ll P^+ = z_h \, {Q}/{\sqrt{2}}$, which is the actual 
condition that allows to consider the outgoing hadron as a collinear particle, according to the power counting rules.

Notice how the expression presented in Eq.~\eqref{eq:fact_theo_bT} beautifully matches with the factorization theorem obtained from a fully perturbative approach in Eq.~\eqref{eq:xs_R2_structure}. In fact, by comparing these two equations, the (subtracted) partonic cross section for the $2$-jet case is given by:
\begin{align}
\label{eq:part_xs_ch3}
&\frac{d \widehat{\sigma}_f^{2\text{-jets}}}{dz \, d\tau} = 
\sigma_B \, N_C \, V \,\delta(1-z)\, e_f^2 
\notag \\
&\quad \times
\int d \tau_{S_+} \, d \tau_{S_-} \, d\tau_B \; J(\tau_B) \, S_{-}(\tau_{S_-})  \mathscr{S}_{+}\left(\tau_{S_+},\,\zeta\right)
\delta(\tau - \tau_{S_+} - \tau_{S_-} - \tau_B).
\end{align}

\bigskip


\subsubsection{Collinear-TMD Factorization \label{sec:collTMD_fact}}


\bigskip

The structure of the factorized cross section obtained in the previous section is analogous to that obtained from a classic collinear factorization theorem, where all the fully perturbative contributions are gathered in a properly subtracted partonic cross section. This represents the full process to the parton level and is completely computable by using perturbation theory techniques. Besides the hard subgraphs, the partonic cross section also includes all the perturbative collinear factors associated to the production of hard jets and the soft factor, all of them thrust-dependent functions predicted, order by order, solely by pQCD. Differently from the usual collinear factorization theorems, however, the final cross section is sensitive to TMD effects, encoded into the collinear subgraph associated to the detected hadron. This contribution is therefore related to a TMD FF.

Therefore, the factorization theorem presented in Eq.~\eqref{eq:fact_theo_bT} can be considered as a hybrid version of collinear and TMD factorization, that we will call \textbf{collinear-TMD}. The analogies with the usual collinear factorization are summarized as follows:
\begin{center}
\begin{tabular}{ c | c }
 Collinear factorization & Collinear-TMD factorization  \\ 
 $\widehat{\sigma}_j \otimes \underbrace{d_{h/j}}_{\text{FF}} $ & $\widehat{\sigma}_j \otimes \underbrace{D_{h/j}}_{\text{TMD FF}} $  \\   
\end{tabular}
\end{center}
Collinear-TMD factorization theorems show a rather simple structure, to the benefit of phenomenological analyses. However, there are two issues that have not be considered so far.

First of all, the TMDs appearing in the factorized cross section are defined by the factorization definition of Eq.~\eqref{eq:sub_coll} instead of the square root definition~\cite{Collins:2011zzd, Aybat:2011zv}, commonly used in the Collins factorization formalism.
Notice that this perfectly matches the naive expectations based on the sole kinematic analysis, as discussed in Section~\ref{sec:kin_Reg}.
This is indeed a general feature of collinear-TMD factorization theorems. In fact, since the soft gluon contribution is totally perturbative, there are no other non-perturbative terms apart from those encoded into the TMDs. In practice, the TMD model extracted from $\epm \to h\,X$ is different from that extracted from $\epm \to h_1 \, h_2$, since the latter contains part of the information associated with the soft gluon emissions that correlate the two collinear parts related to the two hadrons, i.e. a square root of the soft model $M_S$, as explicitly shown in Ref.~\cite{Boglione:2020cwn} and reviewd in Appendix~\ref{app:review_collins}.
This is particularly relevant when performing a phenomenological analysis that combines data of single- and double hadron production, or, more generally, when comparing the TMDs extracted from a collinear-TMD factorization theorem, Eq.~\eqref{eq:fact_theo_2}, to the TMDs defined for the $2$-h class processes.
In these regards we refer, for example, to the two different approaches adopted in Refs.~\cite{DAlesio:2020wjq,Gamberg:2021iat}.

Secondly, in all the formulas provided above, the TMDs have been presented as depending only on transverse momentum and collinear momentum fraction. However, TMDs are also equipped with a rapidity cut-off $y_1$ as explicitly showed in the previous perturbative computations. This is 
required by the subtraction mechanism which removes the overlapping between soft and collinear 
momentum regions, as $y_1$ acts as a lower bound for the rapidity of the particles described by the TMD, which are supposed to be collinear, hence very fast moving along their reference direction. 
On the other hand, physical observables should not depend on the regularization procedure used to regulate the divergences encountered in their computations. Rapidity divergences are not an exception. However, clearly, the factorization theorems presented in this Section do not satisfy this requirement, as  the TMD FFs show a very specific dependence on $\zeta$, given by their Collins-Soper evolution equation, Eq.~\eqref{eq:CSevo_TMD}. The partonic cross section, whih is independent of $b_T$, cannot compensate for this dependence as it cannot evolve in the opposite way.

\bigskip


\subsubsection{The role of rapidity cut-offs \label{sec:role_of_y1}}


\bigskip

The consequence of this residual dependence on the rapidity cut-off left in the final factorized cross section is one of the issues, possibly the most important, related to collinear-TMD factorization theorems.
This might lead to think of an inconsistency\footnote{An inconsistency in the factorization theorem for Region 2 was already pointed out in Ref. \cite{Makris:2020ltr}.} of the factorization theorem. Here, however, we will follow a different philosophy, as we are convinced that the clear signal of the failure of a factorization procedure should be the presence of uncancelled divergences in the final cross section, which here is not the case. 
A consistent interpretation of a factorization theorem in the form of Eq.~\eqref{eq:fact_theo_2} can still be obtained by  reconsidering the role of the rapidity cut-offs. In particular, we interpret their explicit presence in the final result as an indication that they should be promoted from mere computational tools to quantities with a deeper physical meaning. 

An observable sensitive both to TMD effects and to thrust offers a beautiful chance to proceed along this path. In fact, there must be an intrinsic relation between the rapidity cut-off used in the Collins factorization formalism and the (experimentally  accessible) thrust, $T$. There is a simple kinematic argument that naively shows this relation. In fact, if $y_P$ is the rapidity of the detected hadron, then, neglecting all mass corrections, its minimum value is associated to the value of $T$. In fact, there is a kinematic constraint on ${P_T}/{z_h}$, which states that it cannot be larger than $\sqrt{1-T}\,Q$, see Ref.~\cite{Makris:2020ltr}. This follows directly from the definition of thrust in Eq.~\eqref{eq:thrust_def}. Therefore:
\begin{align}
\label{eq:kin_yPy1}
y_P = \frac{1}{2} \log{\frac{P^+}{P^-}} =  \log{\frac{z_h \, Q}{\sqrt{P_T^2 + M_h^2}}} 
\geq -\frac{1}{2} \log{(1-T)} + \mathcal{O}\left( \frac{M_h^2}{P_T^2} \right) ,
\end{align}
Then, thrust acts as a rapidity cut-off, since $y_P$ can be considered a good estimate for the rapidity of all the particles belonging to the same collinear group of the detected hadron.
However, within the Collins formalism, it is not possible to set a precise relation between $y_1$ and $T$. In this regard, Eq.~\eqref{eq:kin_yPy1} should be considered just as a naive argument based only on kinematics, without any claim of formal validity.

\bigskip

In the past, different methods to regulate rapidity divergences have been developed. One of the most elegant is the Rapidity Renormalization Group (RRG), used mainly in SCET-based approaches to factorization, as in Refs.~\cite{Chiu:2011qc,Chiu:2012ir}, where rapidity divergences are regularized applying a procedure totally analogous to that used for UV divergences, i.e. by introducing an auxiliary scale $\nu$, counterpart of $\mu$, and taking the derivative with respect to $\log{\nu}$, which introduces equations analogous to the CS evolution equations. The collinear and TMD factorization theorems obtained with RRG match exactly those derived within the Collins formalism. \cite{Collins:2012uy,Echevarria:2012js}. Therefore, collinear-TMD factorization must produce a cross section which is not RRG-invariant  and 
the nature of the scale $\nu$ must be deeply different from its RG counterpart. 
By extension, wherever the regularization procedure is used to treat rapidity divergences, the regulator must acquire a real physical meaning \cite{RRG:2021}. 
Collinear-TMD factorized observables, like the cross section presented in Eq.~\eqref{eq:fact_theo_bT}, may be the necessary tools to investigate this important feature.

\bigskip

A first attempt to assign more physical significance to the rapidity cut-off can be pursued by introducing a mechanism that allows to consider the TMDs invariant with respect to the choice of the rapidity cut-off.
This approach has been investigated in Ref.~\cite{Boglione:2020cwn}, where TMDs are equipped with a transformation, called Rapidity Dilation, that properly balance their non-perturbative content with the CS-evolution factor that regulates their dependence on $y_1$.
Such procedure should be viewed as a phenomenological tool, useful for the analysis of the factorized cross section of Eq.~\eqref{eq:fact_theo_bT} to lowest order and lowest log (LL).
However, within this (trivial) approximation, the relation between the rapidity cut-off and thrust can hardly be appreciated. Rapidity dilations can give some hint about the physical interpretation of the rapidity cut-off in the Collins factorization formalism, but do not relate $\zeta$ with $T$ explicitly. Furthermore, they do not explain the cross section dependence on the rapidity cut-off beyond the lowest order. 

The relation between $\zeta$ and $T$ becomes evident to NLO-NLL, and it can be made explicit by following the procedure presented in Ref.~\cite{Boglione:2020auc}, where the formalism is modified by introducing a topology cut-off $\lambda$ that forces the partonic cross section to describe the proper final state configuration, in this case a $2$-jet topology. This is a further variable, not included into the original Collins factorization formalism. Without this extra ingredient, a precise relation between the rapidity cut-off and thrust cannot be recovered. 
The topology cut-off $\lambda$ has a double function: on one side it forces the modified partonic cross section to describe the $2$-jet region, on the other side it constrains the total transverse momentum of the radiation collinear to the detected hadron to be in the power counting region. This last feature is crucial, as, for on-shell particles, a constraint on transverse momentum automatically leads to a constraint on rapidity (and vice-versa).
The double nature of the topology cut-off makes it more flexible than $\zeta$. Its direct relation to the topology of the process can be made more explicit by exploiting the kinematic argument of Eq.~\eqref{eq:kin_yPy1}. In fact, on one side $k_T \leq {P_T}/{z_h} \leq \sqrt{\tau}\, Q$, on the other $k_T \leq \lambda$ and the natural choice for the topology cut-off is $\lambda = \sqrt{\tau}\, Q$. Therefore, the limit $\lambda \to 0$ literally corresponds to the $2$-jet limit. 
The more $y_1$ approaches infinity, the narrower the jet in which the hadron is detected, approaching the pencil-like configuration.

\bigskip

Since in the Collins factorization formalism $\zeta$ and $T$ are not explicitly related, the cross section presented in Eq.~\eqref{eq:xs_R2_structure} cannot be properly thrust-resummed, as such operation necessarily needs the correct relation between the rapidity cut-off and thrust.
Consequently, the result obtained in this paper for Region 2 cannot be used for values of thrust extremely close to $T=1$, where the resummation effects become significant \cite{Catani:1991kz,Catani:1992ua}.

A simple way to obtain a formula valid far enough from the pencil-like configuration is the procedure adopted in the formalism described in Ref.~\cite{Boglione:2020cwn}, which consists in dropping the $\tau=0$ terms from the fixed order computation of Eq.~\eqref{eq:partonicW_ch3}. Notice that, differently from Ref.~\cite{Boglione:2020cwn}, this operation still leaves a residual dependence on the rapidity cut-off, i.e. the logarithmic term multiplying the $\tau$-plus distribution in the last line of Eq.~\eqref{eq:partonicW_ch3}. 
Without further constraints $\zeta$ can only be related to $T$ through the naive kinematics argument of Eq.~\eqref{eq:kin_yPy1} and set as $\zeta = \tau Q^2$. 
Of course, this identification can only be done after removing the $\tau=0$ terms, otherwise it would produce ill-defined contributions like $\delta(\tau) \, \log{\tau}$. This is the same kind of trouble that would appear if $y_1 = +\infty$ is set straightforwardly in the final factorization theorem of Eq.~\eqref{eq:fact_theo_2}.

Following this approach, the expression in Eq.~\eqref{eq:partonicW_ch3} leads to the same formula found in the formalism shown in Ref.~\cite{Boglione:2020auc}. In particular from Eqs.~\eqref{eq:partonicW_ch3} and~~\eqref{eq:part_xs_ch3}, the NLO partonic cross section becomes:
\begin{align}
&\frac{d \widehat{\sigma}_f^{2\text{-jets}}}{dz \, dT} \substack{{\text{NLO}}\\{=}}  
-\sigma_B \, e_f^2 \, C_F \, N_C \,  \delta(1-z)  \frac{3+8\log{(1-T)}}{1-T}.
\label{eq:partonic_xs_MODpheno}
\end{align}
As long as the phenomenological analyses do not include data associated to very large values of $T$, this formula for the partonic cross section is expected to be perfectly suitable. An extensive  phenomenological analysis based on this approach will be presented in Ref.~\cite{pheno:2021}.

\bigskip

Finally, notice that setting $\zeta = \tau \, Q^2$ in the final factorized cross section also introduces a thrust dependence into the TMD FFs.
However, this does not undermine their universality properties. In fact, such dependence only regards the rapidity cut-off and does not introduce any $T$-dependence into the functions that describe the non-perturbative behavior of the TMDs, i.e. $g_K$ and the model $M_D$ (see Appendix~\ref{app:review_collins}).

\bigskip


\section{Region 1: TMD factorization \label{sec:reg1}}


\bigskip

In this section we will consider the contribution of the forward radiation in Region 1. All the considerations on the 1-loop decomposition of the partonic tensor are the same derived for Region 2 at the beginning of Section~\ref{sec:reg2}. In fact, also in this case we have to compute the various approximations presented in Eq.~\eqref{eq:1loop_xsf+_structure}. 
In particular, the results obtained for soft-collinear and collinear radiation obtained for Region 2 extend automatically in Region 1, since also for this configurations they are TMD-relevant terms (see Tab.~\ref{tab:kinRef_tmdrel}). However, the contribution of the forward soft radiation is deeply different, since in Region 1 it is a TMD-relevant term as well.
In fact, in this region \hyp{2} is false and hence the size of the transverse momentum of the detected hadron has to be very small in order to be sensitive to soft radiation.
%
\begin{figure}
\centering
\includegraphics[width=6cm]{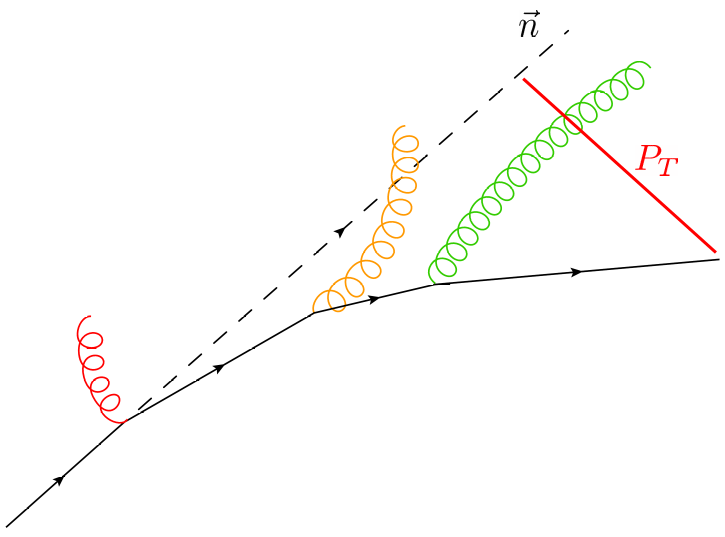}
\caption{Pictorial representation of the effect of a TMD-relevant soft radiation. The final $P_T$, measured with respect to the thrust axis $\vec{n}$ , is the result of several deflections due to emission/absorption of radiation. When the soft gluons (red) are TMD-relevant, they deflect the detected hadron affecting its transverse momentum. This effect must be added to the analogous contributions of soft-collinear gluons (orange) and collinear gluons (green). }
\label{fig:Reg1_pictorial}
\end{figure}
%

Clearly, the structure of the final factorization theorem derived for this region 
will be different from that presented in Eq.~\eqref{eq:fact_theo_2}. In fact, its structure will be rather similar to the standard TMD factorization theorems, see Refs.~\cite{Collins:2011zzd,Boglione:2020auc,Boglione:2020cwn}.

\paragraph{Soft approximation} \hfill

\noindent If the soft radiation is TMD-relevant, the action of $T_S^+$ gives a non-trivial dependence on $\vec{k}_T$:
\begin{align}
\label{eq:1loop_TS+_int}
&T_S^+\left[\widehat{W}_{f,\text{R}}^{\mu \nu, \,[1]}\left(\eps;\,z , \tau ,  k_T\right) \right] =
\int \frac{d \rho}{\rho} 
{}^{\star}\widehat{W}^{\mu \nu,\,[0]}_f ({z}/{\rho},\,Q) \,
\delta (1-\rho ) \Sigma_{+}^{[1]}\left(\eps;\tau, \, k_T\right),
\end{align}
where we have introduced the \textbf{soft thrust factor} $\Sigma$. It is defined exactly as $\ftsoft{2}$, presented in Eq.~\eqref{eq:S2h_bTspace}, but with an explicit dependence on thrust implemented as in the usual soft thrust function.
Therefore, $\Sigma$ depends not only on the total soft transverse momentum and on the rapidity cut-offs associated to the tilted Wilson lines, but also on $\tau$.
In Eq.~\eqref{eq:1loop_TS+_int} only the contribution of the soft thrust factor to the $\mathrm{S}_A$ hemisphere is involved.
To 1-loop order it is defined in momentum space as:
\begin{align}
\label{eq:cssSthrustfactor_plus}
&\Sigma_{+}^{[1]}\left(\eps;\tau, \, k_T,\,y_1,\,y_2\right) =
\notag \\
&=
\int \frac{d l^+ \, d l^-}{(2\pi)^D} \theta\left( l^+ - l^-\right)
\delta\left(\tau - \frac{l^-}{q^-}\right)
\left(
\begin{gathered}
\includegraphics[width=4.cm]{soft_thrust_new.png}
\end{gathered}
+ h.c.
\right) = 
\notag \\
&=
\aS 2 \, C_F \, S_\eps \,
\frac{\Gamma(1-\eps)}{\pi^{1-\eps}} \,
\mu^{2\eps}\,\frac{1}{k_T^2} \,
\int_{0}^{+\infty} \!\!dy \,
\frac{1+e^{-2(y_1-y_2)}}
{\left(1-e^{-2(y_1-y)} \right)
\left(1-e^{2(y_2-y)} \right)}
\delta \left( \tau - \frac{k_T}{Q} e^{-y}\right) + h.c.
\end{align}
Notice how the Heaviside theta enforces the gluon to be emitted in the (+)-hemisphere. Moreover, in analogy to the generalized soft thrust function $\mathscr{S}_{+}$ defined in Section~\ref{sec:backward_rad},  
we should expect that only the rapidity cut-off  relevant for the considered hemisphere (in this case $y_1$) will contribute explicitly to the final result. 
The Fourier transform of the Eq.~\eqref{eq:cssSthrustfactor_plus} can be written in the following form:
\begin{align}
\label{eq:cssSthrustfactor_plusFT_1}
&\widetilde{\Sigma}_{+}^{[1]}\left(\eps;\tau, \, b_T,\,y_1,\,y_2\right) =
\int d^{2-2\eps}\vec{k}_T \, e^{i \vec{k}_T \cdot \vec{b}_T} \,
\Sigma_{+}^{[1]}\left(\eps;\tau, \, k_T,\,y_1,\,y_2\right) =
\notag \\
&\quad=
\aS 2 \, C_F \, S_\eps \, 
\left( \frac{\mu}{Q} \right)^{2\eps}\,
\Gamma(1-\eps) \,
\frac{1+e^{-2(y_1-y_2)}}
{1-e^{-2(y_1-y_2)}} \,
\left( \frac{b}{c1} \right)^\eps \,
e^{-\gamma_E\,\eps} \, \tau^{-1-\eps} \times
\notag \\
&\quad\times\left( 
I_\eps \left( \tau \, b, e^{-2 y_1} \right)-
I_\eps \left( \tau \, b, e^{-2 y_2} \right)
\right) +h.c.,
\end{align}
where $b = b_T Q$ and where we have defined the integral:
\begin{align}
\label{eq:int_difficult}
I_\eps \left(a,\, r \right) =
\int_0^1 dx \frac{x^{{\eps}/{2}}}{x-r} \, J_{-\eps}\left( \frac{a}{\sqrt{x}}\right)
\end{align}
The solution to this integral requires advanced mathematical tools and non-standard techniques. 
From Eqs.~\eqref{eq:difficultint_I1} and~\eqref{eq:difficultint_I2}, the integral appearing in Eq.~\eqref{eq:cssSthrustfactor_plusFT_1} admit the following solutions:
\begin{subequations}
\label{eq:I1_I2_sol}
\begin{align}
&I_\eps(\tau \, b,\,r_1) = 
\left(\frac{\tau \, b}{2}\right)^{-\eps} \frac{1}{\eps \, \Gamma(1-\eps)} 
\pFq{1}{2} \left(-\eps;1-\eps,1-\eps;-\frac{\tau^2 \, b^2}{4}\right) +
\notag \\
&\quad
+\left(\frac{\tau \, b}{2}\right)^\eps \Gamma(-\eps) 
-2\left(-r_1\right)^{\eps/2}K_{-\eps}\left( \frac{\tau \, b}{\sqrt{- r_1}} \right) 
+ \mathcal{O}(r_1) ;
\label{sol_I1} \\
&I_\eps\left(\tau \, b ,\,\frac{1}{r_2}\right) = 
\mathcal{O}(r_2),
\label{sol_I2} 
\end{align}
\end{subequations}
where we set $r_1 = e^{-2 y_1}$ and $r_2 = e^{2 y_2}$. As expected, the dependence on $y_2$ vanishes in the limit $y_2 \to -\infty$ and the final result depends only on $y_1$. 

Inserting the Eqs.~\eqref{eq:I1_I2_sol} into Eq.~\eqref{eq:cssSthrustfactor_plusFT_1}, we obtain the expression for the Fourier transform soft thrust factor to  1-loop order:
\begin{align}
\label{eq:cssSthrustfactor_plusFT_2}
&\widetilde{\Sigma}_{+}^{[1]}\left(\eps;\tau, \, b_T,\,y_1\right)=
\notag \\
&\quad=
\aS 2 \, C_F \, S_\eps \, 
\left( \frac{\mu}{Q} \right)^{2\eps}\,
\Bigg\{
\frac{\tau^{-1-2\eps}}{\eps}\,
\pFq{1}{2} \left(-\eps;1-\eps,\,1-\eps;\,-\frac{\tau^2 \, b^2}{4}\right) +
\notag \\
&\quad+
\frac{1}{\tau}\,
\left( \frac{b}{c1} \right)^{2\eps} \,
e^{-2 \gamma_E\,\eps}\;\Gamma(1-\eps)\Gamma(-\eps)
-2 \tau^{-1-\eps}
\left(-e^{-2y_1}\right)^{\eps/2}
K_{-\eps}\left( \frac{\tau b}{\sqrt{- r_1}} \right)
+ 
\notag \\
&\quad+
\mathcal{O}\left(e^{-2y_1},\,e^{2y_2},\,e^{-2(y_1-y_2)}\right)
\Bigg\} +h.c.,
\end{align}
where $c_1 = 2 e^{-\gamma_E}$. 
The analogous contribution in the opposite hemisphere $\widetilde{\Sigma}_{-}$ can be easily obtained from Eq.~\eqref{eq:cssSthrustfactor_plusFT_2} by replacing $y_1$ with $-y_2$. This is a general property:
\begin{align}
\label{eq:SA_+-}
\widetilde{\Sigma}_{-}\left(\eps;\tau, \, b_T,\,y_2\right) = \widetilde{\Sigma}_{+}\left(\eps;\tau, \, b_T,\,-y_2\right).
\end{align}
Furthermore, a generalization of the factorization theorem in Eq.~\eqref{eq:generalized_S_fact_theo} holds for unintegrated quantities:
\begin{align}
\label{eq:Sigma_fact_theo}
\widetilde{\Sigma}\left(\eps;\tau, \, b_T,\,y_1 - y_2\right)= \widetilde{\Sigma}_{+}\left(\eps;\tau, \, b_T,\,y_1\right) \widetilde{\Sigma}_{-}\left(\eps;\tau, \, b_T,\,y_2\right) .
\end{align}
In Eq.~\eqref{eq:cssSthrustfactor_plusFT_2}, only the first term presents the expected ``soft" behavior for the thrust $\tau^{-1-2\eps}$, i.e. the same dependence shown by the usual soft thrust function at 1-loop order.
This is of no concern, as the extra terms will be canceled after the subtraction of the overlapping soft-collinear contribution, similarly to what was done for the case of the backward radiation.  
In fact, notice that the extra terms match exactly with the soft-collinear thrust factor $\Upsilon_+$ computed in Eq.~\eqref{eq:SCthrustfactor_FT}.
Therefore, only the ``pure soft", i.e. subtracted, term remains after  elimination of the double counting:
\begin{align}
\label{soft-collA_subtraction}
\widetilde{\Sigma}_{+}^{[1]}\left(\eps;\tau, \, b_T,\,y_1\right) - \widetilde{\Upsilon}_{+}^{[1]}\left(\eps;\,\tau,b_T,\,y_1\right) =  \widetilde{\Sigma}_{+}^{[1],\,\text{sub.}}\left(\eps;\tau, \, b_T\right)  .
\end{align}
which is the analogous of Eq.~\eqref{soft-collB_subtraction}. Such result, together with its counterpart in the opposite hemisphere, can be generalized to all orders leading to the following factorization theorems:
\begin{subequations}
\label{eq:Sigma_fact_theo_hemi}
\begin{align}
\widetilde{\Sigma}_{+}^{\text{sub.}}\left(\eps;\tau, \, b_T\right) = 
\frac{\widetilde{\Sigma}_{+}\left(\eps;\tau, \, b_T,\,y_1\right)}{ \widetilde{\Upsilon}_{+}\left(\eps;\,\tau,b_T,\,y_1\right) };
\label{eq:Sigma_fact_theo_minus}\\
\widetilde{\Sigma}_{-}^{\text{sub.}}\left(\eps;\tau, \, b_T\right) = 
\frac{\widetilde{\Sigma}_{-}\left(\eps;\tau, \, b_T,\,y_2\right)}{ \widetilde{\Upsilon}_{-}\left(\eps;\,\tau,b_T,\,y_2\right) }.
\label{eq:Sthr_fact_theo_minus_bis}
\end{align}
\end{subequations}
Together with Eq.~\eqref{eq:Sigma_fact_theo}, these factorization theorems lead to:
\begin{align}
\widetilde{\Sigma}_{+}^{\text{sub.}}\left(\eps;\tau, \, b_T\right)= 
\frac{\widetilde{\Sigma}_{+}\left(\eps;\tau, \, b_T,\,y_1\right)}{ \widetilde{\Upsilon}_{+}\left(\eps;\,\tau,b_T,\,y_1\right) } \, 
\frac{\widetilde{\Sigma}_{-}\left(\eps;\tau, \, b_T,\,y_2\right)}{ \widetilde{\Upsilon}_{-}\left(\eps;\,\tau,b_T,\,y_2\right) }.
\label{eq:SigmaPure_fact_theo}
\end{align}
which is the counterpart of Eq.~\eqref{eq:Sthr_fact_theo} for unintegrated quantities.

\bigskip

We will now focus on the ``purely  soft" term, i.e. the contribution in the first line of Eq.~\eqref{eq:cssSthrustfactor_plusFT_2}, obtained through the subtraction in Eq.~\eqref{soft-collA_subtraction}:
\begin{align}
\label{eq:cssSthrustfactor_pureS}
&\widetilde{\Sigma}_{+}^{[1],\,\text{sub.}}\left(\eps;\tau, \, b_T\right) =
\aS 4 \, C_F \, S_\eps \, 
\left( \frac{\mu}{Q} \right)^{2\eps}\,
\frac{\tau^{-1-2\eps}}{\eps}\,
\pFq{1}{2} \left(-\eps;1-\eps,\,1-\eps;\,-\frac{\tau^2 \, b^2}{4}\right),
\end{align}
where we have also added its complex conjugate counterpart. Notice that without the hypergeometric function $\pFq{1}{2}$, the subtracted soft thrust factor would be equal to the usual soft thrust function $S_+$, restricted to the $\mathrm{S}_A$ hemisphere.

The $2$-jet limit is obtained by studying the large $b_T$-asymptotic of Eq.~\eqref{eq:cssSthrustfactor_pureS}.
%
\begin{figure}
\centering
\includegraphics[width=12cm]{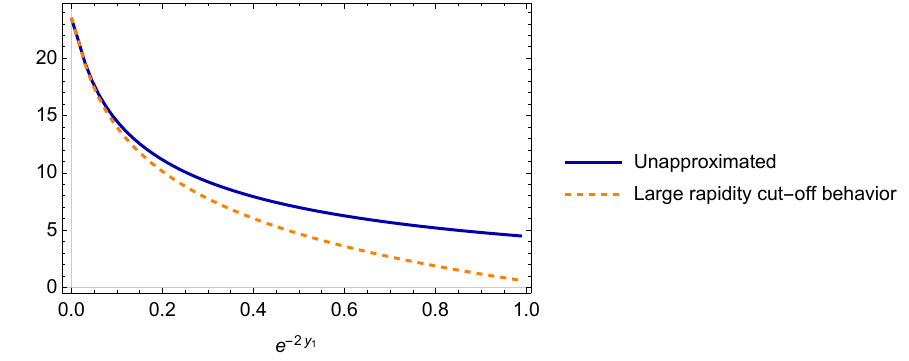}
\caption{The soft thrust factor $\widetilde{\Sigma}_{+}^{[1]}\left(\eps;\tau, \, b_T,\,y_1\right)$, (solid, blue line), Eq. \eqref{eq:cssSthrustfactor_plusFT_1}, is compared to its behavior for large values of the rapidity cut-off, $y_1 \to +\infty$ (orange, dashed line), Eq. \eqref{eq:cssSthrustfactor_plusFT_2}. These lines are obtained by integrating with a test function chosen as $T(\tau)=e^{-\tau}$.}
\end{figure}
%
Applying the trick of Eq.~\eqref{eq:distr_expansion} we can write:
\begin{align}
\label{eq:1F2puresoft_1}
&\frac{\tau^{-1-2\eps}}{\eps}\,
\pFq{1}{2} \left(-\eps;1-\eps,\,1-\eps;\,-\frac{\tau^2 \, b^2}{4}\right) =
\notag \\
&=
\delta(\tau) \, \left[
-\frac{1}{2 \eps^2} \, \pFq{2}{3} \left( -\eps,\,-\eps;\,1-\eps,\,1-\eps,\,1-\eps;\,-\frac{b^2}{4}\right)
\right] +
\notag \\
&+
\left( 
\frac{\tau^{-1-2\eps}}{\eps}\,
\pFq{1}{2} \left(-\eps;1-\eps,\,1-\eps;\,-\frac{\tau^2 \, b^2}{4}\right)
\right)_+ =
\notag \\
&=
\delta(\tau)
\left[
-\frac{1}{2\eps^2} + 
\frac{b^2}{8} \pFq{3}{4} \left( 1,1,1;\,2,2,2,2;\,-\frac{b^2}{4}\right)
\right] + 
\notag \\
&+
\frac{1}{\eps}\tplus{1} -
2\tplus{\log{\tau}} +
\left( \tau 
\frac{b^2}{4} \pFq{2}{3}\left( 1,1;\,2,2,2;\,-\frac{\tau^2 \, b^2}{4}\right)
\right)_+ + 
\mathcal{O}(\eps) =
\notag \\
&=
\delta(\tau)
\left[
-\frac{1}{2\eps^2} + 
\log^2\left( \frac{b}{c_1} \right)
\right]+
\frac{1}{\eps}\tplus{1} +
2\,\log\left( \frac{b}{c_1} \right)\tplus{1}+
\substack{{\mbox{terms suppressed}}\\{\mbox{in the limit }b\to \infty}} +
\mathcal{O}(\eps)
\end{align}
where in the second step we $\eps$-expanded and in the last one we extracted the asymptotic behavior at large-$b$.
%
\begin{figure}
\centering
\includegraphics[width=10cm]{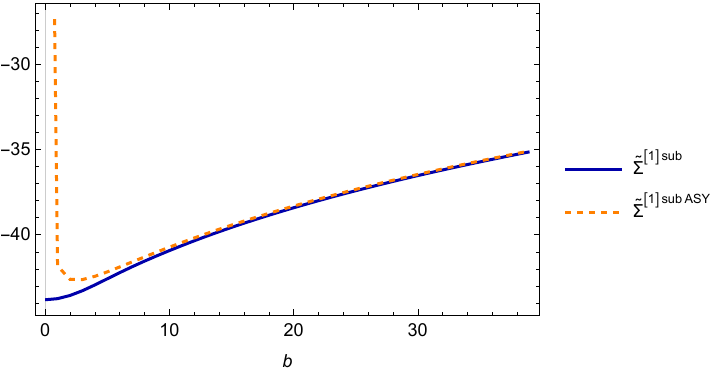}
\caption{
The subtracted soft thrust factor (solid, blue line), Eq.~\eqref{eq:cssSthrustfactor_pureS}, is compared to its asymptotic large-b behaviour (dashed, orange line), Eq.~\eqref{eq:cssSthrustfactor_pureS_2}.  These lines are obtained by integrating 
with a test function chosen as $T(\tau)=e^{-\tau}$.}
\end{figure}
%
Finally:
\begin{align}
\label{eq:cssSthrustfactor_pureS_2}
&\widetilde{\Sigma}_{+}^{[1],\,\text{sub., ASY}}\left(\eps;\tau, \, b_T\right)=
\aS 4 \, C_F \, S_\eps \, 
\Bigg\{
-\frac{1}{2\eps^2}\delta(\tau) +
-\frac{1}{\eps} \left[ 
\delta(\tau) \log{\left( \frac{\mu}{Q}\right)}-\tplus{1}
\right] +
\notag \\
&\quad+
\log\left( \frac{b_T \, \mu}{c_1} \right)
\left[ 
2 \tplus{1} + 
\delta(\tau) \left(
\log \left( \frac{b_T \, \mu}{c_1} \right) - 2 \log\left( \frac{\mu}{Q}\right) \right)
\right]
\Bigg\}.
\end{align}

\bigskip

The most interesting result of this Section, is that combining the result of Eq.~\eqref{eq:cssSthrustfactor_pureS_2} with Eq.~\eqref{eq:SCthrustfactor_FT_3} we can reconstruct the large-$b_T$ behavior of the unsubtracted soft thrust factor:
\begin{align}
\label{eq:uns_Sigma_ASY}
&\widetilde{\Sigma}_{+}^{[1],\,\text{ASY}}\left(\eps;\tau, \, b_T,\,y_1\right) =
\widetilde{\Sigma}_{+}^{[1],\,\text{ASY, sub.}}\left(\eps;\tau, \, b_T\right) +
\widetilde{\Upsilon}_{+}^{[1],\,\text{ASY}}\left(\eps;\,\tau,b_T,\,y_1\right) = 
\notag \\
&\quad=
- \aS \, 4 \, C_F \, \left( \frac{1}{\eps} \, y_1 + 2 y_1 \logbT{} \right) \, \delta(\tau) = 
\delta(\tau) \, \widetilde{\mathbb{S}}_{\text{\small {2}-h},\,+}^{[1],\,(0)} \left( \eps;\,b_T,\,y_1\right),
\end{align}
where $\widetilde{\mathbb{S}}_{\text{\small {2}-h},\,+}^{(0)}$ is the contribution of the $\mathrm{S}_A$-hemisphere to the \emph{bare} $2$-h soft factor, as defined in Eq.~\eqref{eq:S2h_bTspace}. 
From now on, it will be indicated as the (bare) \textbf{forward soft factor}. This result will be crucial in devising a factorization theorem suitable for Region 1. Moreover, thanks to the factorization theorems of Eqs.~\eqref{eq:Sigma_fact_theo} and~\eqref{eq:SigmaPure_fact_theo}, Eq.~\eqref{eq:uns_Sigma_ASY} can be generalized to all orders as:
\begin{subequations}
\label{eq:Sigma_ASY_fact_theo_hemi}
\begin{align}
&\widetilde{\Sigma}_{+}\left(\eps;\tau, \, b_T,\,y_1\right) \hspace{.1cm}\substack{2\text{-jet limit}\\ \sim}  \hspace{.15cm}
\widetilde{\Sigma}_{+}^{\text{ASY}}\left(\eps;\tau, \, b_T,\,y_1\right) = 
\delta(\tau) \widetilde{\mathbb{S}}_{\text{\small {2}-h},\,+}^{(0)} \left( \eps;\,b_T,\,y_1\right);
\label{eq:ASY_Sigmasoft1}\\
&\widetilde{\Sigma}_{-}\left(\eps;\tau, \, b_T,\,y_2\right) \hspace{.1cm}\substack{2\text{-jet limit}\\ \sim}  \hspace{.15cm}
\widetilde{\Sigma}_{-}^{\text{ASY}}\left(\eps;\tau, \, b_T,\,y_2\right) = 
\delta(\tau) \widetilde{\mathbb{S}}_{\text{\small {2}-h},\,-}^{(0)} \left( \eps;\,b_T,\,y_2\right),
\label{eq:ASY_Sigmasoft2}
\end{align}
\end{subequations}
where, by analogy, $\widetilde{\mathbb{S}}_{\text{\small {2}-h},\,-}$ defines the \textbf{backward soft factor}.
Furthermore, combining the theorems above:
\begin{align}
\label{eq:Sigma_ASY_fact_theo}
&\widetilde{\Sigma}\left(\eps;\tau, \, b_T,\,y_1 - y_2\right) \hspace{.1cm}\substack{2\text{-jet limit}\\ \sim}  \hspace{.15cm}
\delta(\tau) \ftsoft{2}^{(0)} \left( \eps;\,b_T,\,y_1 - y_2\right),
\end{align}
where $\ftsoft{2}$ is the \emph{same} soft factor appearing in $2$-h cross sections.
In fact, it is straightforward to show that $\ftsoft{2}$ is given by the product of the forward and the backward soft factors, as shows the following factorization theorem:
\begin{align}
\label{S2h_fact_theo}
\ftsoft{2}(b_T,\,\mu,\,y_1-y_2) = 
\widetilde{\mathbb{S}}_{\text{\small {2}-h},\,+} \left( \eps;\,b_T,\,\mu,\,y_1\right)
\widetilde{\mathbb{S}}_{\text{\small {2}-h},\,-} \left( \eps;\,b_T,\,\mu,\,y_2\right).
\end{align}

\paragraph{Final result for forward radiation in Region 1}  \hfill

\noindent Combining all the results of this section together with the formulae obtained for soft-collinear and collinear forward radiation in Section~\ref{sec:reg2}, we obtain the final expression for the contribution of the radiation emitted in the $\mathrm{S}_A$-hemisphere to the partonic tensor, in Region 1. In $b_T$-space we have:
\begin{align}
\label{eq:1loop_xsf+_structure_after_R1}
&\widetilde{\widehat{W}}_{f,\,\text{forward}}^{\mu \nu, \,[1]}\left(\eps;\,z , \tau ,  b_T\right) =
\notag \\
&= 
\int \frac{d \rho}{\rho} 
{}^{\star}\widehat{W}^{\mu \nu,\,[0]}_f ({z}/{\rho},\,Q) 
\left[ 
\delta (1-\rho ) 
\hspace{-.1cm}
\left( \Sigma_{+}^{[1]}\left(\eps;\tau, \, k_T\right) - 
\widetilde{\Upsilon}_{+}^{[1]}\left(\eps;\,\tau,b_T,\,y_1\right)
\right) +
\widetilde{\Gamma}_{q/q}^{[1]}\left(\eps;\,\tau,\,\rho,\,b_T\right)
\right]
\notag \\
&
\hspace{.1cm}\substack{2\text{-jet limit}\\ \sim}  \hspace{.15cm}
\int \frac{d \rho}{\rho} 
{}^{\star}\widehat{W}^{\mu \nu,\,[0]}_f ({z}/{\rho},\,Q) 
\delta(\tau) \, \left[ 
\delta (1-\rho ) \, \widetilde{\mathbb{S}}_{\text{\small {2}-h},\,+}^{[1],\,(0)} \left( \eps;\,b_T,\,y_1\right)+
\rho \, \widetilde{D}_{q/q}^{[1],\,(0)}\left( \eps;\,\rho,\,b_T,\,y_1 \right)
\right]
\end{align}
Notice that the whole dependence on the rapidity cut-off $y_1$ is washed out in the combination of $\widetilde{\mathbb{S}}_{\text{\small {2}-h},\,+}$ and the TMD FF. 
Moreover, we could have obtain easily the same result by neglecting from the very beginning, already in transverse momentum space, any relation between the thrust and the transverse momentum in the leading momentum regions associated to TMD-relevant contribution, which are the whole forward radiation terms in Region 1. In fact, in this kinematic configuration the detected hadron is so close to the thrust axis that it cannot affect the final state topology by any means.

\bigskip


\subsection{Factorization theorem for Region 1 \label{sec:reg1_fact_theo}}


\bigskip

The whole partonic tensor in Region 1, to 1-loop order, follows easily by summing the contributions of virtual, backward and forward radiation, computed in Eqs.~\eqref{eq:1loop_TH_structure},~\eqref{eq:1loop_xsf-_structure_afterbT} and~\eqref{eq:1loop_xsf+_structure_after_R1} respectively.
In $b_T$-space it is given by:
\begin{align}
\label{eq:1loop_xsR1_structure}
&\widetilde{\widehat{W}}_f^{\mu \nu, \,[1]}\left(\eps;\,z , \tau ,  b_T\right) = 
\notag \\
&= 
H_T^{\mu \nu} \, N_C \, e_f^2 \int \frac{d \rho}{\rho} 
\delta \left(1-{z}/{\rho}\right)  
\Bigg[
\delta (1-\rho ) \delta(\tau) \, V(\eps) 
+ \delta(1-\rho) \left[J^{[1]}\left(\eps;\tau\right) +  S_{-}^{[1]}\left(\eps;\,\tau\right)\right] +
\notag \\
&
+ \delta(\tau) \left[ \delta (1-\rho ) \, \widetilde{\mathbb{S}}_{\text{\small {2}-h},\,+}^{[1],\,(0)} \left( \eps;\,b_T,\,y_1\right)+
\rho \, \widetilde{D}_{q/q}^{[1],\,(0)}\left( \eps;\,\rho,\,b_T,\,y_1 \right) \right]
\Bigg] = 
\notag \\
&=
H_T^{\mu \nu} \, N_C \, e_f^2 
\Bigg[
\delta (1-z ) 
\left( \delta(\tau) \, V(\eps) +J^{[1]}\left(\eps;\tau\right) +  S_{-}^{[1]}\left(\eps;\,\tau\right)  +
\delta(\tau)  \widetilde{\mathbb{S}}_{\text{\small {2}-h},\,+}^{[1],\,(0)} \left( \eps;\,b_T,\,y_1\right) 
\right)+
\notag \\
&+
\delta(\tau) \, z\, \widetilde{D}_{q/q}^{[1],\,(0)}\left( \eps;\,z,\,b_T,\,y_1 \right)
\Bigg].
\end{align}
where we used Eq.~\eqref{eq:subtract_collsoftA_1loop} to rearrange  the combination of the large-$b_T$ asymptotic behavior of the quark-from-quark GFJF and that of the soft-collinear thrust factor into the bare quark-from-quark TMD FF.
Notice that all divergences cancel out among each other, except for the characteristic collinear divergence of the TMD FF, which cannot be dealt with in pQCD.
This can be readily verified by substituting the 1-loop expressions for each of the terms appearing in Eq.~\eqref{eq:1loop_xsR1_structure}.

\bigskip

This result can easily be generalized to all orders, analogously to what we did for Region 2 in Section~\ref{sec:reg2_topdown}. Because of the divergence cancellation, we can drop the $\eps$ dependence from the various contributions appearing in the final result and also the label ``(0)" from the TMD and the forward soft factor. 
Therefore, the factorization theorem for Region 1 is:
\begin{align}
\label{eq:xs_R1_structure}
&\frac{d \sigma_{R_1}}{dz_h \, d P_T^2 \, dT} = 
\sigma_B \, \pi N_C \, V \, 
\int d \tau_S \, d\tau_B \; J(\tau_B) \, S_{-}(\tau_S) \delta(\tau - \tau_S - \tau_B) \, \times
\notag \\
&\quad \times
\int \frac{d^2 \vec{b}_T}{(2\pi)^2} \, e^{i \, \frac{\vec{P}_T}{z_h} \cdot \vec{b}_T} \, 
  \widetilde{\mathbb{S}}_{\text{\small {2}-h},\,+} \left(b_T,\,\zeta \right) \;
\sum_f e_f^2 \;\widetilde{D}_{h/f}\left( z_h,\,b_T,\,\zeta \right) .
\end{align}
The rapidity cut-off has been recast into the variable $\zeta$, defined as in Appendix \ref{app:review_collins}. 
Region 1 is one of the cases where the all-orders generalization has to be performed with special care. The reason is that the contribution of the soft radiation is intrinsically asymmetric in this region, as the soft gluons emitted backward affects the value of thrust but not the transverse momentum of the detected hadron, while for the soft gluons emitted forwardly the situation is exactly the opposite. If there are more than two gluons, they can be emitted either in one hemispher or in the other. The asymmetry between these configurations produces new logarithmic terms correlating the emissions in the different hemispheres. Such terms are usually called Non-Global Logarithms (NGLs). The problem with NGLs is related to their resummation, which involves non-perturbative effects. Their contribution can be included into Eq.~\eqref{eq:xs_R1_structure} as an extra factor, see for instance Refs.~\cite{Dasgupta:2001sh,Banfi:2002hw,Larkoski:2015zka,Becher:2017nof}.
A correct treatment of NGLs is beyond the purpose of this article and they will not be included in our final factorization theorems. 

The same factorized cross section has been obtained within the framework of SCET in Ref.~\cite{Makris:2020ltr}, adopting a completely different approach. A factorization scheme similar to that presented in this Section for Region 1, was addressed  in Ref.~\cite{Kang:2020yqw}.

\bigskip

\subsubsection{The role of soft radiation \label{subsubsec:role-soft-rad}}

The factorization theorem of Eq.~\eqref{eq:xs_R1_structure} is profoundly different from that devised in Section~\ref{sec:reg2}.  
First of all, the TMD FFs are not the only factors that encode non-perturbative effects. In fact, the forward soft factor $\widetilde{\mathbb{S}}_{\text{\small {2}-h},\,+}$ has a non-trivial long-distance behavior as well. 
It can be written explicitly in terms of its perturbative and non-perturbative contributions in analogy to the procedure adopted for the $2$-h soft factor \cite{Boglione:2020cwn}. In fact, a CS-evolution equation for $\widetilde{\mathbb{S}}_{\text{\small {2}-h},\,+}$ can readily be written as:
\begin{align}
\label{eq:soft+_CSevo}
\frac{\partial}{\partial \log{\sqrt{\zeta}}}\widetilde{\mathbb{S}}_{\text{\small {2}-h},\,+} \left(b_T,\,\mu,\,\zeta \right) = 
-\frac{1}{2} \widetilde{K}\left( b_T,\,\mu\right).
\end{align}
This is nothing but the evolution equation with respect to $y_1$ of the $2$-h soft factor, Eq. \eqref{eq:S2hevo_1}.
Notice that $\widetilde{K}$ is the same Collins-Soper kernel appearing in the CS-evolution of the TMDs.
Therefore, by making use of the $b^\star$ prescription~\cite{Collins:2011zzd}, the forward soft factor can be written as a solution of Eq.~\eqref{eq:soft+_CSevo}:
\begin{equation} \label{eq:S2h+_final}
\widetilde{\mathbb{S}}_{\text{\small {2}-h},\,+} \left(b_T,\,\mu,\,\zeta \right)= 
e^{\frac{1}{4}\loggen{\zeta}{Q^2}{} \left[ \int_{\mu_0}^{\mu}\, \frac{d \mu'}{\mu'}\, \gamma_K\left(\alpha_S(\mu')\right) \; - 
\widetilde{K}(b_T^\star;\,\mu_0)\right]}
\sqrt{M_S}(b_T) \, e^{\frac{1}{4}\loggen{\zeta}{Q^2}{} \, g_K(b_T)},
\end{equation}  
which must be compared to the corresponding expression obtained for $\ftsoft{2}$ in Eq.~\eqref{eq:S2h_final}.
In the previous equation, the soft model of $\widetilde{\mathbb{S}}_{\text{\small {2}-h},\,+}$ is assumed to be insensitive to the hemisphere into which the soft radiation is emitted. In fact, it depends only on $b_T$, i.e. (simplifying) only on the transverse momentum of the soft radiation. Therefore it is completely unaware of the plus and minus components, which encode the selection of the emission direction. For this reason, in Eq.~\eqref{eq:S2h+_final} we have simply $M_{S_{+}} \equiv \sqrt{M_S}$.

\bigskip

Another crucial difference with collinear-TMD factorization theorems concerns the dependence on the rapidity cut-off. In fact, the whole dependence on $\zeta$ disappears in the combination $\widetilde{\mathbb{S}}_{\text{\small {2}-h},\,+} \, \widetilde{D}_{h/f}\left( z_h,\,b_T,\,\zeta \right)$ of Eq.~\eqref{eq:xs_R1_structure}, making the cross section in Region 1 CS-invariant. This can be directly checked by considering the solutions to the evolution equations for the forward soft factor and the TMD FF, i.e. Eq.~\eqref{eq:S2h+_final} and~\eqref{eq:tmd_evosol}, respectively. A direct check gives:
\begin{align} \label{eq:S2h+TMDFF_combo}
&\widetilde{\mathbb{S}}_{\text{\small {2}-h},\,+} \left(b_T,\,\mu,\,\zeta \right)
\widetilde{D}_{h/f}\left( z_h,\,b_T,\,\mu,\,\zeta \right) = 
\left(\widetilde{\mathcal{C}}_{j/f}(b_{T}^\star; \, \mu_b, \,\zeta_b) \otimes d_{j/h} (\mu_b)\right) (z_h) \times 
\notag \\
& \times 
\text{exp} \left\{
\frac{1}{4} \, \widetilde{K}(b_T^\star;\,\mu_b) \, \log{\frac{Q^2}{\zeta_b}} 
+ 
\int_{\mu_b}^{\mu} \frac{d \mu'}{\mu'} \, \left[ \gamma_C(\alpha_S(\mu'),\,1) - \frac{1}{4} \, 
\gamma_K(\alpha_S(\mu')) \, \log{\frac{Q^2}{\mu'^2}}\right]
\right\} \times 
\notag \\
& \times 
\left(M_C\right)_{j,\,h}(z_h,\,b_T) \sqrt{M_S(b_T)}
\text{exp} \left \{
-\frac{1}{4} \, g_K(b_T) \, \log{z_h^2 \, \frac{Q^2}{M_h^2}}.
\right\}
\end{align}  
Despite this equation has been written for an unpolarized TMD FF, it is totally general and can be applied to any TMD.
Notice that the neat effect induced by  $\widetilde{\mathbb{S}}_{\text{\small {2}-h},\,+}$ is a modification of the TMD model, which is multiplied by  $\sqrt{M_S(b_T)}$.
This is the same modification as  that introduced in $2$-h cross section in order to absorb the $2$-h soft factor into the definition of the TMDs.
Such operation leads to the square root definition of TMDs \cite{Collins:2011zzd, Aybat:2011zv}. However, in this case the same trick cannot be applied, despite the final results look the same.
In fact, naively, one might expect that the square root definition would correspond to the combination appearing in the cross section relative to  Region 1. This clearly cannot be possible, as the combination of Eq.~\eqref{eq:S2h+TMDFF_combo} is CS-invariant, while the square root definitions obey to the CS-evolution equations that regulate the behavior with respect to the rapidity cut-off $y_n$. A direct comparison shows that:
\begin{align}
\label{eq:comparison_hemi_sqrt}
&\widetilde{D}_{h/f}^{sqrt}\left( z_h,\,b_T,\,\mu,\,y_n \right) = 
\widetilde{\mathbb{S}}_{\text{\small {2}-h},\,+} \left(b_T,\,\mu,\,y_1 \right)
\widetilde{D}_{h/f}\left( z_h,\,b_T,\,\mu,\,y_1 \right) \, \text{exp}\left( -\frac{y_n}{2} \widetilde{K}(b_T,\,\mu \right).
\end{align}
Therefore, the square root definition and the combination of Eq.~\eqref{eq:S2h+TMDFF_combo} coincide only if $y_n=0$.

\bigskip

From the point of view of the phenomenological analyses, Eq.~\eqref{eq:S2h+TMDFF_combo} is particularly relevant. The TMD model $M_D^{\text{sqrt}} = M_D \times \sqrt{M_S}$ as defined in the square root definition can easily be experimentally accessed as a whole, but its two inner components cannot be disentangled. As for the $2$-h class cross sections, there are in total three unknown non-perturbative functions, $g_K$, $M_D$ and $M_S$. 
The square root definition is useful in that it effectively decreases the number of unknowns reducing them to $M_D^{\text{sqrt}}$ and $g_K$,  as usual in the standard TMD factorization theorems. Due to these analogies, we will refer to the cross section for Region 1, presented in Eq.~\eqref{eq:xs_R1_structure}, as a \textbf{TMD factorization theorem}.

TMD and collinear-TMD factorizations do not differ only in their phenomenological applications. They are different in  spirit. In fact, the rapidity cut-off is totally irrelevant in TMD factorization, even when thrust is measured. In this regard, notice that the kinematic argument of Eq.~\eqref{eq:kin_yPy1} widely used to discuss the role of the rapidity cut-off in collinear-TMD factorization brakes down when $P_T$ becomes too small, which is the main feature of Region 1. 
TMD factorization can play a leading role in the investigation of the role of soft physics, as all the interesting dependence on the long-distances effects induced by soft radiations is encoded in the TMD factorized cross section. However, it is totally blind to the effects associated to the rapidity cut-off which,  as  discusses in Section~\ref{sec:role_of_y1}, it 
should be assigned a proper physical meaning, given its relation to the measured value of thrust.

\bigskip


\section{Region 3: generalized collinear factorization \label{sec:reg3}}


\bigskip

The last configuration to be considered corresponds to the Region 3, where the kinematic requirements \hyp{1} is false and hence the transverse momentum of the detected hadron is large enough to significantly affect the topological configuration of the final state.
In particular, $P_T$ is large enough to be insensitive also to soft-collinear emissions, other than to soft radiation. Therefore, similarly to Region 2, the soft contribution is TMD-irrelevant, but here the overlapping region, associated to soft-collinear radiation, is TMD-irrelevant too.
%
\begin{figure}
\centering
\includegraphics[width=6.5cm]{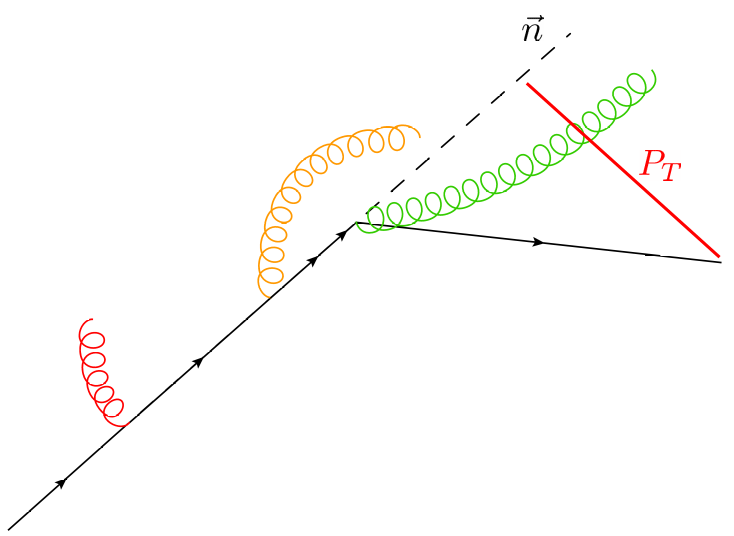}
\caption{Pictorial representation of the effect of the three kinds of radiation in Region 3. The final $P_T$, measured with respect to the thrust axis $\vec{n}$ , is not affected by the emission/absorption of soft  and soft-collinear radiations as the soft (red) and the soft-collinear (orange) gluons are TMD-irrelevant. The transverse momentum of the detected hadron is generated only by collinear radiations (green). }
\label{fig:Reg3_pictorial}
\end{figure}
%
As a consequence, only collinear radiation produces significant TMD effects.

\bigskip

This kinematic configuration will inevitably produce yet another factorization theorem, different from those obtained for Region 1 and 2.  Most importantly, this is the only kinematic region where the TMD effects are not described by TMD FFs, but rather by GFJFs, which are as universal as the TMDs but have a further explicit dependence on the invariant mass of the jet to which they are associated. In this case, the invariant mass of the jet is related to thrust. 

\bigskip

The explicit computation to 1-loop is straightforward, as all the necessary ingredients have already been worked out in the previous Sections. In particular, the action of $T_S^+$ gives the same result found for Region 2, Eq.~\eqref{eq:1loop_TS+_reg2}. On the other hand, the soft-collinear term, given by the action of $T_S T_A$, is totally analogous to its opposite hemisphere counterpart, and can be easily obtained through Eq.~\eqref{eq:SB_+-}.

The main difference with the other two kinematic regions regards the only $b_T$-dependent quantity, i.e. the quark-from-quark GFJF defined in Eq.~\eqref{eq:Cthrustfactor_kT}, obtained through the action of $T_A$. 
When the detected hadron causes the jet spreading, then the whole final state is inevitably far from the ideal pencil-like configuration, making the value of the thrust to decrease. 
If this is the case, then the relation between $k_T$ and $\tau$ in Eq.~\eqref{eq:thurstCollFact_gluon} cannot be neglected anymore. Moreover, the large-$b_T$ limit is not a faithful representation of this kinematics configuration, as the thrust cannot reach the ideal limit $\tau=0$ because of the size of $k_T$, large enough to forbid a pencil-like final state. In this case, the final result in $b_T$-space is given by the Fourier transform of the gluon-from-quark GFJF of Eq.~\eqref{eq:FTthurstCollFact_gluon}. 

\bigskip


\subsection{Factorization theorem for Region 3 \label{sec:reg3_fact_theo}}


\bigskip

The 1-loop order of the partonic tensor in Region 3 is given by:
\begin{align}
\label{eq:1loop_xsR3_structure}
&\widetilde{\widehat{W}}_f^{\mu \nu, \,[1]}\left(\eps;\,z , \tau ,  k_T\right) = 
\notag \\
&= 
H_T^{\mu \nu} \, N_C \, e_f^2 \int \frac{d \rho}{\rho} 
\delta \left(1-{z}/{\rho}\right)  
\Bigg[
\delta (1-\rho ) \delta(\tau) \, V(\eps) 
+ \delta(1-\rho) \left[  J^{[1]}\left(\eps;\tau\right) + S_{-}^{[1]}\left(\eps;\,\tau\right) \right] +
\notag \\
&
+ \delta(1-\rho) \,  S_{+}^{[1]}\left(\eps;\tau \right) +
\widetilde{\Gamma}_{q/q}^{[1]}\left( \eps;\,\rho,\,b_T,\,\tau \right) \Bigg] = 
\notag \\
&=
H_T^{\mu \nu} \, N_C \, e_f^2  
\Bigg[
\delta (1-z ) \left( 
\delta(\tau) \, V(\eps) +J^{[1]}\left(\eps;\tau\right) + S^{[1]}\left(\eps;\tau\right) 
\right)
+ \Gamma_{q/q}^{[1]}\left( \eps;\,z,\,k_T,\,\tau \right)  \Bigg],
\end{align}
where we used the $\mathrm{S}_A$-hemisphere counterpart of Eq.~\eqref{soft-collB_subtraction} to reorganize the
soft and the soft-collinear terms into the forward radiation contribution to the usual thrust function, which is then combined with its counterpart in the opposite hemisphere. Therefore, the usual soft thrust function appears as a fundamental ingredient of the partonic tensor in  Eq.~\eqref{eq:1loop_xsR3_structure}. 
Furthermore, notice that in this region there is no trace of rapidity cut-offs, as all the soft and soft-collinear terms are integrated over the transverse momentum.

\bigskip

The cross section for Region 3 is obtained by generalizing the 1-loop result for the partonic tensor of Eq.~\eqref{eq:1loop_xsR3_structure} to all orders. 
Then the final cross section can be written as:
\begin{align}
\label{eq:xs_R3_structure}
&\frac{d \sigma_{R_3}}{dz_h \, d P_T^2 \, dT} = 
\notag \\
&\quad=
\sigma_B \, \pi N_C \, V \, z_h \!
\int \! d \tau_{S}  \, d\tau_A \, d\tau_B\; J(\tau_B) \, S(\tau_{S})  \, 
\! \sum_f \! e_f^2 \Gamma_{h/f}\!\left( z_h,\,\frac{P_T}{z_h},\,\tau_A \right) 
\delta(\tau - \tau_{S} - \tau_A - \tau_B), 
\end{align}
where the presence of the $z_h$ factor is due to the unconventional normalization of the GFJFs. Notice that the same factorized cross section has been obtained in Ref.~\cite{Makris:2020ltr}, within the framework of SCET, adopting a completely different approach.
The structure of this cross section is similar to the factorization theorem devised for Region 2. In fact, naively,  Eq.~\eqref{eq:xs_R3_structure} can be obtained from Eq.~\eqref{eq:1loop_xsR3_structure} by removing the rapidity cut-off and replacing the TMD FFs with the corresponding GFJFs. 
It is interesting to notice that here the hybrid nature of the collinear-TMD factorization theorem is ``transferred" from the structure of the cross section to its TMD part. In fact, the GFJFs have the same features of the TMD FFs (they depend on the transverse momentum of the fragmenting parton, as well as on the momentum fraction $z_h$), but follow the DGLAP evolution equations tipical of the collinear FFs.
However, differently from the TMDs, GFJFs do not require a  rapidity cut-off, since all the rapidity divergences are regulated by the additional dependence on the invariant mass of the jet (a role played by thrust in this case).
Therefore, they should more properly be considered like a ``generalized" version of the usual collinear FFs, rather than an extension of the TMD FFs.
For this reasons, the cross section presented in Eq.~\eqref{eq:xs_R3_structure} will be referred to as the  \textbf{generalized collinear factorization theorem}.

Despite the similarities, the factorization theorems of Region 2 and Region 3 are remarkably different.
In fact, the factorized cross section of Eq.~\eqref{eq:xs_R3_structure} is not of any use for the investigation of the physical meaning of the rapidity cut-off, as they are defined without any explicit rapidity regulator.
Further considerations on GFJFs are beyond the purpose of this work. For more details we refer the reader to Ref.~\cite{Jain:2011iu} and references therein.

\bigskip


\section{ Algorithm for Region selection \label{sec:reg_algorithm}}


\bigskip

So far, three different factorization theorems corresponding to as many different kinematics regions have been developed: TMD factorization for Region 1 in Eq.~\eqref{eq:xs_R1_structure}, collinear-TMD factorization for Region 2 in Eq.~\eqref{eq:xs_R2_structure} and generalized collinear factorization for Region 3 in Eq.~\eqref{eq:xs_R3_structure}.
Each region corresponds to a different physical configuration and, in particular, the transverse momentum of the detected hadron increases as we move from region 1 to region 3.
In fact, in region 1 the soft radiation contributes actively to the transverse deviation of the hadron with respect to the thrust axis, which  must then have a low transverse momentum, otherwise it would not be sensitive to these tiny corrections.
On the other hand, in Region 3, the detected hadron has a transverse momentum large enough to be among the causes of the spread of the jet in which it is detected, inevitably decreasing the value of thrust.
However, lacking a proper criterion to discriminate among the regions, these three factorization theorems are hardly useful for phenomenological analyses.
Moreover, the boundaries of the three regions are not sharply defined, making the description of data  difficult, especially in the overlapping regions.

\bigskip

In order to define a suitable algorithm for selecting each individual region, it is useful to review the approximations that lead to the three factorization theorems presented in this Chapter. Depending on the initial assumptions~\hyp{1} and~\hyp{2}, organized\footnote{Notice that the combination where both the initial assumptions are false does not correspond to any kinematic regions as it is kinematically forbidden.} as in Tab.~\ref{tab:kinRef_hyp}, the leading momentum regions associated to the forward radiation contribute differently to the observed TMD effects and hence the soft, soft-collinear and collinear terms can be classified on the basis of their TMD-relevance, as in Tab.~\ref{tab:kinRef_tmdrel}.  
In other words, Tab.~\ref{tab:kinRef_hyp} implies Tab.~\ref{tab:kinRef_tmdrel}.
Moreover, we have shown how the final results could have been found by neglecting the correlation between thrust and transverse momentum from the very beginning, at the level of definition of the various factors contributing to the forward radiation. Such further approximations make the perturbative computations much easier, as all the issues related to the $2$-jet limit in $b_T$-space clearly disappear when $\tau$ and $k_T$ are independent variables.

In Region 1, this is equivalent to implementing the following approximations in the definitions of the various factors in transverse momentum space, Eqs.~\eqref{eq:cssSthrustfactor_plus},~\eqref{eq:SCthrustfactor_kT} and~\eqref{eq:Cthrustfactor_kT}:
\begin{subequations}
\label{eq:reg1_replacements}
\begin{align}
&\delta\left( \tau - \frac{l^-}{q^-} \right) = \delta\left( \tau - \frac{k_T}{Q}  e^{-y} \right) \sim \delta(\tau) \quad\text{both in }\Sigma_+\text{ and in }\Upsilon_+ ;
\label{eq:reg1_replacements_S} \\
&\delta\left( \tau - \frac{z}{1-z}\frac{k_T^2}{Q^2} \right) =\delta\left( \tau - (1-z)\,z \, e^{-2 y} \right) \sim \delta(\tau) \quad\text{in }\Gamma_{q/q},
\label{eq:reg1_replacements_C}
\end{align}
\end{subequations}
where $y$ is the rapidity of the emitted gluon.
Analogously, the result of Region 2 could have been obtained by leaving $\mathscr{S}^+$ unchanged and setting:
\begin{subequations}
\label{eq:reg2_replacements}
\begin{align}
&\delta\left( \tau - \frac{l^-}{q^-} \right) = \delta\left( \tau - \frac{k_T}{Q}  e^{-y} \right) \sim \delta(\tau) \quad\text{in }\Upsilon_+ ;
\label{eq:reg2_replacements_S} \\
&\delta\left( \tau - \frac{z}{1-z}\frac{k_T^2}{Q^2} \right) =\delta\left( \tau - (1-z)\,z \, e^{-2 y} \right) \sim \delta(\tau) \quad\text{in }\Gamma_{q/q}.
\label{eq:reg2_replacements_C}
\end{align}
\end{subequations}
Finally, in Region 3 all the relations between thrust and transverse momentum have to be kept into the definitions of the factors:
\begin{align}
\label{eq:reg3_replacements}
&\delta\left( \tau - \frac{z}{1-z}\frac{k_T^2}{Q^2} \right) =\delta\left( \tau - (1-z)\,z \, e^{-2 y} \right)  \quad\text{not approximated in }\Gamma_{q/q}.
\end{align}
Eqs.~\eqref{eq:reg1_replacements},~\eqref{eq:reg2_replacements} and~\eqref{eq:reg3_replacements} follow directly from the classification of Tab.~\ref{tab:kinRef_tmdrel}.

\bigskip

\begin{figure}
\centering
\includegraphics[width=7.5cm]{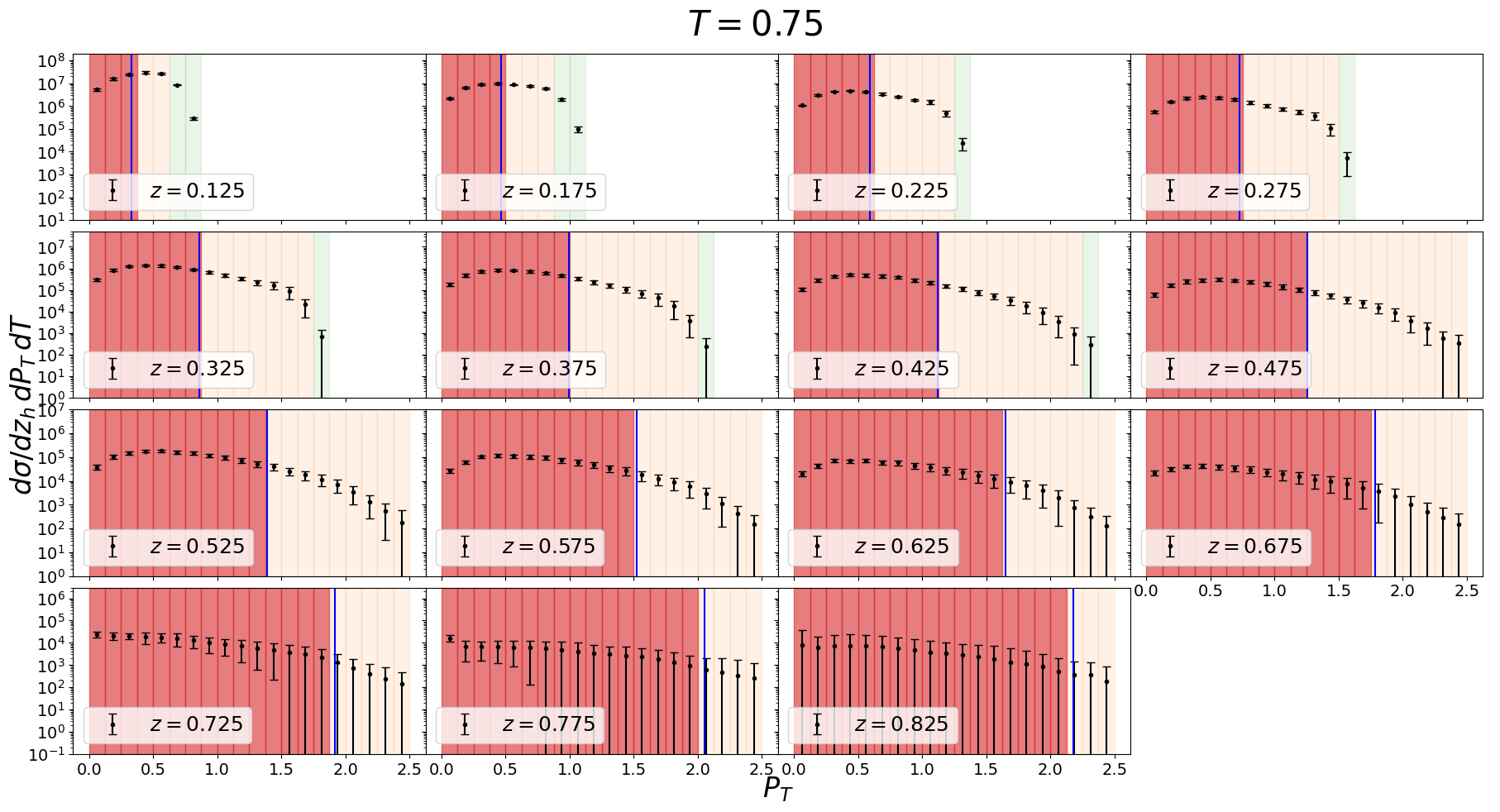}
\includegraphics[width=7.5cm]{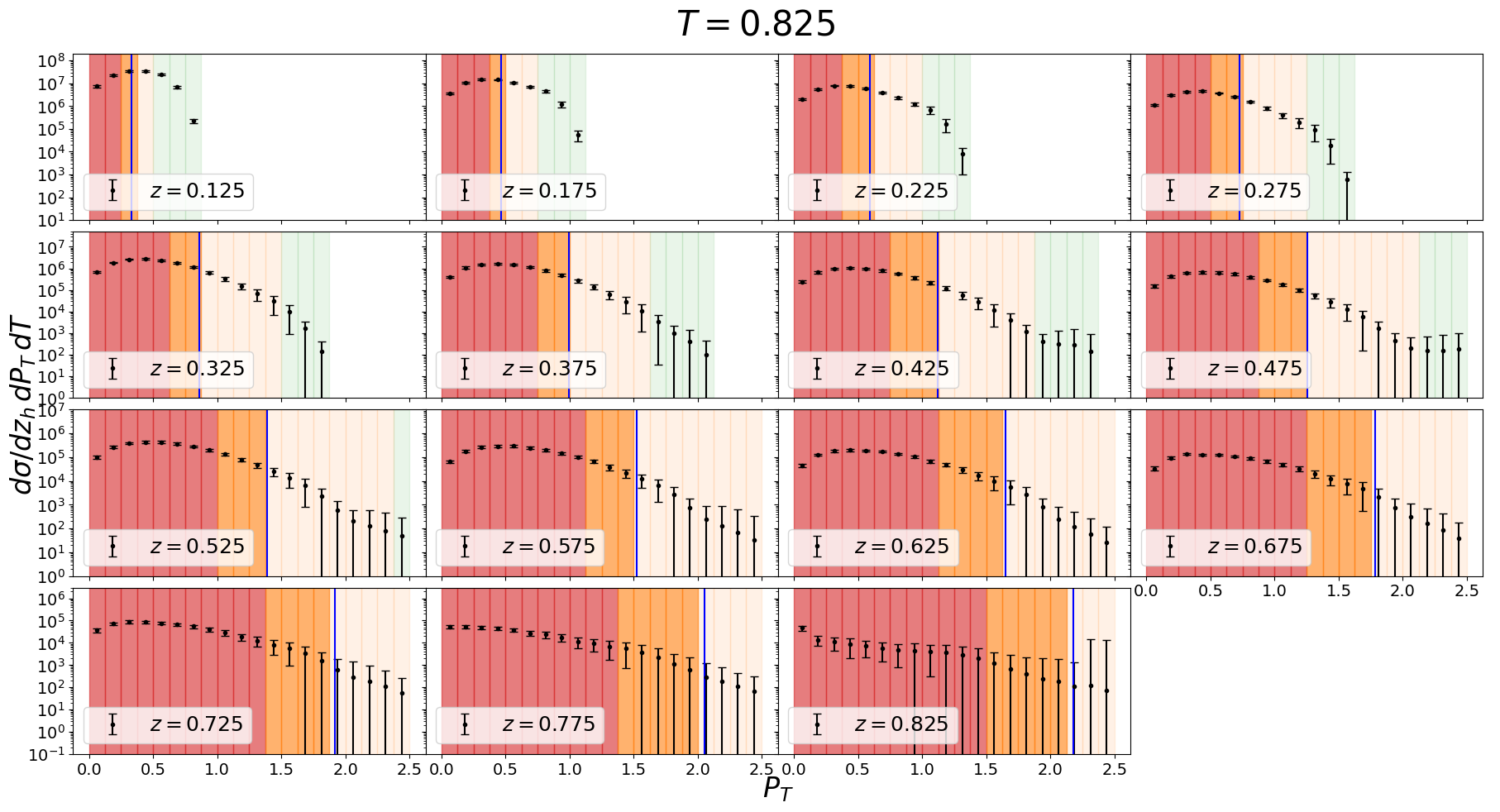}\\
\includegraphics[width=7.5cm]{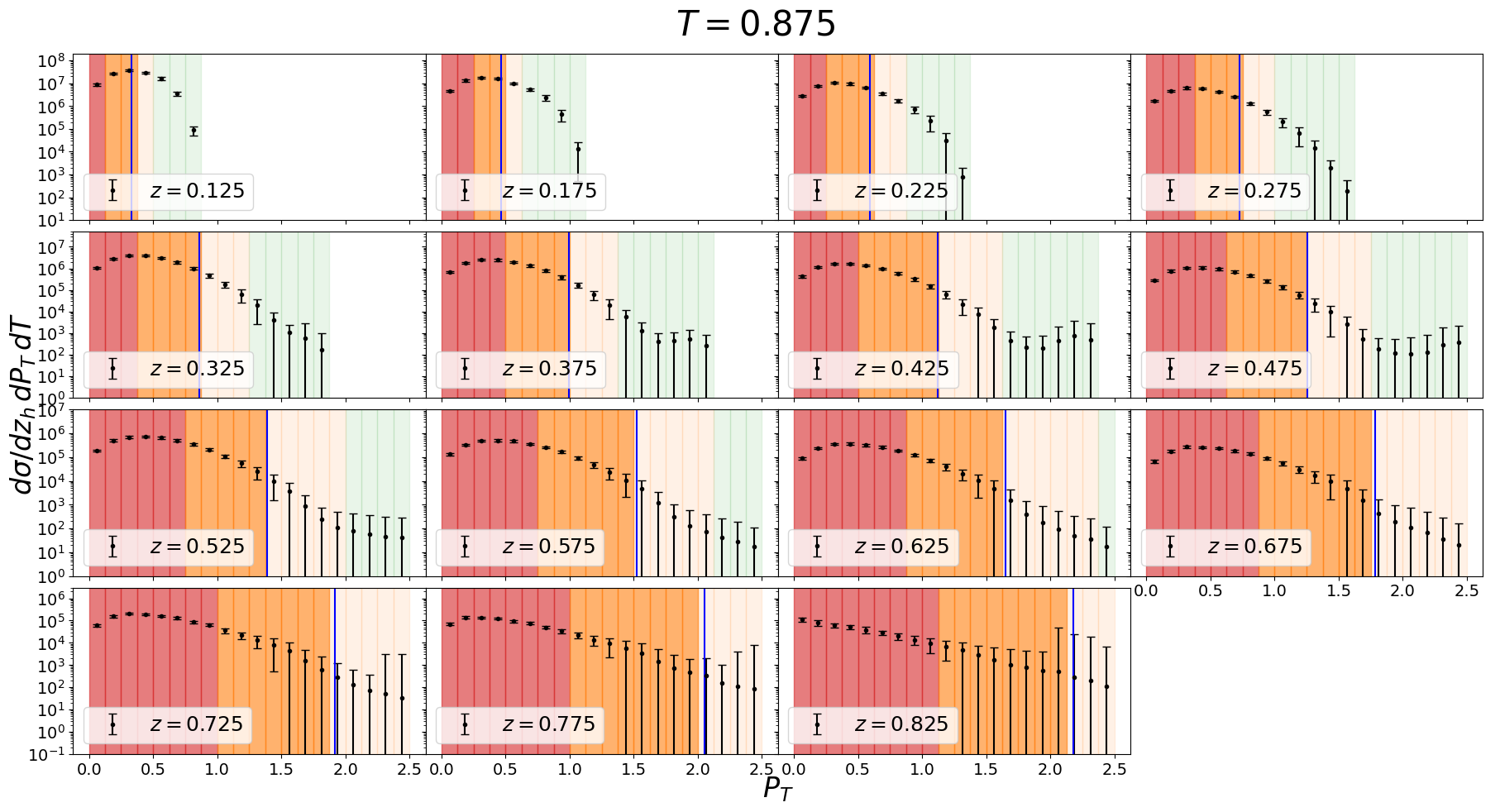}
\includegraphics[width=7.5cm]{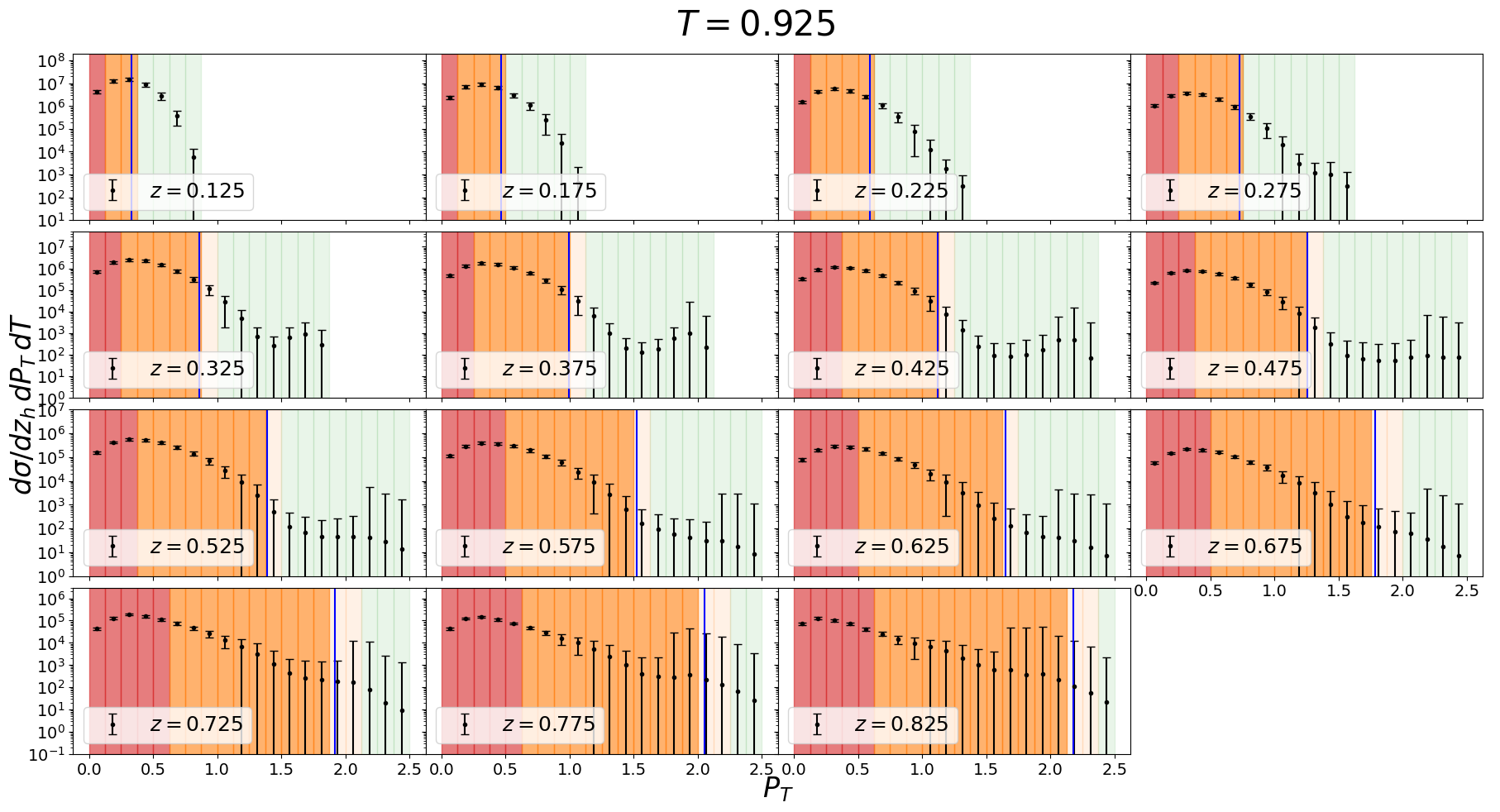}\\
\includegraphics[width=7.5cm]{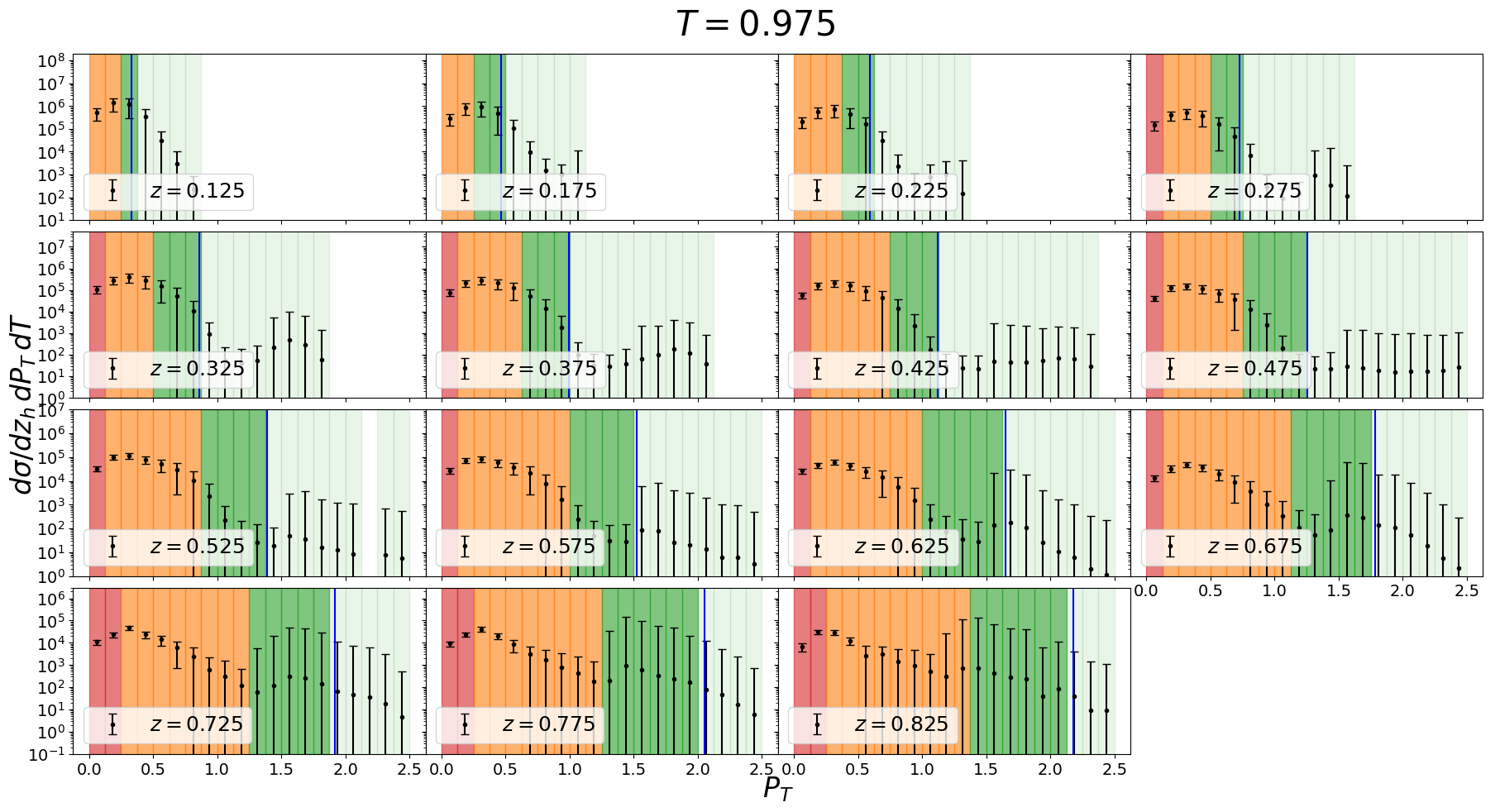}
\caption{BELLE data~\cite{Seidl:2019jei} selected according to the criteria of Eqs.~\eqref{eq:makris_crit}. Red bins correspond to Region 1, orange bins to Region 2 and green bins to Region 3. 
The shaded areas correspond to bins outside the TMD-regime, where Eq.~\eqref{eq:TMDcut} is not satisfied.
The purpose of this representation is to capture at a glance how the three kinematic regions are distributed through the whole thrust spectrum, for a $2$-jet topology. Here we do not focus on the details of each thrust bin. A more detailed representation, together with a thorough description of each panel, can be found in Appendix~\ref{app:canvas}.} 
\label{fig:makris_canvas}
\end{figure}

The simplest criterion consists in comparing the size of $k_T$ to the typical scale set according to the value of the  thrust, $T$.
In particular, in Region 1, Eq.~\eqref{eq:reg1_replacements_S} can be interpreted as $\tau \, Q \ll k_T$, while, since in Region 3 the only TMD-relevant quantity is left unapproximated, Eq.~\eqref{eq:reg3_replacements} can be interpreted as $k_T \sim \sqrt{\tau} Q$, which is also the maximum value kinematically allowed for ${P_T}/{z_h}$.
As a consequence, Region 2 covers the whole intermediate configuration, satisfying $\tau \, Q \lesssim k_T \lesssim \sqrt{\tau} Q$. 
Since $k_T$ is directly related to the transverse momentum $P_T$ of the detected hadron by the relation $P_T=z_h k_T$, 
these interpretations can be transferred to hadronic quantities. In particular:
\begin{subequations}
\label{eq:makris_crit}
\begin{align}
&R_1 \longrightarrow \frac{P_T}{z_h} \ll \tau \, Q ;\\
&R_2 \longrightarrow \tau \, Q \lesssim \frac{P_T}{z_h} \lesssim \sqrt{\tau} Q ;\\
&R_3 \longrightarrow \frac{P_T}{z_h} \sim \sqrt{\tau} Q .
\end{align}
\end{subequations}
This criteria uses only the typical scales associated to the value of   thrust, i.e. $\tau \, Q$ as a soft scale and $\sqrt{\tau} \, Q$ as a collinear scale. In fact, these values are commonly used as reference scales in thrust-resummed quantities, see for instance Ref.~\cite{Schwartz:2014sct}. Notice that this has been proposed as a selection criterion for BELLE $e^+e^- \to h X$ data in Ref.~\cite{Makris:2020ltr}.
Moreover, since the factorized cross sections of Regions 1 and 2 involve TMD FFs, an additional cut in $P_T$ is necessary, since the Fourier transform acts as an analytic continuation that unnaturally extends the TMD beyond the (physical) small transverse momentum region. In this case, the requirement comes directly from power counting:
\begin{align}
\label{eq:TMDcut}
&P_T \ll P^+ = z_h \frac{Q}{\sqrt{2}}.
\end{align}
The application of this algorithm to the BELLE data~\cite{Seidl:2019jei} in the $2$-jet region, corresponding to $0.7 \leq T \leq 1$, produces the results shown in Fig.~\ref{fig:makris_canvas}, where Regions 1, 2 and 3 have been color coded to red, orange and green, respectively.
The cut in $P_T$ for constraining the range of applicability of TMDs is indicated by the blue vertical line.
From the phenomenological point of view this kind of data selection inevitably raises the issue of matching different regions, since  Fig.~\ref{fig:makris_canvas} shows that there are at least two different overlapping regions in each panel. 
The problem of matching different kinematic regions, each corresponding to a  different factorization theorem, is not new in the context of TMD physics. This, in fact, has recently been one of the most debated issues when dealing with phenomenological applications of $2$-h cross sections (like SIDIS and  Drell-Yan), that have two distinct regimes associated to two factorization theorems: collinear factorization at large-$q_T$ and TMD factorization at small-$q_T$,~\cite{Boglione:2014oea,Echevarria:2018qyi,Collins:2016hqq,Collins:2017oxh}. In the case of $\epm \to h\, X$ there are \emph{three} different regions, making the matching even more problematic.
%
%

\bigskip

The set of rules devised above is not the only possible choice to obtain a valid criterion to perform a data selection.
In particular, it oversimplifies the complex structure of the three kinematic regions, as it only considers the typical thrust scale associated to the two leading momentum regions, soft and collinear. 
Most importantly, the approximations of Eq.~\eqref{eq:makris_crit} do not take into account the rapidity of the detected hadron, which, remarkably, is the crucial information to discriminate between a configuration where the transverse deflection is due to soft and soft-collinear radiation and one in which only soft-collinear emissions play an active role in generating TMD effects. This is of course strictly connected to the boundary between Region 1 and Region 2.

Therefore, in the following we will propose a different criterion which takes into account also the information encoded in the hadron rapidity.
It is devised on the basis of the 1-loop computation presented in the previous Sections (note that the argument of the deltas constraining thrust and transverse momentum would be different at higher order in pQCD). 

The approximations of Eqs.~\eqref{eq:reg1_replacements},~\eqref{eq:reg2_replacements} and~\eqref{eq:reg3_replacements} are considered the fundamental tool to implement the selection algorithms. The partonic quantities in the deltas are promoted to their hadronic equivalent, i.e $z\mapsto z_h$ and $k_T \mapsto {P_T}/{z_h}$. Then, denominating $y_P$  the rapidity of the detected hadron, Eq.~\eqref{eq:kin_yPy1}, we introduce the following reference ratios:
\begin{itemize}
\item \textbf{Soft Ratio} $r_S$, defined as:
\begin{align}
\label{eq:rS_def}
r_S = \frac{P_T}{z_h \, Q} e^{-y_P}.
\end{align}
\item  \textbf{Collinear Ratio} $r_C$, defined as:
\begin{align}
\label{eq:rC_def}
r_C = z_h \, (1-z_h) \,  e^{-2 \, y_P}.
\end{align}
\end{itemize}
Then, by comparing these ratios to thrust, we devise an algorithm that takes into account also the role of soft-collinear radiation. In fact, we can write the analogue of Eqs~\eqref{eq:reg1_replacements},~\eqref{eq:reg2_replacements} and~\eqref{eq:reg3_replacements} to hadronic level, as follows:
\begin{align}
&\textbf{Region 1:}\quad
\begin{aligned}
&r_S(z_h,\,P_T) \ll \tau \quad\text{for both soft and soft-collinear radiations} ; \\
&r_C(z_h,\,P_T) \ll \tau \quad\text{for collinear radiation}.
\end{aligned}
\label{eq:reg1_replacements_h}\\[10pt]
&\textbf{Region 2:}\quad
\begin{aligned}
&r_S(z_h,\,P_T) \sim \tau \quad\text{for soft radiation} ; \\
&r_S(z_h,\,P_T) \ll \tau \quad\text{for soft-collinear radiation} ;  \\
&r_C(z_h,\,P_T) \ll \tau \quad\text{for collinear radiation}.
\end{aligned}
\label{eq:reg2_replacements_h}\\[10pt]
&\textbf{Region 3:}\quad
\begin{aligned}
&r_S(z_h,\,P_T) \sim \tau \quad\text{for both soft and soft-collinear radiation};\\
&r_C(z_h,\,P_T) \sim \tau \quad\text{for collinear radiation}.
\end{aligned}
\label{eq:reg3_replacements_h}
\end{align}
Therefore, Region 3 is the only kinematic region where $r_S$ is never small enough to be neglected. This constitutes the first rule:
\begin{enumerate}
\item If $r_S \ll \tau$ we can either be in Region 1 or  in Region 2, but not in Region 3.
\end{enumerate}
Next, a small soft ratio characterizes both Region 1 and Region 2, which can be discriminated according to the rapidity of the detected hadron. In particular, if the soft ratio is small because of the size of ${P_T}/{z_h}$, regardless of the rapidity $y_P$, then $r_S$ will be \textbf{momentum dominated}. In this case we are in Region 1, because the soft ratio is neglected even if the rapidity is large, i.e. for both soft and soft-collinear radiation.
If instead the smallness of the soft ratio is due to the largeness of the rapidity $y_P$, then $r_S$ will be denoted as \textbf{rapidity dominated}. In this case we are in Region 2, as $r_S$ can be neglected only when the rapidity is large, i.e. only for soft collinear radiation.

In order to discriminate between these two configurations, we can compare the contributions of the transverse momentum and the rapidity to the soft ratio and write the second rule:
\begin{enumerate}
\setcounter{enumi}{1}
\item If $\frac{P_T}{z_h \, Q} \ll e^{-y_P}$ the soft ratio is momentum dominated and we are in Region 1. Otherwise the soft ratio is rapidity dominated and we are in Region 2.
\end{enumerate}
This exhausts all the possibilities implied by the first rule.
When $r_S \sim \tau$ the first rule is violated. In this case Region 1 is automatically excluded as it is the only kinematic region where the soft ratio is always considered small, for both soft and soft-collinar radiation. Therefore, the third rule reads:
\begin{enumerate}
\setcounter{enumi}{3}
\item If $r_S \sim \tau$ we can only be either in Region 2 or  in Region 3, but not in Region 1.
\end{enumerate}
It is here that the collinear ratio comes into play, discriminating between Region 2 and Region 3. In fact, if the collinear ratio is not small  enough to be neglected, i.e. $r_C \sim \tau$, then we are in Region 3. Otherwise we are in Region 2. This generates the last rule:
\begin{enumerate}
\setcounter{enumi}{4}
\item If $r_C \ll \tau$ we are in Region 2,  otherwise we are in Region 3.
\end{enumerate}
A set of criteria based on kinematic ratios was proposed independently for SIDIS in Refs.~\cite{Boglione:2016bph,Boglione:2019nwk}.

\begin{figure}
\centering
\includegraphics[width=10cm]{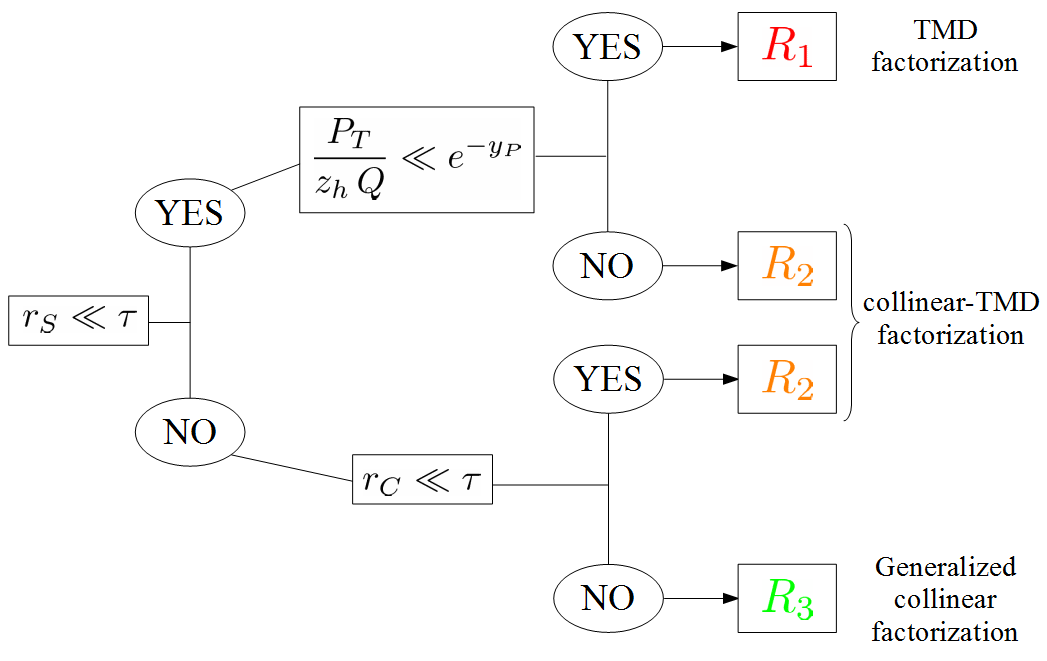}
\caption{Flow-chart representation of the algorithm based on soft and collinear ratios.}
\label{fig:AS_algorithm}
\end{figure}

\begin{figure}
\centering
\includegraphics[width=7.5cm]{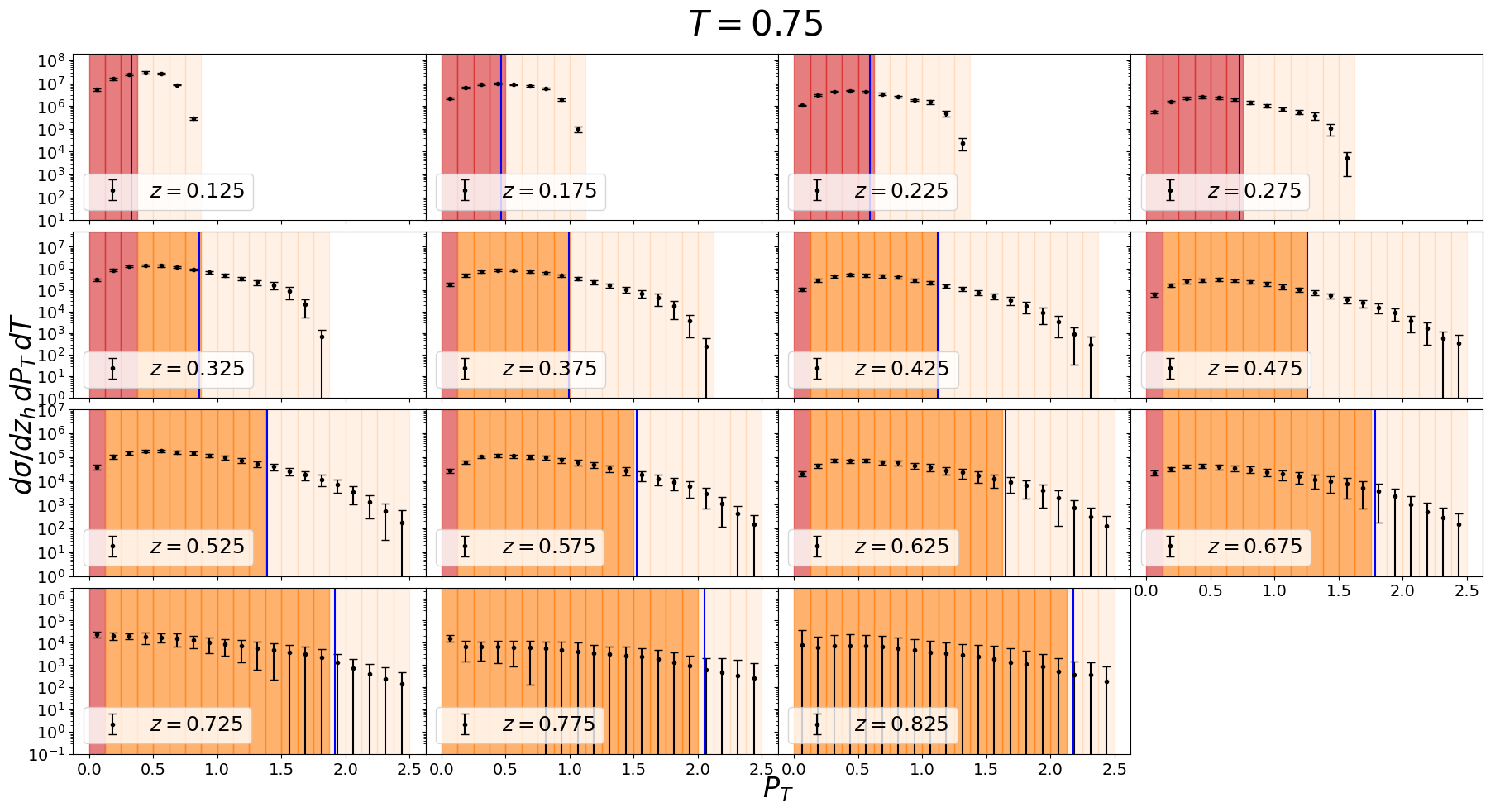}
\includegraphics[width=7.5cm]{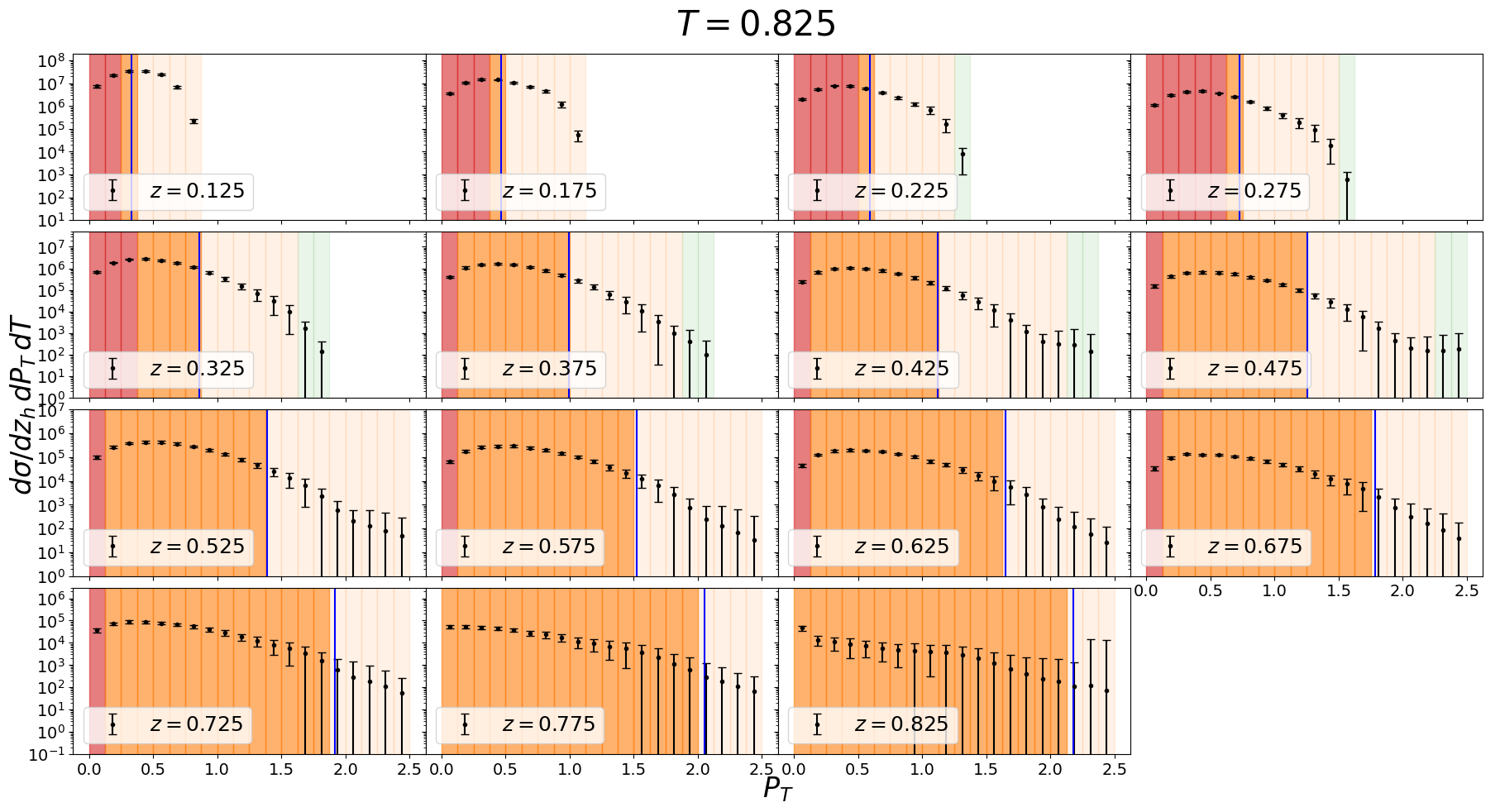}\\
\includegraphics[width=7.5cm]{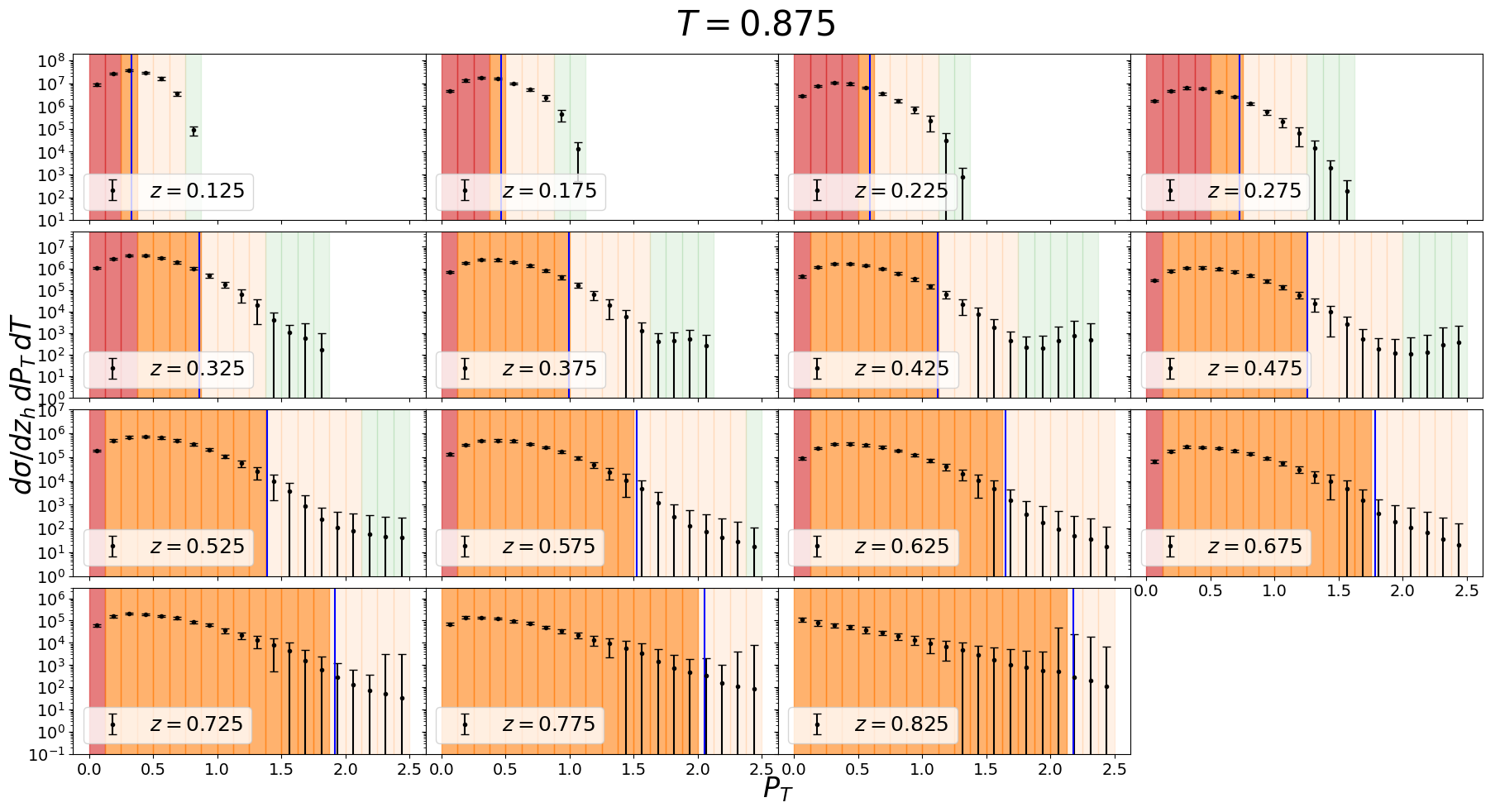}
\includegraphics[width=7.5cm]{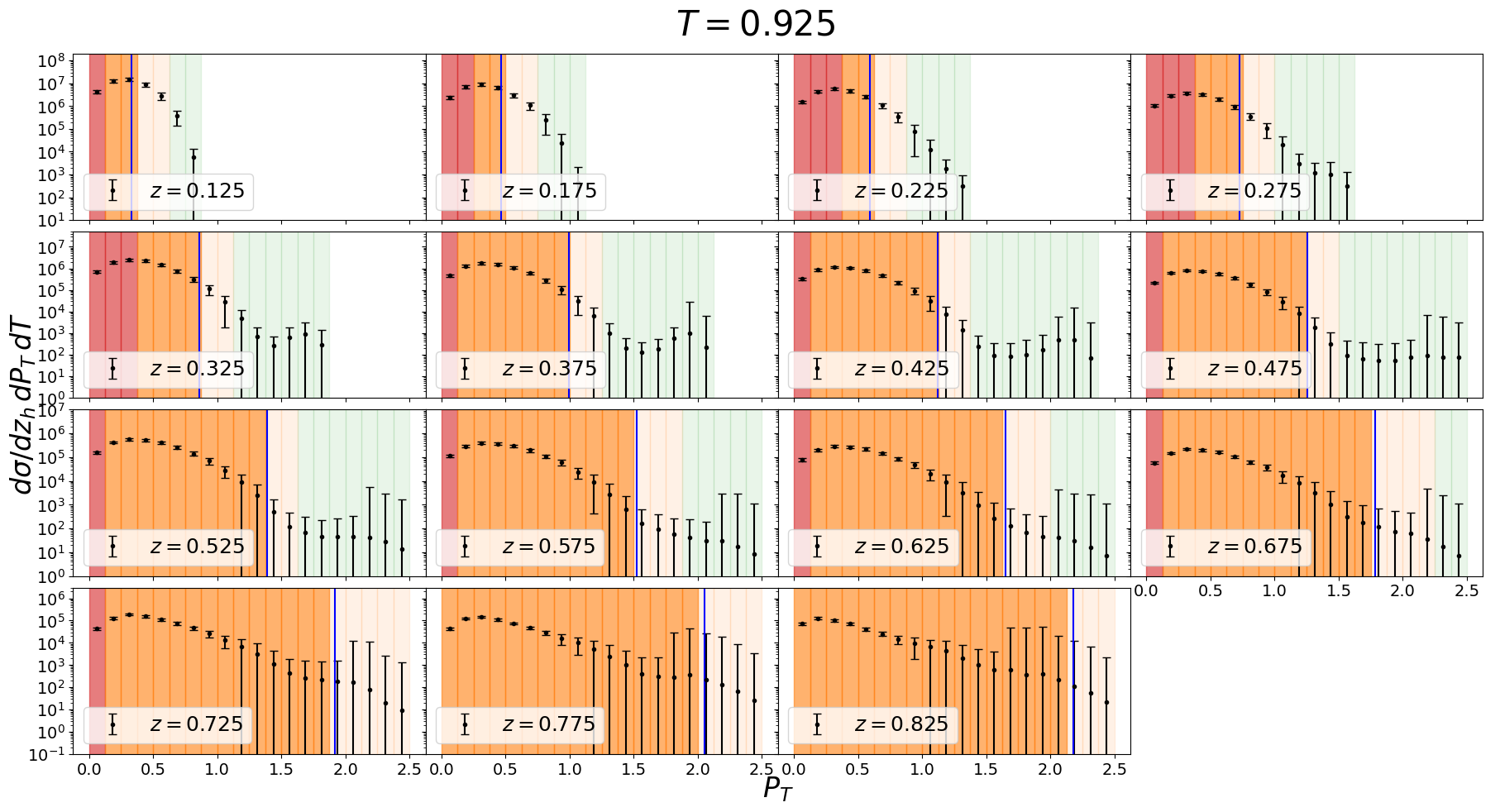}\\
\includegraphics[width=7.5cm]{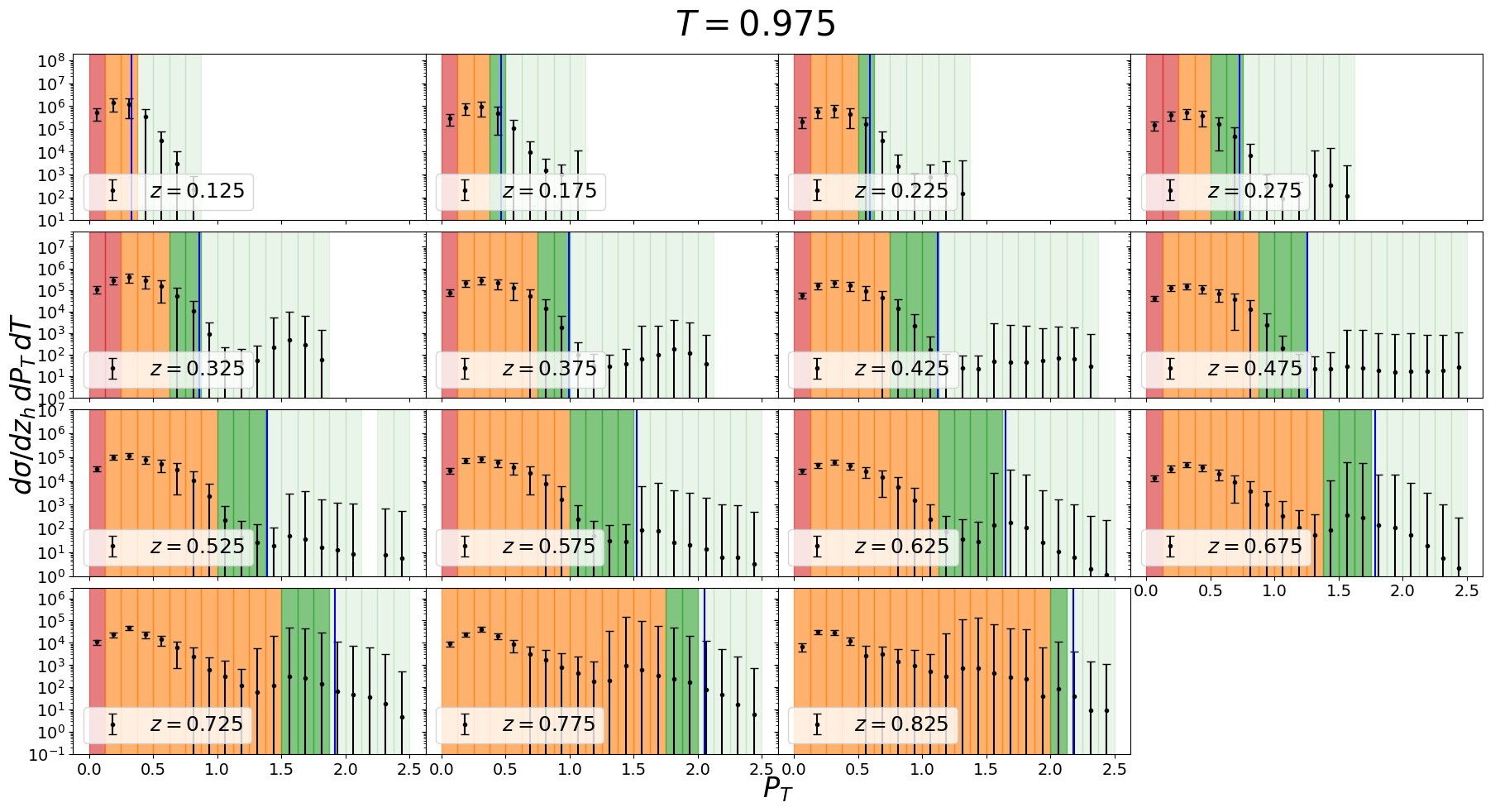}
\caption{BELLE data~\cite{Seidl:2019jei} selected according represented in Fig.~\ref{fig:AS_algorithm}. Red bins correspond to Region 1, orange bins to Region 2 and green bins to Region 3. 
The shaded areas correspond to bins outside the TMD-regime, where Eq.~\eqref{eq:TMDcut} is not satisfied.
The purpose of this representation is to capture at a glance how the three kinematic regions are distributed through the whole thrust spectrum, for a $2$-jet topology. Here we do not focus on the details of each thrust bin.
A more detailed representation together with a thorough description of each panel can be found in Appendix~\ref{app:canvas}.} 
\label{fig:AS_canvas}
\end{figure}

This algorithm is represented graphically in Fig.~\ref{fig:AS_algorithm}. The purpose of this representation is to capture at a glance how the three kinematic regions are distributed through the whole thrust spectrum, for a $2$-jet topology. Here we do not focus on the details of each thrust bin. A more detailed representation together with a thorough description of each panel can be found in Appendix~\ref{app:canvas}, dedicated to the comparison of the results obtained using out algorithm to those obtained applying the criteria proposed in Ref.~\cite{Makris:2020ltr}.
%
%
Notice how, within this criterion, Region 2 can be reached following two different routes, while Region 1 and Region 3 can only be reached through one path (in the sense that it corresponds to one unique selection rule). This is in agreement  with the naive expectation that Region 2 corresponds to the widest kinematic range, as it describes the ``intermediate" situation where $P_T$ is neither extremely small, as in Region 1, nor sizeable  as in Region 3.
This automatically affects the data selection, showed in Fig.~\ref{fig:AS_canvas}. Differently from the coarse algorithm of Fig.~\ref{fig:makris_canvas}, here there is a neat prevalence of (orange) bins  corresponding  to Region 2. Moreover, and most importantly, this selection scheme shows the presence of several \emph{monocromatic panels}, where only one kinematic configuration is realized. This offers an extraordinary  advantage for phenomenological analyses, as it allows to overcome the problem  of matching, which concerns only the panels where more than one Region (i.e. more than one color) appears.
%
%

\bigskip

Leaving aside Region 3 (green bins) which is not described by factorization theorems involving TMD FFs, the great opportunity offered by the algorithm presented above is in its application for the comparison between Region 1 and Region 2, corresponding to red and orange bins, respectively. In fact, as discussed at the end of Section~\ref{sec:reg1_fact_theo}, the TMD factorization theorem devised for Region 1 allows to access the square root model, $M_D^{\text{sqrt}}=M_D \sqrt{M_s}$, while the collinear-TMD factorization theorem obtained for Region 2 gives the chance to extract directly the TMD model $M_D$ alone. A direct comparison can shed light on the (still unknown) soft model $M_S$, which generate a deformation of the TMD defined by the square root definition compared to the that defined by the factorization definition. 
Since the algorithm presented in this Section allows to bypass the matching issues in a rather large number of bins, this analysis can safely be carried out, and is currently under study.

Finally, it is important to stress again that the soft model is the same unknown function which appears in the $2$-h cross sections of SIDIS, Drell-Yan and $e^+e^- \to h_1 h_2 X$ processes. Its independent extraction from a process which belongs to another hadron class is promisingly one of the most powerful phenomenological tools for future studies. Indeed, the ultimate goal is the application of the above formalism to a global analysis, combining data from all available $1$-h and $2$-h processes. 

\bigskip


\section{ Conclusions \label{sec:conclusions_ch3}}


\bigskip

Explicit perturbative computations require to extend the definitions of the soft factors and of the TMDs in order to take into account the dependence on thrust. In fact, this operation produces a large variety of integrated and unintegrated objects that not only generalize the definitions of soft factors and TMDs, but also include and extend the usual thrust-dependent functions, commonly encountered in the study of $\epm$ annihilation processes. All these new thrust-dependent operators have been defined and computed to 1-loop accuracy in Sections~\ref{sec:reg2},~\ref{sec:reg1} and~\ref{sec:reg3}.

Within this general approach we were able to recover the same collinear-TMD factorization theorem obtained in Ref.~\cite{Boglione:2020auc} by means of an explicit perturbative computations, in a completely natural way.
This scheme, however, is realized in a more general context and leads to a much richer kinematic structure, where three different regions are generated, each corresponding to a different factorization theorems.
The proof of factorization has been founded on two assumptions: the transverse momentum of the detected hadron is neither too large to affect significantly the topology of the final state (\hyp{1}), nor too small to be affected by the emission/absorption of soft radiation  (\hyp{2}). 
Modifying this initial hypothesis inevitably leads to a different factorization theorem, which corresponds to a different kinematic region.
Since these two assumptions cannot be false at the same time, there are in total three different kinematic regions, defined as:
\begin{itemize}
\item Region 1, corresponding to set~\hyp{1} true and~\hyp{2} false, treated in Section~\ref{sec:reg1}. In this region the soft radiation participates  actively to TMD effects. In fact, the resulting factorization theorem is very  similar to the standard TMD factorized cross section known to hold in the $2$-h class for processes like SIDIS, Drell-Yan and $e^+e^- \to h_1 h_2 X$. In fact, the TMD FFs describing the non-perturbative hadronization process that generates the detected hadron appears in the cross section according to the square root definition. 
Their TMD model, which describes their tipical long-distance behavior, is contaminated by the non-perturbative content of the soft radiation by a square root of the soft model, $\sqrt{M_S}$, confirming that in Region 1 soft emissions play a leading role in generating TMD effects. Due to these similarities, we  identify the factorized cross section obtained for Region 1 as a TMD factorization theorem.
\item Region 2, corresponding to the case in which both the hypotheses hold true, treated in Section~\ref{sec:reg2}. 
The resulting factorization theorem has the structure of collinear factorized cross section but involves TMD FFs. Because of its hybrid nature, this factorization theorem has been called collinear-TMD. It presents two important issues, which are totally new features for TMD obsrvables.
First of all, the TMDs appearing into the collinear-TMD factorized cross section are defined by the factorization definition and hence they are not contamined by any soft radiation contribution. 
Secondly, such cross section requires to assign a real physical meaning to the rapidity cut-off, as it is strictly related to the measured value of the thrust. The role of rapidity regulators is widely discussed in Section~\ref{sec:role_of_y1}.
\item Region 3, corresponding to set~\hyp{1} false and~\hyp{2} true, treated in Section~\ref{sec:reg3}. Here all the effects of soft and even soft-collinear radiation are irrelevant for TMD effects, as their contribution can be neglected, given the (relatively large) size of the transverse momentum of the detected hadron. Since in this case the measured value of thrust takes part  in the collinear radiation contribution, the final factorization theorem cannot involve TMD FFs. In fact, they are replaced by the corresponding GFJFs, defined similarly to the (unsubtracted) TMDs, but with a further dependence on the invariant mass of the jet to which they are associated. The hybrid nature of the collinear-TMD factorization theorems is totally encoded in  these functions, which share both characteristics of the TMD FFs, and also of the usual FFs. However, they are very  different from TMDs, as in their definition there is no trace of any rapidity cut-off, as all the rapidity divergences are naturally regulated by their additional dependence on the jet invariant mass. For this reason, GFJFs should be considered more as a generalized version of the usual FFs than an extended counterpart of the TMDs. Therefore, the factorized cross section devised for Region 3 has been denoted as a generalized collinear factorization theorem.
\end{itemize}
As these three kinematic regions contain different kind of information on TMD physics, it is extremely important to devise a solid methodology to identify them unequivocally within the large set of data provided by the BELLE Collaboration~\cite{Seidl:2019jei}. In Section~\ref{sec:reg_algorithm} we showed how a standard algorithm based solely on the typical soft and collinear energy scales associated to the value of thrust is unable to capture all the features encoded into the rich structure of a process like $\epm \to h\,X$. Therefore, we propose a finer algorithm that allows to perform a more refined data selection, taking into account not only soft and collinear radiation, but also the contribution of soft-collinear emissions.
These new criteria shows that a rather large amount of data are actually  described by a single factorization theorem, bypassing all the issues related to the matching procedure to  describe data at the boundaries of the corresponding kinematic regions.
This is a very promising phenomenological tool, as the direct comparison between an extraction made in Region 1 and another made in Region 2 would shed light on the effects of the soft radiation in the standard TMD factorization. 

\newpage

\section*{Acknowledgements}

We are grateful to J.O. Gonzalez-Hernandez, A. Bacchetta and L. Gamberg for interesting discussions, useful comments and suggestions. \\    
This project has received funding from the European Union’s Horizon 2020 
research and innovation programme under grant agreement No 824093.

\vspace{2cm}

\appendix


\section{Review of the factorization formalism \label{app:review_collins}}


\bigskip

This Appendix is a short review of the factorization formalism and of the methodologies used throughout this paper.

We start by introducing the kinematics approximators as defined in Ref.~\cite{Collins:2011zzd}. Their definition is based on the power counting rules:
\begin{enumerate}
    \item Given the typical (large) energy scale $Q$ of a process, the hard, collinear and soft momenta are weighted as:
    \begin{align}
    \label{eq:power_counting}
    \begin{split}
        &P_{\text{hard}} \sim (Q, Q, Q);\\
        &P_{\text{coll.}} \sim (Q, \lambda_S, \lambda);\\
        &P_{\text{soft}} \sim (\lambda_S, \lambda_S, \lambda_S);
    \end{split}
    \end{align}
    where $\lambda << Q$ is some IR energy scale and $\lambda_S = {\lambda^2}/{Q}$.
    Such scaling allows to classify sets of subgraphs inside a generic Feynman diagram. According to this classification, the hard subgraph will contain particles carrying hard momenta and so on. 
    \item Any \emph{extra} collinear line attached to the hard subgraph gives a suppression. Here \emph{extra} means any line besides the minimal number of fermions required by the kinematics of the process and any number of scalar polarized gluons.
    \item Any soft line attached to the hard subgraph gives a suppression.
    \item Any fermionic line connecting collinear and soft subgraphs gives a suppression.
\end{enumerate}
All the gluons connecting soft and collinear subgraphs and hard and collinear subgraphs are collected into Wilson lines (or gauge links). They are path-ordered exponential operators defined as:
\begin{align}
W_\gamma = P \left \{ \exp \left[
-i g_0 \, \int_0^1 ds \,
\dot{\gamma}^\mu(s) A^a_{(0)\,\mu}(\gamma(s)) t_a 
\right]
\right \},
\label{eq:wilsdef_1}
\end{align}
where $\gamma$ is a generic path and $P$ denotes the path ordering (i.e. when the exponential is expanded the fields corresponding to higher values of $s$ are to be placed to the left).
The coupling constant and the gluon field are bare quantities, as indicated by the label ``0".
In the previous formula, $t_a$ are the generating matrices of the gauge group, in the appropriate representation.
The Wilson lines guarantee that PDFs and FFs (in both collinear and TMD cases) are gauge invariant, by linking the quark to the anti-quark fields. If this direction $\gamma^\mu$ is a straight line the Wilson line depends only on the endpoints of the 
path and can be written in a compact way as:
\begin{align}
W_n\left(x_2,\,x_1,\,n\right) = P \left \{ 
\exp \left[
-i g_0 \, \int_{x_1}^{x_2} d\lambda \,
n^\mu A^a_{(0)\,\mu}(\lambda n) t_a 
\right]
\right \},
\label{eq:wilsdef_2}
\end{align}
If the strongly boosted particle is a quark, the associated Feynman rules are:
\begin{align}
\begin{gathered}
\includegraphics[width=1.5cm]{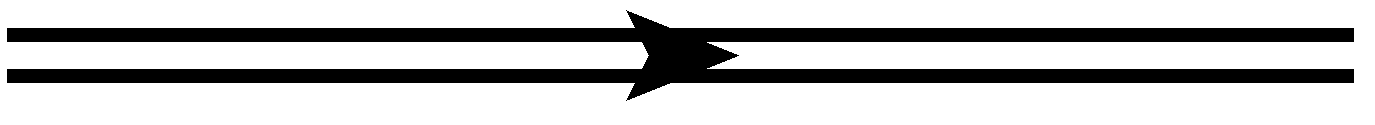}
\end{gathered} 
&= \frac{i}{k \cdot n + i 0} ; \\
\begin{gathered}
\includegraphics[width=1.5cm]{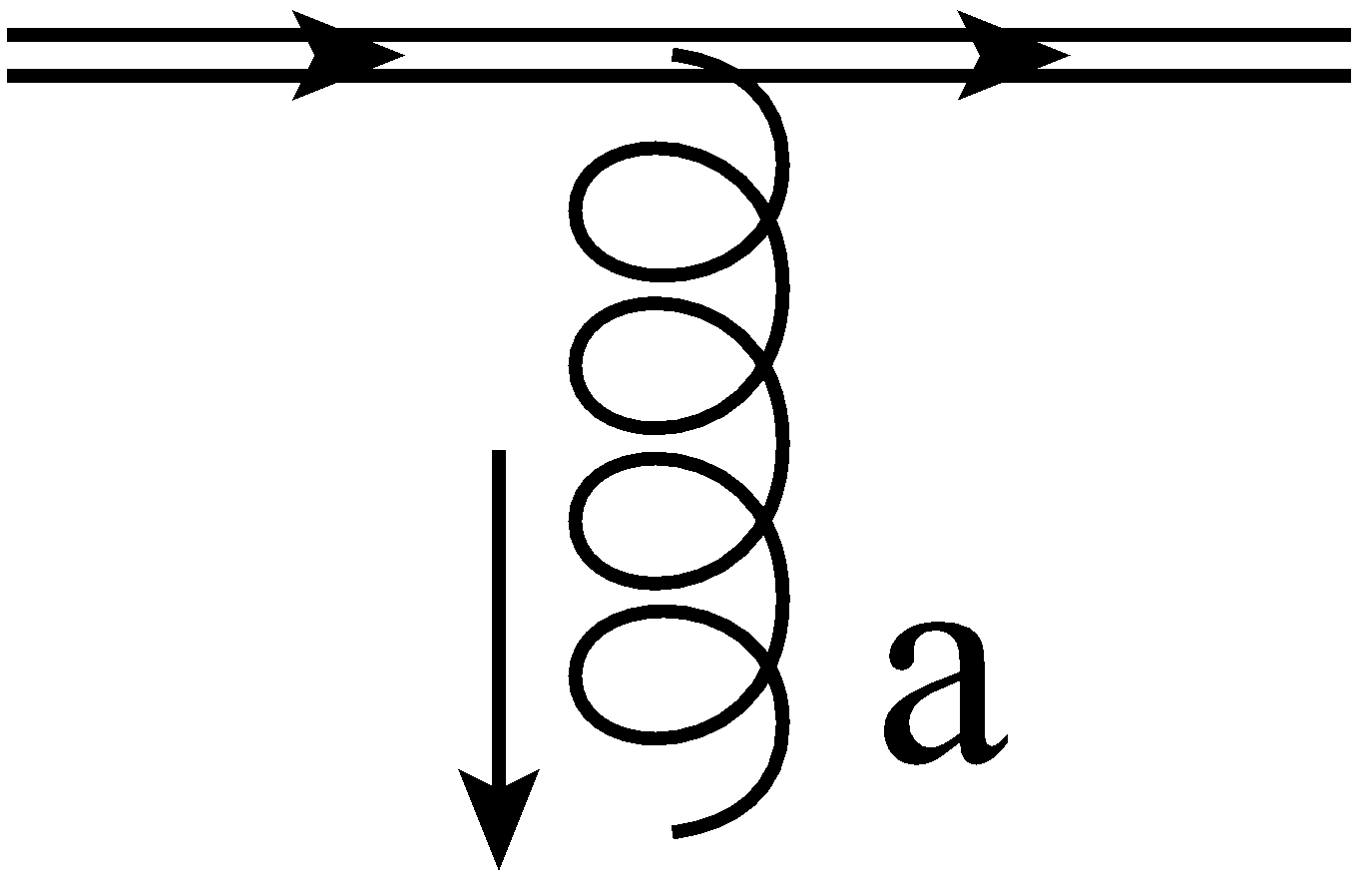}
\end{gathered} 
&= -i g_0 \, n^\mu \, t_a .
\label{eq:wils_feynrules}
\end{align}
In the Collins factorization formalism, the Wilson lines associated to soft contributions are tilted off the light-cone. In the following, $w_1 = (1,0,\vec{0}_T)$ and $w_2 = (0,1,\vec{0}_T)$  represent the plus- and minus-directions respectively, while the tilted directions are defined as:
\begin{align}
\label{eq:tilted_dir}
    &n_1 = (1,-e^{-2 y_1},\vec{0}_T),
    \hspace{2.5cm}
    n_2 = (-e^{2 y_2},1,\vec{0}_T).
\end{align}
The auxiliary parameters $y_1$ and $y_2$ acts as rapidity cut-offs and  regulates the rapidity divergences due to the Wilson lines associated to the collinear contributions, which are straight along the light-cone. 

\bigskip

Since we are interested in $2$-jet topologies, we label the hemisphere associated with the plus-direction as ``A" and the opposite hemisphere as ``B" (backward emission, along minus-direction). Therefore, in each Feynman diagram we identify an hard subgraph, labeled as ``H", a soft subgraph, labeled as ``S" and two collinear subgraphs, labeled either ``A" or ``B" depending on the leading direction of the collinear particles that flow inside them. 
The classification is not unique, but depends on the kinematic configuration (or region R) associated to the particles involved. 
The operation of choosing a certain kinematic region ``R" is defined to be the action of the kinematic approximator $T_R$. 
In practice, any generic $T_R$ is based on the following recipe:
\begin{enumerate}
    \item In the collinear subgraphs approximate the circulating soft momenta as:
    \begin{align}
    \label{eq:StoC_mom}
        &k \sim w_2 \, \frac{k \cdot n_1}{w_2 \cdot n_1} \quad \text{for A}, 
        \hspace{2.1cm}
        k \sim w_1 \, \frac{k \cdot n_2}{w_1 \cdot n_2} \quad \text{for B}.
    \end{align}
    \item In the hard subgraph neglect all the masses and approximate the circulating collinear momenta as:
    \begin{align}
    \label{eq:CtoH_mom}
        &k \sim w_1 \, \frac{k \cdot w_2}{w_1 \cdot w_2} \quad \text{for A}, 
        \hspace{2.1cm}
        k \sim w_2 \, \frac{k \cdot w_1}{w_1 \cdot w_2} \quad \text{for B}.
    \end{align}
    Notice that the circulating soft momenta are totally neglected in the hard subgraph.
    \item The attachment of a soft gluon to a collinear subgraph is approximated as (Grammer-Yennie approximation):
    \begin{align}
    \label{eq:StoC_num}
    \begin{split}
        &A(\dots,k,\dots)^\mu S(\dots,k,\dots)_\mu \sim 
        A(\dots,\widehat{k},\dots)^\mu \ddfrac{\widehat{k}_\mu \; n_{1,\nu}}{k \cdot n_1 + i 0} S(\dots,k,\dots)_\nu, \\
        &B(\dots,k,\dots)^\mu S(\dots,k,\dots)_\mu \sim 
        B(\dots,\widehat{k},\dots)^\mu \ddfrac{\widehat{k}_\mu \; n_{2,\nu}}{k \cdot n_2 + i 0} S(\dots,k,\dots)_\nu.
    \end{split}
    \end{align}
    where $\widehat{k}$ are the approximated momenta defined in Eq.~\eqref{eq:StoC_mom} and the $i0$-prescription is correct when the momentum $k$ flows out of the collinear subgraph.
    \item The attachment of a collinear gluon to the hard subgraph is approximated as (Gram\-mer-Yen\-nie approximation):
    \begin{align}
    \label{eq:CtoH_num}
    \begin{split}
        &H(\dots,k,\dots)^\mu A(\dots,k,\dots)_\mu \sim 
        H(\dots,\widehat{k},\dots)^\mu \ddfrac{\widehat{k}_\mu \; w_{2,\nu}}{k \cdot w_2 + i 0} A(\dots,k,\dots)_\nu, \\
        &H(\dots,k,\dots)^\mu B(\dots,k,\dots)_\mu \sim 
        H(\dots,\widehat{k},\dots)^\mu \ddfrac{\widehat{k}_\mu \; w_{1,\nu}}{k \cdot w_1 + i 0} B(\dots,k,\dots)_\nu.
    \end{split}
    \end{align}
    where $\widehat{k}$ are the approximated momenta defined in Eq.~\eqref{eq:CtoH_mom} and the $i0$-prescription is correct when the momentum $k$ flows out of the hard subgraph.
    \item For a Dirac line leaving the hard subgraph and entering in the collinear-A subgraph, insert the projector $\mathrm{P} = \frac{1}{2} \gamma^+ \gamma^-$. The same rule applies for a Dirac line leaving the collinear-B subgraph and entering in the hard subgraph.
    For a quark line in the reverse direction, insert $\overline{\mathrm{P}} = \frac{1}{2} \gamma^- \gamma^+$.
\end{enumerate}
The application of the kinematic approximators allows to define each of the ingredients appearing in final factorized cross sections.
In particular, soft factors and TMDs are of special importance. 

The $2$-h soft factor appearing in the TMD factorized cross sections of the three benchmark processes (Drell-Yan, SIDIS, $\epm \to h_1 \, h_2 \, X$ ) is defined in $b_T$-space as:
  \begin{align}
& \ftsoft{2}(b_T;\, \mu, \,y_1 - y_2) =
Z_S(\mu,\,y_1 - y_2)
\times \notag \\
&\quad \times \,  
\TrC \,
\langle 0 | 
W(-{\vec{b}_T}/{2}, \,\infty; \, n_1(y_1)\,)^\dagger \, 
W({\vec{b}_T}/{2}, \,\infty; \, n_1(y_1)\,) \,
\times \notag \\
&\quad \times \,  
W({\vec{b}_T}/{2}, \,\infty; \, n_2(y_2)\,)^\dagger \, 
W(-{\vec{b}_T}/{2}, \,\infty; \, n_2(y_2)\,) 
| 0 \rangle \, \vert_{\text{NO S.I.}},
\label{eq:S2h_bTspace} 
\end{align}  
where $N_C$ is the number of colors available for quarks and antiquarks (3 in QCD). The factor $Z_S$ is the UV renormalization factor that cancels the poles generated when the $k_{S,\,T}$-integration associated to the Fourier transform stretches outside of the soft momentum region. 
The rapidity-independent kernel $\widetilde{K}(b_T;\,\mu)$, often referred to as ``Collins-Soper kernel" or ``soft kernel" defined as~\cite{Collins:2011zzd} rules the evolution equation for $\soft{2}$:
\begin{subequations}
\label{eq:S2hevo}
\begin{align}
&\lim_{y_2 \to -\infty} \frac{\partial \log{\ftsoft{2}(b_T; \, \mu, \, 
y_1 - y_2)}}{\partial y_1} = 
\frac{1}{2}\, \widetilde{K}(b_T;\,\mu), 
\label{eq:S2hevo_1}\\
&\lim_{y_1 \to +\infty} \frac{\partial \log{\ftsoft{2}(b_T; \, \mu, \, 
y_1 - y_2)}}{\partial y_2} = 
- \frac{1}{2}\, \widetilde{K}(b_T;\,\mu) \,.
\label{eq:S2hevo_2}
\end{align}  
\end{subequations}
It has an anomalous dimension $\gamma_K$:
  \begin{equation} \label{eq:gammaK_evo}
\frac{d  \widetilde{K}(b_T;\,\mu)}{d \log{\mu}} = - 
\gamma_K(\alpha_S(\mu)),
\end{equation}  
where $\gamma_K$ depends on $\mu$ through the strong coupling $\alpha_S$ and is independent of $b_T$. The solution to Eqs.~\eqref{eq:S2hevo} can be properly written by separating out the perturbative content of $\soft{2}$ from what cannot be predicted by the sole perturbative QCD. Following Ref.~\cite{Boglione:2020cwn} we have:
\begin{equation} \label{eq:S2h_final}
\ftsoft{2}(b_T; \, \mu,\,y_1 - y_2) = 
e^{- \frac{y_1 - y_2}{2} \left[ \int_{\mu_0}^{\mu}\, \frac{d \mu'}{\mu'}\, \gamma_K\left(\alpha_S(\mu')\right) \; - 
\widetilde{K}(b_T^\star;\,\mu_0)\right]}
M_S(b_T) \, e^{-\frac{y_1 - y_2}{2} \, g_K(b_T)},
\end{equation}  
where the separation has been obtained by introducing the $b^\star$-prescription as in Ref.~\cite{Collins:2011zzd}.
The functions $M_S(b_T)$ and 
$g_K(b_T)$ have to be committed to a 
phenomenological model. The first one describes the long-distance behavior of the soft kernel and hence it is strictly related to evolution. On the other hand, the \textbf{soft model} $M_S(b_T)$ encodes the long-distance behavior of the non-exponential part of the soft factor and can be considered as the true fingerprint of the soft radiation contribution in the non-perturbative regime.

\bigskip

The TMDs are defined in order to describe the collinear radiation contributions. The factorization procedure lead to a collinear part that overlaps with the soft momentum region. Therefore, TMDs are defined by subtracting out the double-counting soft-collinear terms. A generic TMD $C$ in $b_T$-space is then defined as:
\begin{align}
\widetilde{C}_{j,\,h} (\xi,\, \vec{b}_{T}; \, \mu, \, y_P - y_1) &= 
Z_C(\mu,\, y_P - y_1) \,
Z_2 \left(\alpha_S(\mu)\right) \, 
\lim_{y_{u_2} \to -\infty}
\dfrac{\widetilde{C}_{j,\,H}^{(0),\,\text{uns.}} (\xi, \,\vec{b}_{T}; \, \mu, \, 
y_P - y_{u_2})}
{\ftsoft{2}^{(0)} (b_{T}; \, \mu, \,y_1 - y_{u_2})} \,.
\label{eq:sub_coll}
\end{align}
where the label ``(0)" means that the corresponding factors are bare functions of bare fields and $Z_C$ is the UV counterterm. It depends only on variables which are ``intrinsic" to the pair constituted by the reference hadron $h$ and the reference parton of type $j$. In fact, they depends only on their relative transverse momentum, expressed by its Fourier conjugate variable $b_T$ in Eq.~\eqref{eq:sub_coll}, and by the collinear momentum fraction $\xi$. Moreover, the subtraction mechanism introduces a dependence on the rapidity cut-off $y_1$, which acts as a lower bound for the rapidity of the particles described by the TMD.

The evolution equations for $C$ are written  with respect to both $\mu$ (RG-evolution) and $y_1$ (CS-evolution), recast as $\zeta=2 (k^+)^2 e^{-2 y_1}$. We have:
\begin{subequations}
\label{eq:tmd_evo}
\begin{align}
&\dfrac{\partial \log{\widetilde{C}_{j,\,h}(\xi,\, b_{T}; \, \mu, 
\,\zeta)}}{\partial \log{\sqrt{\zeta}}} = 
\frac{1}{2} \widetilde{K}(b_T;\,\mu) , 
\label{eq:RGevo_TMD}\\
&\dfrac{\partial \log{\widetilde{C}_{j,\,h}(\xi,\, b_{T}; \, \mu, \, 
\zeta)}}{\partial \log{\mu}} = 
\gamma_C \left(\alpha_S(\mu),\, \frac{\zeta}{\mu^2} \right) .
\label{eq:CSevo_TMD}
\end{align}
\end{subequations}
Moreover, the anomalous dimension of the TMD $\gamma_C$ obeys:
\begin{align}
\frac{\partial \gamma_C \left(\alpha_S(\mu),\, {\zeta}/{\mu^2} \right)}
{\partial \log{\sqrt{\zeta}}} = -\frac{1}{2}\gamma_K(\alpha_S(\mu)) ,
\label{eq:gammaC_evo}
\end{align}
Finally, it is a standard result that the solution of TMDs evolution equations reads  \cite{Collins:2011zzd,Aybat:2011ge,Aybat:2011zv}
\begin{align}
\label{eq:tmd_evosol}
&\widetilde{C}_{j,\,h}(\xi,\, b_{T}; \, \mu, \,\zeta) = 
\underbrace{\left(
\widetilde{\mathcal{C}}_j^{\;k}(b_{T}^\star; \, \mu_0, \,\zeta_0) \otimes c_{k,\,h} (\mu_0) 
\right) (\xi)}_{\mbox{TMD at reference scales}} \times 
\notag \\
& \times \,
\underbrace{\text{exp} \left\{
\frac{1}{4} \, \widetilde{K}(b_T^\star;\,\mu_0) \, \log{\frac{\zeta}{\zeta_0}} 
+ 
\int_{\mu_0}^{\mu} \frac{d \mu'}{\mu'} \, \left[ \gamma_C(\alpha_S(\mu'),\,1) - \frac{1}{4} \, 
\gamma_K(\alpha_S(\mu')) \, \log{\frac{\zeta}{\mu'^2}}\right]
\right\}}_{\mbox{Perturbative Sudakov Factor}} \times \notag \\
& \times \underbrace{\left(M_C\right)_{j,\,h}(\xi,\,b_T) \mbox{ exp} \left \{
-\frac{1}{4} \, g_K(b_T) \, \log{\frac{\zeta}{\overline{\zeta}_0}}
\right\} }_{\mbox{Non-Perturbative content}}
\end{align} 
where the the $b\star$-prescription has been used to separate out perturbative from non-per\-tur\-ba\-ti\-ve content and the standard choices for the reference values of the scales are:
\begin{subequations}
\label{eq:standard_evo_choices}
\begin{align}
&\mu_0 = \mu_b = \frac{2 e^{-\gamma_E}}{b_T^\star} ;  \\
&\zeta_0 = \mu_b^2 ;  \\
&\hspace{-.3cm}
\begin{cases}
	\overline{\zeta}_0 = \left(M_h \, x \right)^2
	&\mbox{ initial state}; \\
	\overline{\zeta}_0 = \left(\frac{M_h}{z} \right)^2
	&\mbox{ final state}. 
\end{cases}
\end{align}
\end{subequations}
The non-perturbative behavior of the TMD is described by two functions.
The first is $g_K$, the same function that appears in Eq.~\eqref{eq:S2h_final},  describing the long-distance behavior of the Collins-Soper kernel.
The second is the \textbf{TMD model} function  $\left(M_C\right)_{j,\,h}(\xi,\,b_T)$, that embeds the 
genuine non-perturbative behavior of the TMD. 
It is the collinear counterpart of the soft model  but, in contrast, $M_C$ does not depend only on $b_T$, but also on the collinear momentum fraction $\xi$,  the flavor of the reference parton and 
the type of reference hadron associated to the collinear part. 

\bigskip

The definition given for TMDs in Eq.~\eqref{eq:sub_coll} is referred as \textbf{factorization definition} in Ref.~\cite{Boglione:2020cwn}. It differs from the commonly used definition~\cite{Collins:2011zzd,Aybat:2011zv}, that combines the TMDs with the soft factor of Eq.~\eqref{eq:S2h_bTspace} by naively absorbing a square root of it. For this reason, this definition is referred as \textbf{square root definition} in Ref.~\cite{Boglione:2020cwn}. It beautifully simplifies the factorized cross sections of the $2$-h class as it makes the whole soft radiation contribution to disappear in the final result. However, the factorization properties of processes different from the benchmark cases involve naturally the factorization definition, where the TMDs are not contaminated by any soft contributions.
A direct comparison between the two TMD definitions shows that the corresponding TMD models are related in the following way~\cite{Boglione:2020cwn}:
\begin{equation} \label{eq:models_comparison}
M_C^{\; \text{sqrt}} (\xi,\,b_T)= M_C(\xi,\,b_T) \times \sqrt{M_S(b_T)}\,.
\end{equation}
%


\section{A benchmark study: the fragmenting gluon case \label{app:frag_gluon}}

\bigskip

The perturbative approach to the factorization procedure discussed in this paper involves the solution of some very tough integrals, due to the non-trivial interplay between thrust and transverse momentum dependence. The solution of such integrals require non-standard techniques, even just for a NLO approximation. Therefore, it is convenient to show the procedure and some of this advanced mathematical tools in the simple case of a fragmenting gluon. This can be considered as a benchmark study for the treatment of the more relevant case of a fragmenting fermion, which actively contributes to the final result.
Clearly, we have to recover the expected result also in this ``bottom-up" approach. For a $2$-jet final state, one jet is generated by a quark, the other by its corresponding antiquark, while the chance of a jet being produced by a gluon is strongly suppressed.
\begin{figure}[t]
\centering
\includegraphics[width=3.cm]{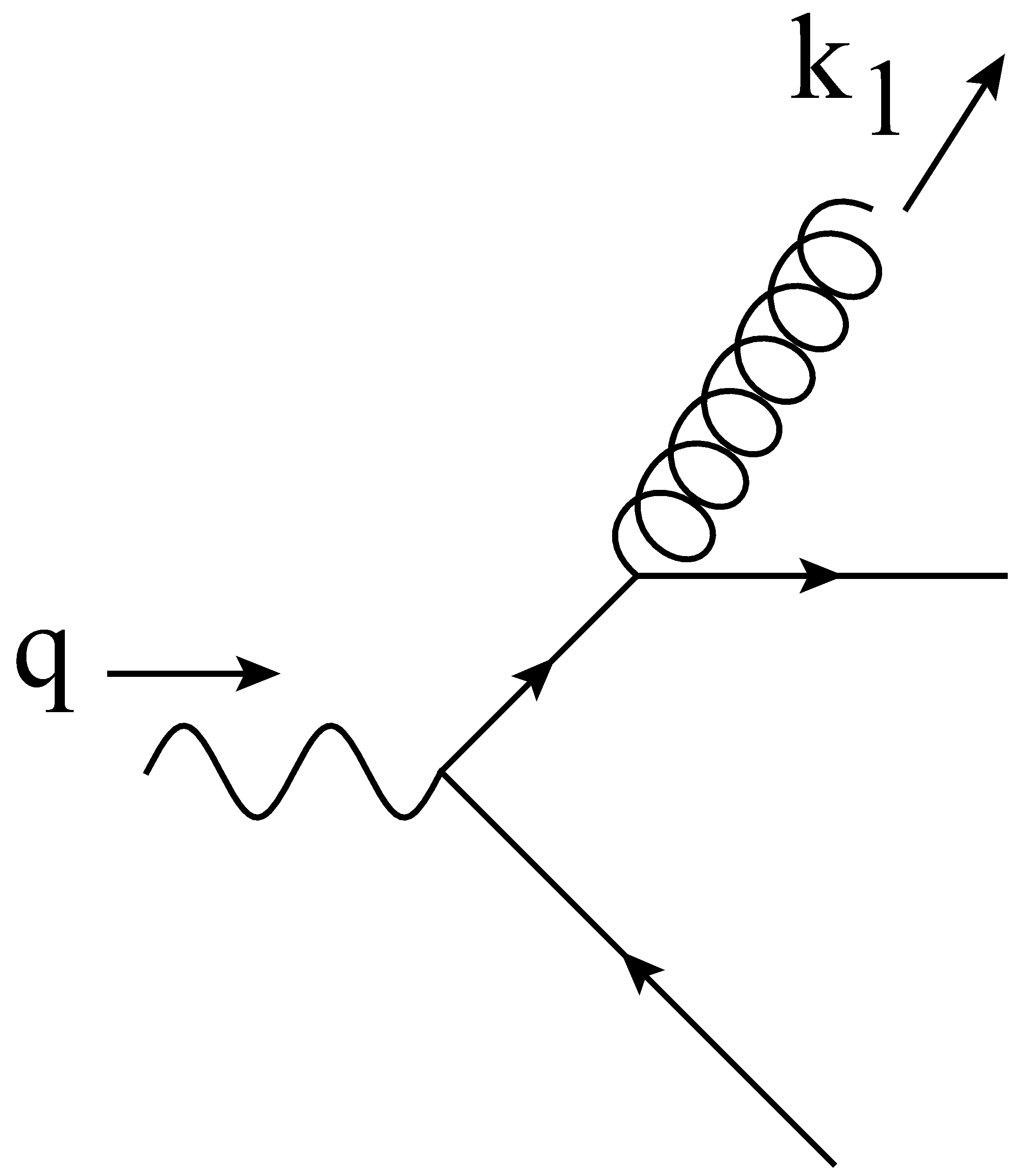}
\caption{The only 1-loop Feynman graph contributing to $\widehat{W}_g^{\mu \nu}$, when the gluon is emitted by the quark.
The emission from the antiquark line is analogous.}
\label{fig:frag_gluon}
\end{figure}
%
Moreover, the power counting suppresses both the configurations in which the emitting fermion reflects backward (action of $T_B$) and when it turns soft (action of $T_S$), regardless of the hemisphere in which it is directed. Therefore, the only leading momentum region is realized by the fermion being collinear to the fragmenting gluon, obtained through the action of $T_A$. 
Therefore we have:
\begin{align}
\label{eq:1loop_xsg_structure}
&\widehat{W}_g^{\mu \nu, \,[1]}\left(\eps;\,z , \tau ,  k_T\right) = 
T_A \left[ \widehat{W}_g^{\mu \nu, \,[1]}\left(\eps;\,z , \tau ,  k_T\right) \right] +  
\substack{\mbox{power suppressed}\\\mbox{corrections}},
\end{align}
where the $T_A$ approximator gives:
\begin{align}
&T_A \left[ \widehat{W}_g^{\mu \nu, \,[1]}\left(\eps;\,z , \tau ,  k_T\right) \right] = 
\sum_f \int \frac{d k^+}{k^+} 
{}^{\star}\widehat{W}^{\mu \nu,\,[0]}_f(k^+,\,Q) \, 
\Gamma_{g/q}^{[1]}\left(\eps;\,{k^+}/{k'^+},\,k_T,\,\tau\right) =
\notag \\
&\quad=
\sum_f \int \frac{d \rho}{\rho} 
{}^{\star}\widehat{W}^{\mu \nu,\,[0]}_f ({z}/{\rho},\,Q) \, 
\Gamma_{g/q}^{[1]}\left(\eps;\,\rho,\,k_T,\,\tau\right),
\end{align}
where, the function $\Gamma_{g/q}$ is the 1-loop  gluon-from-quark generalized fragmentation jet function (GFJF), defined as:
\begin{align}
&\Gamma_{g/q}^{[1]}\left(\eps;\,z,\,k_T,\,\tau\right) = 
\int \frac{d k^-}{(2\pi)^{4-2\eps}}
\frac{\text{Tr}_C}{N_C}\frac{\text{Tr}_D}{4}\left( \gamma^+
\begin{gathered}
\includegraphics[width=3.0cm]{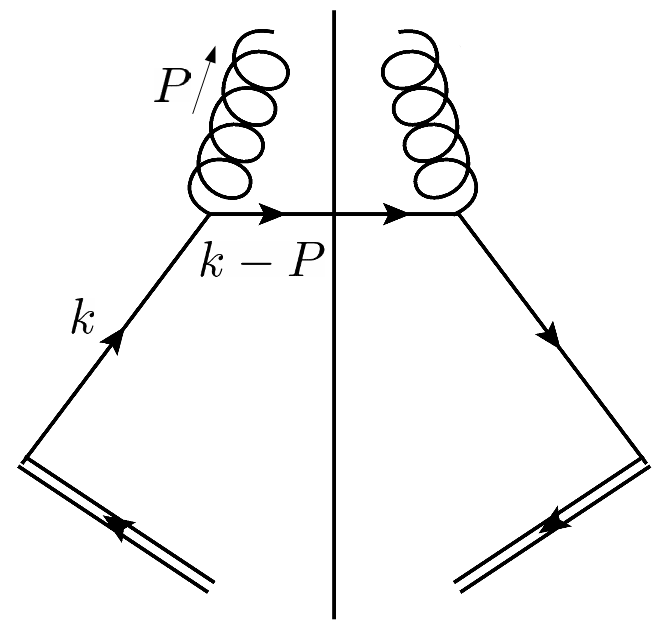}
\end{gathered}
\right) \,\delta\left(\tau - \frac{z}{1-z} \frac{k_T^2}{Q^2}\right) = 
\notag \\
&\quad=
\frac{\alpha_S}{4\pi} 2 C_F S_\eps 
\frac{\Gamma(1-\eps)}{\pi^{1-\eps}}
\frac{\mu^{2\eps}}{k_T^2} 
\theta(1-z) 
\frac{1+(1-z)^2-\eps z^2}{z^2}
\delta\left( \tau - \frac{z}{1-z} \frac{k_T^2}{Q^2}\right).
\label{eq:thurstCollFact_gluon}
\end{align}
The Fourier transform gives:
\begin{align}
&\widetilde{\Gamma}_{g/q}^{[1]}\left(\eps;\,z,\,b_T,\,\tau\right) = 
\notag \\
&=
\frac{\alpha_S}{4\pi} 2 C_F S_\eps
\left(\frac{\mu}{Q} \right)^{2\eps}
\hspace{-.05cm}
\frac{1+(1-z)^2-\eps z^2}{z^2}
\left( \frac{1-z}{z} \right)^{-\eps} 
\hspace{-.13cm}
\theta(1-z) \;
\tau^{-1-\eps} \;\;
\hspace{-.15cm}
\pFq{0}{1} \hspace{-.05cm}\left( 1-\eps;\,-\tau \frac{1-z}{z}  \frac{b^2}{4} \right)
\label{eq:FTthurstCollFact_gluon}
\end{align}
where $b = b_T \, Q$ and the hypergeometric function can also be written as:
\begin{align}
\label{eq:0F1}
\pFq{0}{1} \left( 1-\eps;\,-\tau \frac{1-z}{z}  \frac{b^2}{4} \right) = 
\Gamma(1-\eps)
\left(\frac{b}{2} \sqrt{\tau \frac{1-z}{z}}\right)^{\eps}
J_{-\eps}
\left(b \sqrt{\tau \frac{1-z}{z}}\right)
\end{align}
We may be tempted to expand the hypergeometric function in powers of $\tau$ in Eq.~\eqref{eq:FTthurstCollFact_gluon} and 
assume the lowest order in $\tau$ provides a good description of the $2$-jet region.
However, this also implies that $b << 1$, which naively does not correspond to the power counting region of small $k_T$.
This confirms that in the Fourier conjugate space, the $2$-jet limit is much less trivial than in the transverse momentum space.
In fact, a proper treatment of this issue involves dealing with the $\eps$-expansion of Eq.~\eqref{eq:FTthurstCollFact_gluon} in terms of $\tau$-distributions.
In order to accomplish this, we will make use of a rather simple trick, that can always be exploited in presence of functions of some variable $x$ that varies in the range $[0,1]$,  divergent at most as simple poles when $x$ approaches $0$. In fact, if $f(x)$ is a  function that behaves at most as $\sim {1}/{x}$ when $x \to 0$, then we can recast it as:
\begin{align}
\label{eq:distr_expansion}
f(x) = \delta(x) \int_0^1 d\alpha f(\alpha) +
\left(f(x)\right)_+
\end{align}
With this technique, we can reorganize the dependence on $\tau$ in Eq.~\eqref{eq:FTthurstCollFact_gluon} and disentangle it from the dependence on $b$ when $\tau = 0$. After this operation we can then  safely perform the $\eps$-expansion.
\begin{align}
&\tau^{-1-\eps} 
\pFq{0}{1} \left( 1-\eps;\,-\tau \frac{1-z}{z}  \frac{b^2}{4} \right) = 
\notag \\
&\quad=
-\frac{1}{\eps} \pFq{1}{2} (-\eps;\,1-\eps,1-\eps;\,-\frac{1-z}{z} \frac{b^2}{4}) \delta(\tau) +
\left( \tau^{-1-\eps} 
\pFq{0}{1} \left( 1-\eps;\,-\tau \frac{1-z}{z}  \frac{b^2}{4}\right)
\right)_+ =
\notag \\
&\quad=
\delta(\tau) \left(
-\frac{1}{\eps}  
-\frac{1-z}{z}  \frac{b^2}{4}
\pFq{2}{3}\left(1,1;\,2,2,2;\,
-\frac{1-z}{z}  \frac{b^2}{4}\right)
\right)+ 
\left( 
\frac{J_0\left(b \sqrt{\tau \frac{1-z}{z}}\right)}{\tau}
\right)_+ +
\mathcal{O}\left(\eps\right)
\label{eq:coll0F1_exp}
\end{align}
where $\text{Re}(\eps) < 0$ is required for convergence. We have pushed the expansion up to $\mathcal{O}(\eps^0)$ since the remaining $\eps$-dependent terms in Eq.~\eqref{eq:FTthurstCollFact_gluon} do not present any pole in $\eps$.
%
\begin{figure}
\centering
\includegraphics[width=12cm]{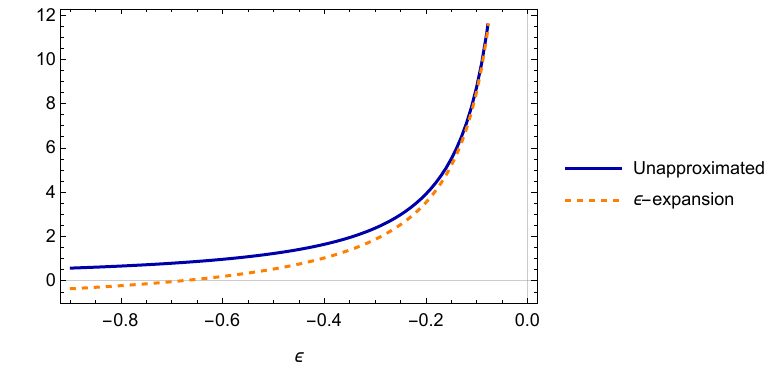}
\caption{The term  $\tau^{-1-\eps} \pFq{0}{1} \left( 1-\eps;\,-\tau \frac{1-z}{z} \frac{b^2}{4} \right)$ in the first line of Eq.~\eqref{eq:coll0F1_exp}, (solid, blue line) is compared with its small $\epsilon$-expansion (orange, dashed line) in the last line of   Eq.~\eqref{eq:coll0F1_exp}. 
These lines are obtained by integrating the r.h.s and l.h.s of Eq.~\eqref{eq:coll0F1_exp} with a test function chosen as $T(\tau)=e^{-\tau}$.}
\end{figure}
%

\bigskip

Now we are ready to consider the large-$b_T$ limit, in order to bring the result back to the $2$-jet approximation. Let's define for simplicity $a = \frac{b^2}{4} \frac{1-z}{z}$. Then, since $z$ in the collinear region cannot be too close to $1$ (large values of $z$, $z\to 1$, can only be reached in the soft approximation), the limit $a \to \infty$ will correspond to the asymptotic behavior for large values of $b$. 
Finding the asymptotic behavior of the term multiplying the $\delta(\tau)$ in the last line of Eq.~\eqref{eq:coll0F1_exp} is quite easy. In fact we have:
\begin{align}
&a \, \pFq{2}{3}\left(1,1;\,2,2,2;\,-a\right) =
\notag \\
&\quad=
\log{\left( a \, e^{2\gamma_E}\right)} +
\frac{1}{a^{3/4}} \, 
\frac{1}{\sqrt{\pi}} \, \cos{\left(2\sqrt{a} + \frac{\pi}{4}\right)}
+
\mathcal{O}
\left( \frac{1}{a^{5/4}} 
\times 
\parbox{4.5em}{oscillating\\ function}
\right).
\label{2F3_asy}
\end{align}
On the other hand, a proper estimation of the asymptotic behavior of the plus distribution in Eq.~\eqref{eq:coll0F1_exp} is much less trivial.
Clearly, the Bessel function $J_0$ behaves as $\sim a^{-1/2}$ for large-$a$. However, such a rough estimation compromises the $\tau$ dependence, which becomes $\sim \tau^{-5/4}$, not integrable anymore for \emph{any} test function $T(\tau)$.
Such an operation should therefore be performed more carefully. With the help of a test function $T(\tau)$, the following asymptotic series can be obtained by integrating $N$ times by parts:
\begin{align}
&\int_0^1 d\tau \, T(\tau)
\tplus{J_0
\left( 2 \sqrt{a \tau} \right)} = 
S_N(\sqrt{a}) + R_N(\sqrt{a}).
\label{eq:J0plus_int}
\end{align}
where we have introduced:
\begin{align}
&S_N(\sqrt{a}) = 
\sum_{j=0}^{N-1} (-1)^j 
\frac{J_{j+1}(2\sqrt{a})}{a^{(j+1)/2}}
\left.\frac{d^j}{d\tau^j} \left(
\frac{T(\tau) - T(0)}{\tau}
\right)\right\vert_{\tau=1};
\label{eq:Sn}
\\
&R_N(\sqrt{a}) = 
(-1)^N a^{-N/2}
\int_0^1 d\tau
\frac{d^N}{d\tau^N} \left(
\frac{T(\tau) - T(0)}{\tau}
\right) \,
\tau^{N/2} J_{N}(2\sqrt{a \, \tau}).
\label{eq:Rn}
\end{align}
The series $S_{N\to\infty}$ diverges for any value of $a$, however its partial sums (for finite $N$) can be used to compute the integral of Eq.~\eqref{eq:J0plus_int} to any desired accuracy. In fact, $R_N$ is of order $a^{-(N+1)/2-1/4}$ and can be made small at will.
Furthermore, the derivative of the test function in Eq.~\eqref{eq:Sn} can be rewritten as:
\begin{align}
\frac{d^j}{d\tau^j} \left(
\frac{T(\tau) - T(0)}{\tau}
\right) = 
\frac{(-1)^j j!}{\tau^j}
\left(
\frac{T(\tau)-T(0)}{\tau} +
\sum_{k=1}^j \frac{(-1)^{k}}{k!} \,
\frac{T^{[j]}(\tau)}{\tau^{1-k}}
\right).
\label{eq:deriv_FTest}
\end{align}
Hence, we can recast $S_N$ in the following form:
\begin{align}
&S_N(\sqrt{a}) = 
\sum_{j=0}^{N-1}
j! \, \frac{J_{j+1}(2\sqrt{a})}{a^{(j+1)/2}}
\left[
-T(0) + 
\sum_{k=0}^j \frac{(-1)^{k}}{k!} \,
\frac{T^{[j]}(1)}{\tau^{1-k}}
\right],
\label{eq:Sn_2}
\end{align}
which at level of distribution becomes: 
\begin{align}
\tplus{J_0
\left( 2 \sqrt{a \tau} \right)} = 
\sum_{j=0}^{N-1}
j! \, \frac{J_{j+1}(2\sqrt{a})}{a^{(j+1)/2}}
\left[
-\delta(\tau) + 
\sum_{k=0}^j \frac{1}{k!} \,
\delta^{[k]}(1-\tau)
\right] + 
\mathcal{O}\left( a^{-\frac{(3+2N)}{4}} \right),
\label{eq:J0plus}
\end{align}
where $\delta^{[k]}(1-\tau)$ is the $k$-th distributional derivative of $\delta(1-\tau)$, which, setting $\tau=1$, produces (boundary) terms that are not contributing to the $2$-jet limit and hence that will be thrown away.
Notice that the integration of each side of Eq.~\eqref{eq:J0plus} gives zero as required.
Now, we have a correct asymptotic expansion of the plus distribution appearing in the last line of Eq.~\eqref{eq:coll0F1_exp}. In the $2$-jet approximation, its crudest estimation is given by:
\begin{align}
&\tplus{J_0
\left( 2 \sqrt{a \tau} \right)} = 
- \delta(\tau)  \frac{J_{1}(2\sqrt{a})}{\sqrt{a}} + 
\mathcal{O}
\left( \frac{1}{a} 
\times 
\parbox{4.5em}{Bessel\\ function}
\right)
\notag \\
&\quad=
\delta(\tau) 
\frac{1}{a^{3/4}} \, 
\frac{1}{\sqrt{\pi}} \, \cos{\left(2\sqrt{a} + \frac{\pi}{4}\right)} + 
\mathcal{O}
\left( \frac{1}{a^{5/4}} 
\times 
\parbox{4.5em}{oscillating\\ function}
\right)
\label{eq:J0plus_crudest}
\end{align}
Notice that this result cancels exactly the first correction to the logarithm in Eq.~\eqref{2F3_asy}. This does not happen by chance; it can be verified at any order  $\mathcal{O}\left(a^{-k/4}\right)$, for $k=3,\dots,2 N$, for any $N$.
%
\begin{figure}
\centering
\includegraphics[width=12cm]{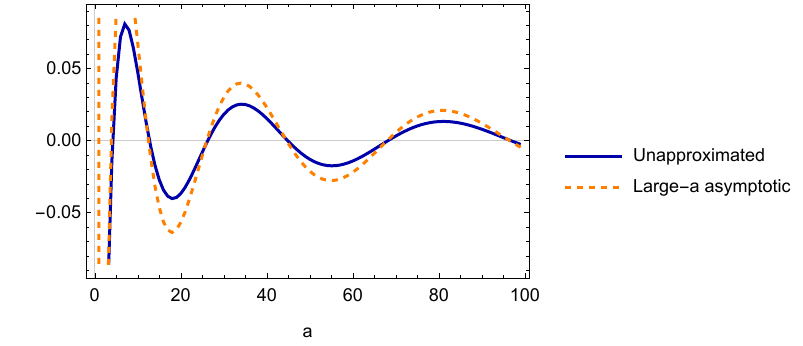}
\caption{The distribution   $\tplus{J_0
\left( 2 \sqrt{a \tau} \right)}$ in the first line of Eq.~\eqref{eq:J0plus_crudest} (solid, blue line) is compared with its large $a$-expansion (orange, dashed line) as obtained in the last line of Eq.~\eqref{eq:J0plus_crudest}. These curves are obtained by integrating 
with a test function chosen as $T(\tau)=e^{-\tau}$.}
\end{figure}
%
Finally, the $\eps$-expansion in the $2$-jet limit for the l.h.s. of Eq.~\eqref{eq:coll0F1_exp} is given by:
\begin{align}
&\tau^{-1-\eps} 
\pFq{0}{1} \left( 1-\eps;\,-\tau \frac{1-z}{z}  \frac{b^2}{4} \right) 
\stackrel{2-\text{jet}}{=}
\delta(\tau) 
\left(
-\frac{1}{\eps}
- 2\log{\left( \frac{b}{c_1} \right) - 
\log\left(\frac{1-z}{z} \right)}
\right)
+\notag \\
&\quad+
\mathcal{O}(\eps) +
\mathcal{O}\left( b^{-\frac{3+2N}{2}} \right),
\quad \forall \; N = 0,1,\dots
\label{eq:coll0F1_exp_2}
\end{align}
where $c_1=2 e^{-\gamma_E}$.
Inserting this result in Eq.~\eqref{eq:FTthurstCollFact_gluon}, we can write the large-$b$ asymptotic behavior of $\widetilde{\Gamma}_{g/q}$, which has to be considered as its $2$-jet approximation. We have:
\begin{align}
&\widetilde{\Gamma}_{g/q}^{[1],\text{ASY}}\left( \eps;\,z,\,b_T,\,\tau \right) = 
\delta(\tau) \, z \, \widetilde{D}_{g/q}^{[1]}\left( 
\eps;\,z,\,b_T
\right) +
\mathcal{O}\left( b^{-\frac{3+2N}{2}} \right),
\quad \forall \; N = 0,1,\dots.
\label{eq:FTthurstCollFactASY_gluon}
\end{align}
where $\widetilde{D}_{g/q}^{}[1]$ the 1-loop gluon-from-quark TMD FF in $b_T$-space.
Notice how the whole dependence on the thrust has been washed away in the $2$-jet approximation.
Finally:
\begin{align}
\label{eq:gWpartonic_ch3}
&\widetilde{\widehat{W}}_g^{\mu \nu, \,[1]}\left(\eps;\,z , \tau ,  b_T\right) =
\notag \\
&=
\sum_f \int \frac{d \rho}{\rho} 
\sum_f \int \frac{d \rho}{\rho} 
\widehat{W}^{\mu \nu,\,[0]}_f ({z}/{\rho},\,Q,\,\tau) \, 
\left( \rho \, \widetilde{D}_{g/q}^{[1]}\left( \eps;\,z,\,b_T \right) \right) + 
\substack{\mbox{power suppressed}\\\mbox{corrections}}.
\end{align}
Notice that the $\delta(\tau)$ in Eq.~\eqref{eq:FTthurstCollFactASY_gluon} recreates the LO partonic tensor. The power suppresed terms involve both the errors associated to the factorization procedure  of Eq.~\eqref{eq:1loop_xsg_structure}, and also the errors associated to the $2$-jet limit of Eq.~\eqref{eq:FTthurstCollFactASY_gluon}.

%
\begin{figure}
\centering
\includegraphics[width=10cm]{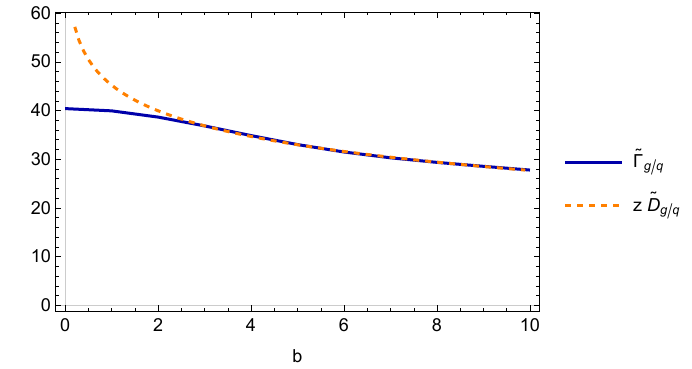}
\caption{The gluon-from-quark GFJF
in Eq.~\eqref{eq:FTthurstCollFact_gluon}, (solid, blue line) is compared to its 2-jet limit (orange, dashed line) in the r.h.s. of Eq.~\eqref{eq:FTthurstCollFactASY_gluon}. 
The plotted lines are obtained by integrating 
with a test function chosen as $T(\tau)=e^{-\tau}$.}
\end{figure}
%
Eq.~\eqref{eq:gWpartonic_ch3} is a crucial result. 
In the $2$-jet limit, the 1-loop contribution of the fragmenting gluon turns out to be simply the gluon-from-quark TMD FF, plus a remnant which is power suppressed. 
Therefore, in $b_T$-space, the overlapping with the contribution of the TMD FFs is self-evident. In fact, the subtraction mechanism described in Refs.~\cite{Boglione:2020auc,Boglione:2020cwn} simply returns the power suppressed terms of Eq.~\eqref{eq:FTthurstCollFactASY_gluon}. This means that the subtracted partonic tensor, which describes the ``core" of the process,  is power suppressed in the $2$-jet limit. 

\bigskip


\section{Solution of Integrals through the Mellin transform technique \label{app:solutionsint}}


\bigskip

In this section, we present the solution of the non-trivial integral appearing in the Fourier transform of the soft thrust factor in Section~\ref{sec:reg1}. The Mellin transform trick is taken from Ref.~\cite{Fikioris:2006}.

Let's considerthe integral of Eq.~\eqref{eq:int_difficult}.
Its solution can be obtained by exploiting the convolution property of the Mellin transforms:
\begin{align}
\label{eq:MellinTransf_conv}
\int_0^\infty dy \, h(y)\,g(a y) = \int_{\delta-i \infty}^{\delta+i \infty}\frac{d u}{2\pi i} a^{-u} \widehat{h}(1-u)\widehat{g}(u)
\end{align}
where $\widehat{f}$ denotes the Mellin transform of the function $f$. The off-set $\delta$ in Eq.~\eqref{eq:MellinTransf_conv} is a real number that lies in the intersection of the convergence region of $\widehat{h}$ and $\widehat{g}$. 

This property can be applied to $I_\eps{a,\,r}$ if we first change variable $x \mapsto y^{-2}$. Then:
\begin{align}
\label{eq:int_difficult_1}
I_\eps \left(a,\, r \right) =
2 \int_0^\infty dy \underbrace{\frac{y^{-1-\eps}}{1-r\,y^2}\theta (1-y)}_{h_{\eps,\,r}(y)} \, \underbrace{J_{-\eps}\left( a y \right)}_{g_\eps (a y)}
\end{align}
The Mellin transforms are:
\begin{align}
\label{eq:int_difficult_MellinT}
&\widehat{h}_{\eps,\,r}(u) = 
-\frac{1}{r} \frac{1}{3+\eps-u} 
\pFq{2}{1} \left(1,\,\frac{3+\eps-u}{2};\,\frac{5+\eps-u}{2};\,\frac{1}{r} \right),
\quad \text{for } \text{Re}(u) < 3+\text{Re}(\eps);\\
&\widehat{g}_\eps(u) = 
2^{-1+u} \frac{\Gamma\left(-\frac{\eps}{2}+\frac{u}{2}\right)}{\Gamma\left( 1-\frac{\eps}{2}-\frac{u}{2}\right)},
\quad \text{for } \text{Re}(\eps) <\text{Re}(u) <\frac{3}{2}.
\end{align}
Therefore, we can choose $\text{Re}(\eps) < \delta < {3}/{2}$. 
Finally:
\begin{align}
\label{eq:int_difficult_2}
&I_\eps \left(a,\, r \right) =
\notag \\
&\quad=
-\frac{1}{r} \int_{\delta-i \infty}^{\delta+i \infty}\frac{d u}{2\pi i} \left(\frac{a}{2}\right)^{-u} \,
\frac{\Gamma\left(-\frac{\eps}{2}+\frac{u}{2}\right)}{\Gamma\left( 1-\frac{\eps}{2}-\frac{u}{2}\right)}
\frac{1}{2+\eps+u}
\pFq{2}{1} \left(1,\,1+\frac{\eps}{2}+\frac{u}{2};\,2+\frac{\eps}{2}+\frac{u}{2};\,\frac{1}{r} \right) =
\notag \\
&\quad=
S_1(\eps,\,a,\,r)+S_2(\eps,\,a,\,r)
\end{align}
where we have defined the two series:
\begin{align}
\label{eq:int_difficult_Series}
&S_1(\eps,\,a,\,r) = -\frac{1}{r} 
\sum_{k=0}^{\infty} \substack{\text{\large Res} \\ u = -2k + \eps} 
\, f_{\eps}(u,\,r), \\
&S_2(\eps,\,a,\,r) = -\frac{1}{r} 
\sum_{k=1}^{\infty} \substack{\text{\large Res} \\ u = -2k - \eps} 
\; f_{\eps}(u,\,r)
\end{align}
with:
\begin{align}
\label{eq:int_difficult_integrand}
&f_{\eps}(u,\,r) = 
\left(\frac{a}{2}\right)^{-u} \,
\frac{\Gamma\left(-\frac{\eps}{2}+\frac{u}{2}\right)}{\Gamma\left( 1-\frac{\eps}{2}-\frac{u}{2}\right)}
\frac{1}{2+\eps+u}
\pFq{2}{1} \left(1,\,1+\frac{\eps}{2}+\frac{u}{2};\,2+\frac{\eps}{2}+\frac{u}{2};\,\frac{1}{r} \right).
\end{align}
%
%
\begin{figure}
\centering
\includegraphics[width=10cm]{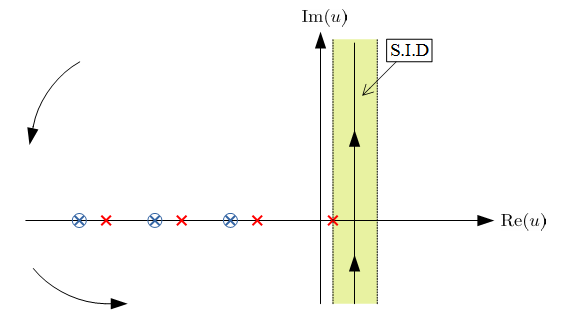}
\caption{Graphical representation of the Mellin conjugate space. The integrand function in Eq.~\eqref{eq:int_difficult_integrand} has two sets of poles in $u = -2k + \eps$, for $k\geq0$ and in $u = -2k - \eps$, for $k\geq1$. The first set is due associated with the Gamma function $\Gamma\left(-\frac{\eps}{2}+\frac{u}{2}\right)$, the second to the hypergeometric function. The green strip between $\eps$ and $3/2$ is the "strip of initial definition" (SID) and coincides with the intersection of the convergence regions of the two functions $g_\eps$ and $h_{\eps,r}$. The integration path must lie into the SID. Finally, since the hypergeometric function produces an essential singularity in $u \to +\infty$, we must close the contour to the left.}
\end{figure}
Let's start by considering $S_1$: 
%
%
%
\begin{align}
\label{eq:S1_1}
S_1(\eps,\,a,\,r) = 
-\left( \frac{a}{2} \right)^{-\eps} r^\eps \sum_{k=0}^\infty \, \frac{(-1)^k}{k!} \, \left( \frac{a}{2} \right)^{2 k} r^{-k} \frac{1}{\Gamma(1+k-\eps)}\,B_{1/r}(1+\eps-k,0)
\end{align}
If $r = r_1 \equiv e^{-2 y_1}$, as in the first term of Eq.~\eqref{eq:cssSthrustfactor_plusFT_1}, we are interested in the limit $r_1 \to 0$. Since:
\begin{align}
\label{eq:S1_r1_Beta}
&B_{1/{r_1}}(1+\eps-k,0) = 
\notag \\
&=
- (-1)^{-\eps+k} \frac{\Gamma(2+\eps-k)\Gamma(-\eps+k)}{1+\eps-k}- r_1^{-1-\eps+k} \left( \frac{\Gamma(\eps-k)\Gamma(2+\eps-k)}{(1+\eps-k)\Gamma(1+\eps-k)^2} r + \mathcal{O}\left(r_1^2\right)\right),
\end{align}
we have:
\begin{align}
\label{eq:S1_r1}
&S_1(\eps,\,a,\,r_1) = 
- (-1)^{-\eps} \frac{\pi}{\sin(\eps \pi)} r_1^{\eps/2} J_{-\eps}\left( \frac{a}{\sqrt{r_1}} \right) -
\notag \\
&\quad-
\left( \frac{a}{2} \right)^{-\eps} \frac{1}{\eps^2 \Gamma(-\eps)} \pFq{1}{2}\left( -\eps;\,1-\eps,1-\eps;\,-\frac{a^2}{4}\right) 
+ \mathcal{O}\left(r_1^2\right)
\end{align}
On the other hand, if $r = {1}/{r_2} \equiv e^{-2 y_2}$, as in the second term of Eq.~\eqref{eq:cssSthrustfactor_plusFT_1} we are interested in the limit $r_2 \to 0$. Since:
\begin{align}
\label{eq:S1_r2_Beta}
&B_{r_2}(1+\eps-k,0) = 
r_2^{1+\eps-k} \left( \frac{1}{1+\eps-k} + \mathcal{O}\left(r_2\right)\right),
\end{align}
Therefore:
\begin{align}
\label{eq:S1_r2}
&S_1\left(\eps,\,a,\,\frac{1}{r_2}\right) = 
\notag \\
&\quad=
- \left( \frac{a}{2} \right)^{-\eps} \frac{1}{(1+\eps)\Gamma(1-\eps)} \pFq{1}{2}\left( -1-\eps;\,1-\eps,-\eps;\,-\frac{a^2}{4}\right)r_2 
+ \mathcal{O}\left(r_2^2\right) = 
\mathcal{O}\left(r_2\right).
\end{align}
We can now move to $S_2$:
\begin{align}
\label{eq:S2}
&S_2(\eps,\,a,\,r) = 
-\frac{a}{4 r} \left( \frac{a}{2} \right)^\eps \Gamma(-1-\eps) - \left(\frac{a}{2}\right)^\eps
\sum_{k=2}^\infty
\frac{1}{k!}\left(\frac{a}{2}\right)^{2 k}r^{-k}\Gamma(-\eps-k) = 
\notag \\
&\quad=
\left( \frac{a}{2} \right)^\eps \Gamma(-\eps) + 
\frac{\pi}{\sin(\eps \pi)} r^{\eps/2}J_\eps\left( \frac{a}{\sqrt{r}}\right).
\end{align}
If $r = r_1$, we cannot expand anymore the previous result around $r_1 \sim 0$. However, if $r = {1}/{r_2}$, then $S_2(\eps,\,a,\,{1}/{r_2}) = \mathcal{O}(r_2)$ and it can be neglected.
Notice that all the contributions involving $y_2$ are suppressed. Hence, only $y_1$, the leading rapidity cut-off in the $\mathrm{S}_A$-hemisphere, will survive in the final result.
In fact, using Eqs.~\eqref{eq:S1_r1}~\eqref{eq:S2}, we have:
\begin{align}
\label{eq:difficultint_I1}
&I_\eps(a,\,r_1) = 
\frac{\pi}{\sin{(\eps \pi)}} r_1^{\eps/2}
\left[
J_\eps\left( \frac{a}{\sqrt{r_1}}\right) - (-1)^{-\eps} J_{-\eps}\left( \frac{a}{\sqrt{r_1}}\right)
\right] + 
\left(\frac{a}{2}\right)^\eps \Gamma(-\eps) -
\notag \\
&\quad-
\left(\frac{a}{2}\right)^{-\eps} \frac{1}{\eps^2 \Gamma(-\eps)} \pFq{1}{2} \left(-\eps;1-\eps,\,1-\eps;\,-\frac{a^2}{4}\right) + \mathcal{O}(r_1) 
\end{align}
where the combination of the Bessel-$J$ functions can be rearranged as:
\begin{align}
\label{eq:Jeps_rearrangement}
\frac{\pi}{\sin{(\eps \pi)}} r_1^{\eps/2}
\left[
J_\eps\left( \frac{a}{\sqrt{r_1}}\right) - (-1)^{-\eps} J_{-\eps}\left( \frac{a}{\sqrt{r_1}}\right)
\right] = 
-2\left(-r_1\right)^{\eps/2}K_{-\eps}\left( \frac{a}{\sqrt{- r_1}} \right).
\end{align}
The other integral instead is suppressed as $r_2 \to 0$:
\begin{align}
\label{eq:difficultint_I2}
&I_\eps\left(a,\,\frac{1}{r_2}\right) = 
\mathcal{O}(r_2).
\end{align}

\noindent
The solution of the integral of Eq.~\eqref{eq:SCthrustfactor_FT} can be obtained through the same procedure used to solve the integration in Eq.~\eqref{eq:int_difficult}. The result is:
\begin{align}
\label{eq:int_notsodifficult}
\int_0^\infty 
\frac{x^{\eps/2}}{x-r_1}\,J_{-\eps}\left( \frac{a}{\sqrt{x}} \right) = 
-2\left(-r_1\right)^{\eps/2}K_{-\eps} \left( \frac{a}{\sqrt{-r_1}} \right) + 
\left( \frac{a}{2} \right)^{\eps} \Gamma(-\eps),
\end{align}
Notice that, differently from Eqs.~\eqref{eq:difficultint_I1} and~\eqref{eq:difficultint_I2}, this is an exact result.

\bigskip


\section{Color-coded representation of the kinematic regions \label{app:canvas}}


\bigskip

In this Appendix we will present the same plots shown in Figs.~\ref{fig:makris_canvas} and \ref{fig:AS_canvas}, but from a different perspective. Each panel corresponding to the set of criteria presented in Ref.~\cite{Makris:2020ltr}, see Eq.~\eqref{eq:makris_crit}, will be directly compared to the analogue panel obtained by applying the more refined algorithm proposed in this paper, shown in Fig.~\ref{fig:AS_algorithm}. 
As the size of these figures is augmented, all labels should result more visible and easier to read.
%
%
In the following plots, BELLE data~\cite{Seidl:2019jei} are presented according to the color-coding associated to the three kinematic regions of $\epm \to h\,X$.
Red bins correspond to Region 1, orange bins to Region 2 and green bins to Region 3. 
Clearly, only thrust values corresponding 
to a $2$-jet topology are considered, namely all bins with 
$0.75 \leq T \leq 1.0$. As far as $z_h$ is concerned, all available bins are included.
%
The shaded areas correspond to bins for which the value of $P_T$ falls outside of the TMD-regime. The cut-off in $P_T$ of Eq.~\eqref{eq:TMDcut} is represented by a vertical blue line in each panel. In the implementation of this cut, the symbol $\ll$ ``much smaller than" is rendered as "less than $25\%$". 
Instead,
for the algorithm of Fig.~\ref{fig:AS_algorithm}, the symbol $\ll$ ``much smaller than" is evaluated as "less than $30\%$".


Some general features of the two different set of criteria for the data selections become evident from the direct comparison. First of all, the criteria of Ref.~\cite{Makris:2020ltr} suggest a strong dominance of Region 1. Instead,  with the more refined algorithm presented in this paper, most of the BELLE data turn out to belong to Region 2, as expected, while only the lower $z_h$ bins correspond to Region 1. This fits perfectly with the physical expectation that  
in Region 1 it is much easier for soft radiation to transversely deflect a low-energetic hadron than in Region 2.
On the other hand, the distribution of the green bins, associated to Region 3, seem not to be affected by the two different kinds  of data selection. In both cases, Region 3 starts becoming relevant for TMD studies only at very large values of thrust. In fact, green bins appear on the left of the cut in $P_T$ only in the very last bin, for $T=0.975$.
This is in agreement with the physical expectation that 
in Region 3 a hadron detected near the jet boundary can hardly be associated to the ``pure" TMD-regime, unless the jet is extremely narrow, i.e. at very large values of thrust.

%
\begin{figure}
\centering
\includegraphics[width=15cm]{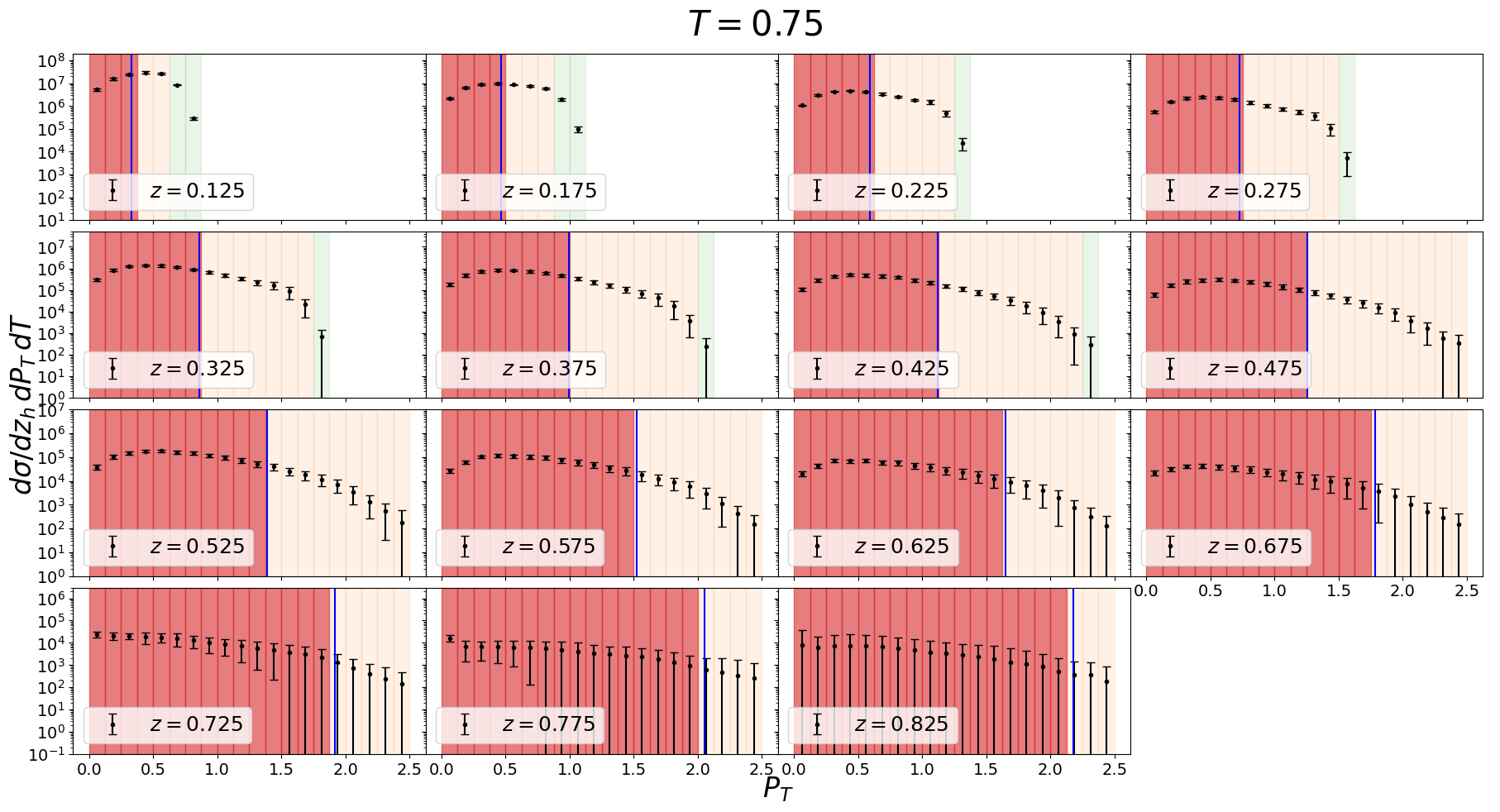}\\
\includegraphics[width=15cm]{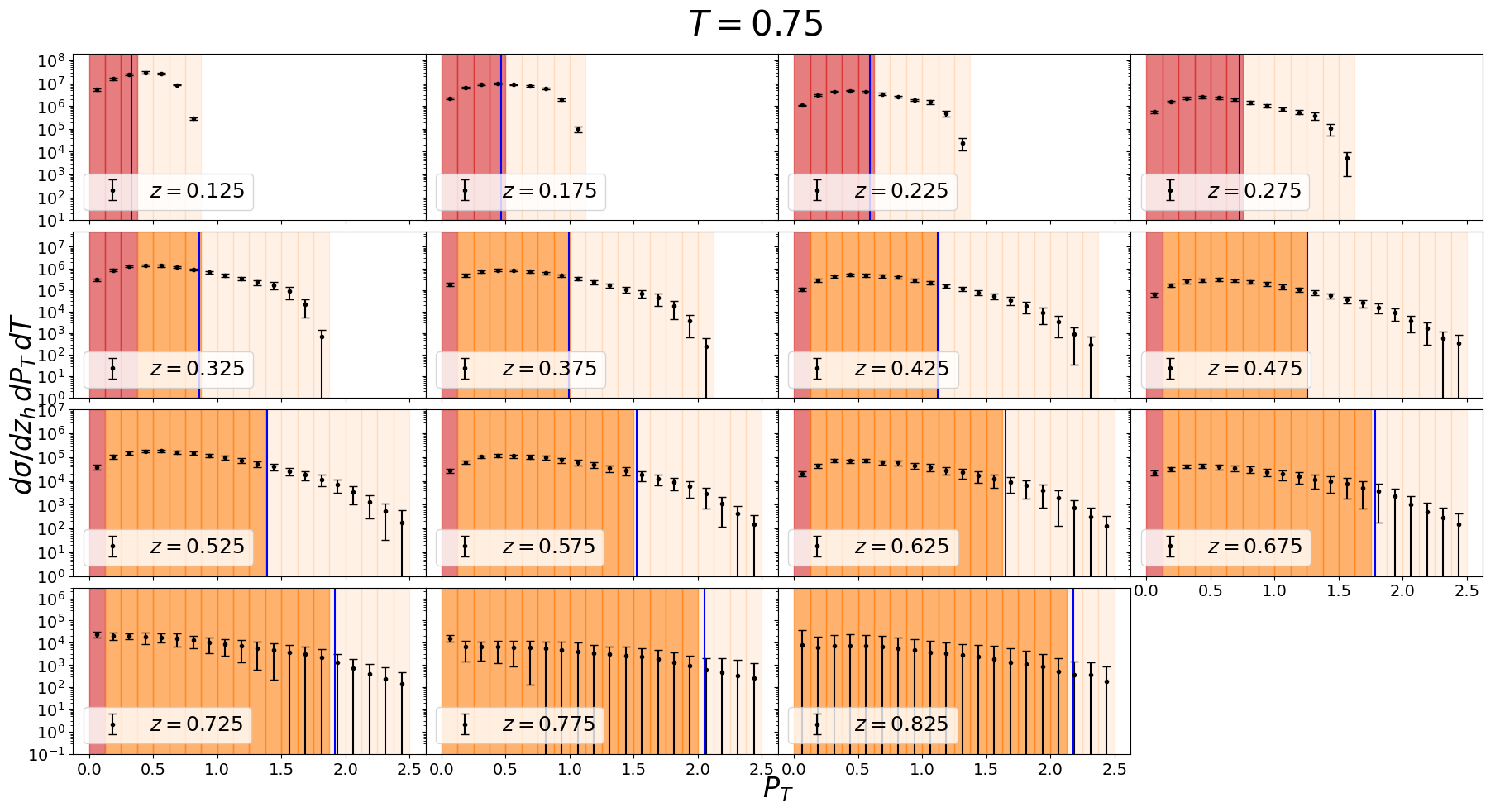}\\
\caption{BELLE data for thrust $T=0.750$ selected according to the criteria of Eq.~\eqref{eq:makris_crit} (upper panel) and to the algorithm of  Fig.~\ref{fig:AS_algorithm} (lower panel). 
Here Region 3 (green bins) is not realized. 
The criteria of Ref.~\cite{Makris:2020ltr} suggest a total dominance of Region 1. Instead,  with the more refined algorithm proposed in this paper, we find that only the lower $z_h$ bins correspond to Region 1, while most of the BELLE data belong to Region 2.} 
\label{fig:canvas_T_750_compare}
\end{figure}

\begin{figure}
\centering
\includegraphics[width=14cm]{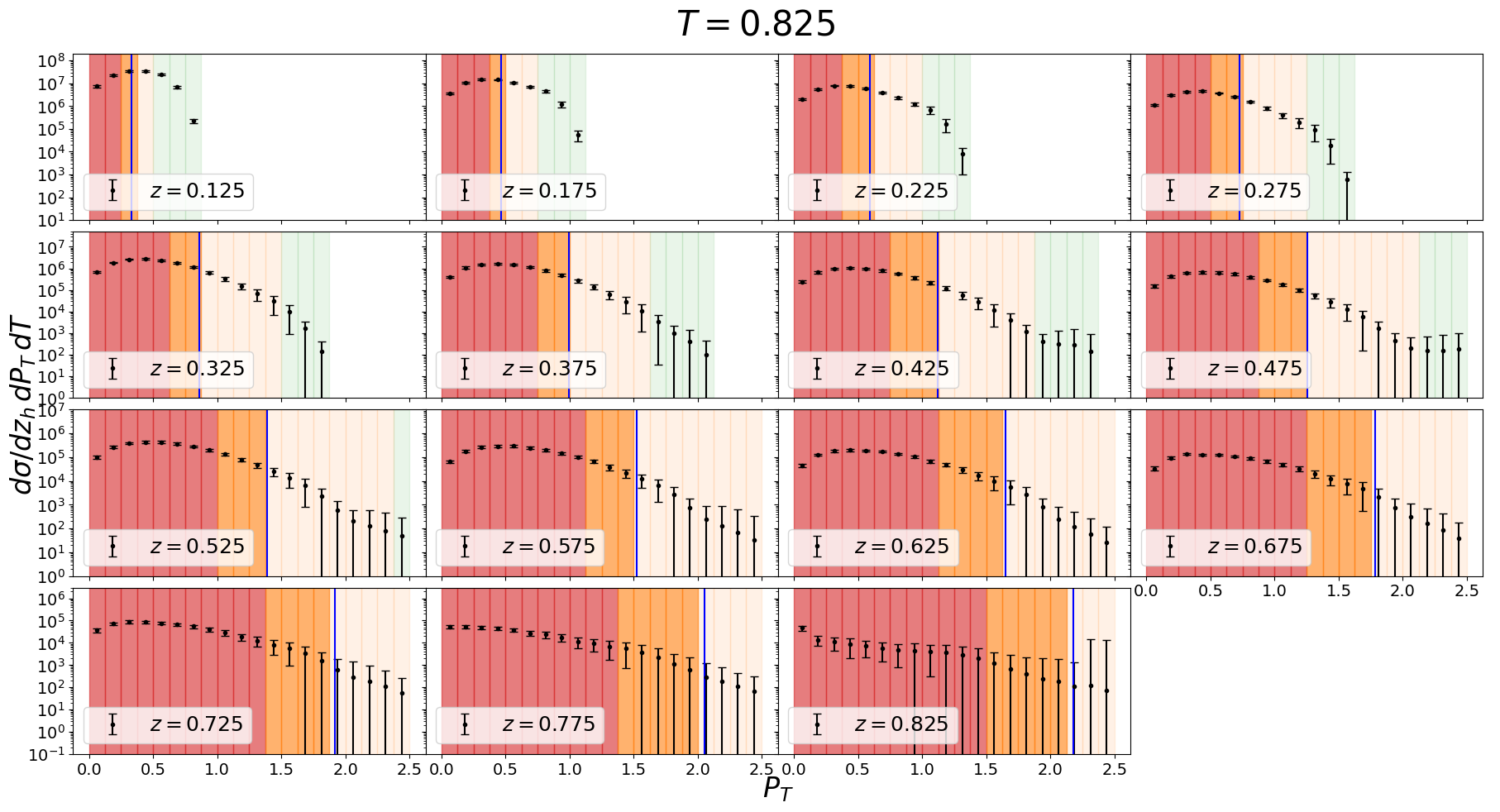}\\
\includegraphics[width=14cm]{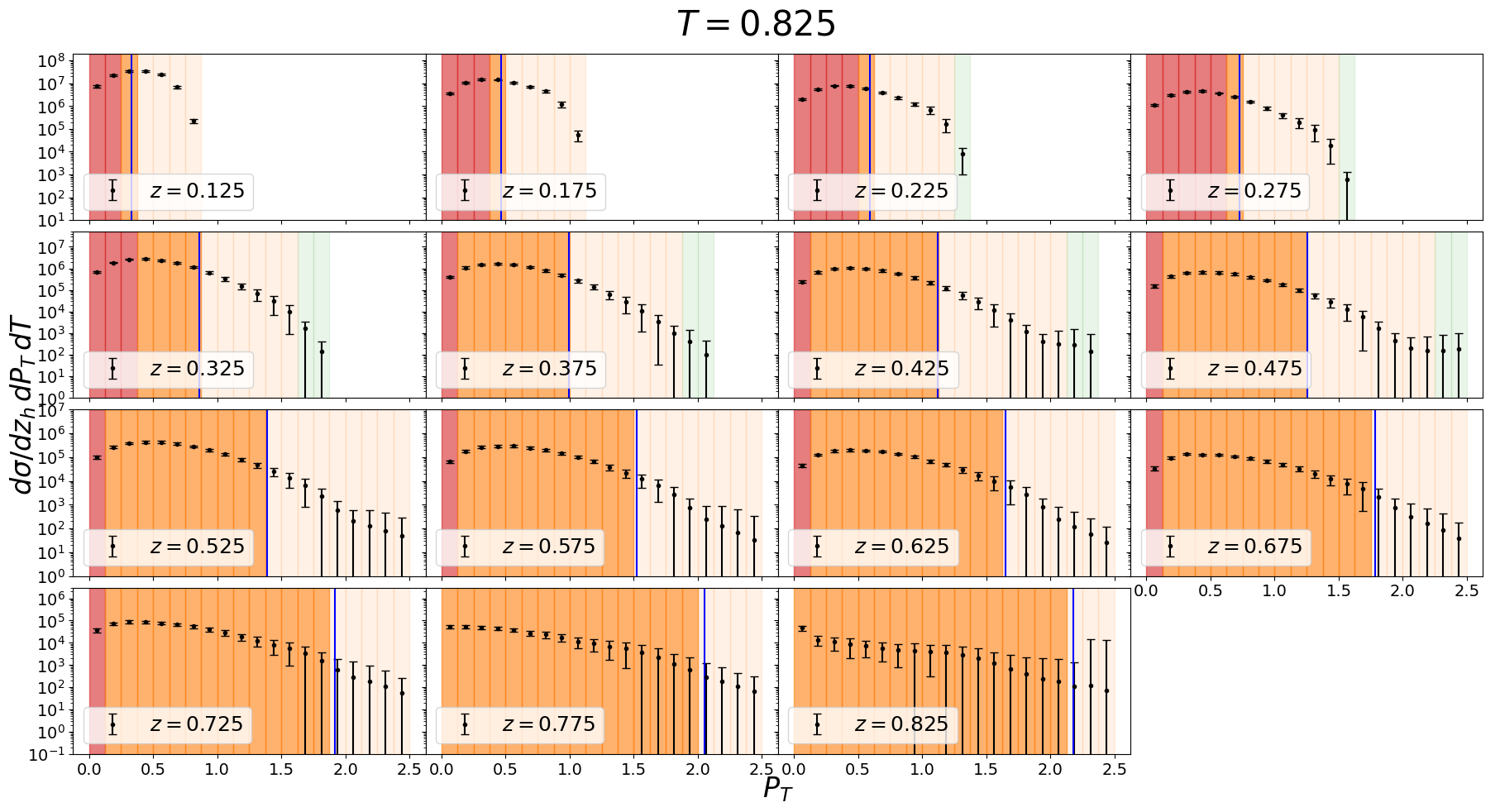}
\caption{BELLE data for thrust $T=0.825$ selected according to the criteria of Eq.~\eqref{eq:makris_crit} (upper panel) and to the algorithm of  Fig.~\ref{fig:AS_algorithm} (lower panel). 
Here Region 3 (green bins) is not realized.
The criteria of Ref.~\cite{Makris:2020ltr} suggest a total dominance of Region 1. Instead,  with the more refined algorithm proposed in this paper, we find that only the lower $z_h$ bins correspond to Region 1, while most of the BELLE data belong to Region 2.}
\label{fig:canvas_T_825_compare}
\end{figure}

\begin{figure}
\centering
\includegraphics[width=14cm]{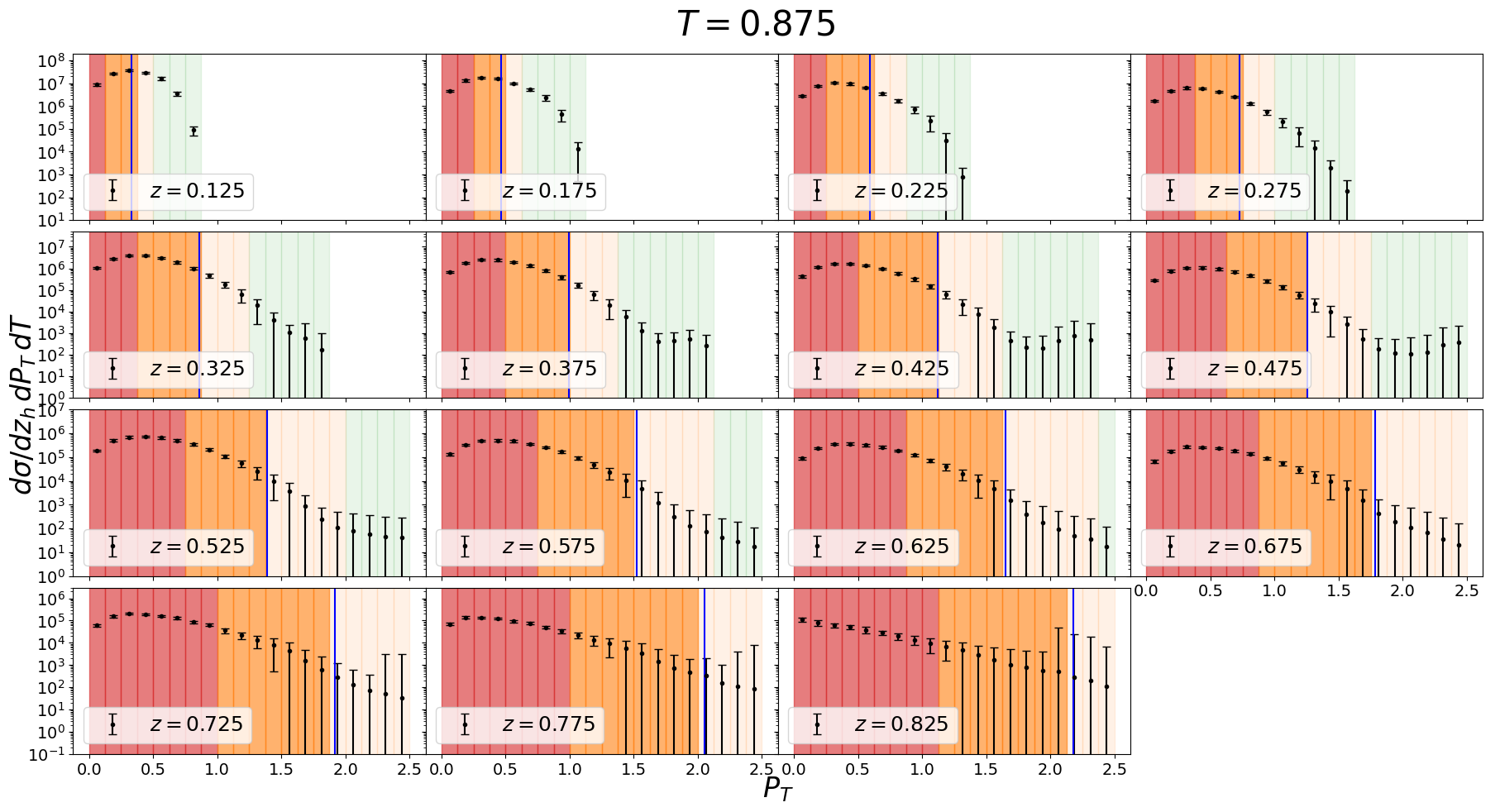}\\
\includegraphics[width=14cm]{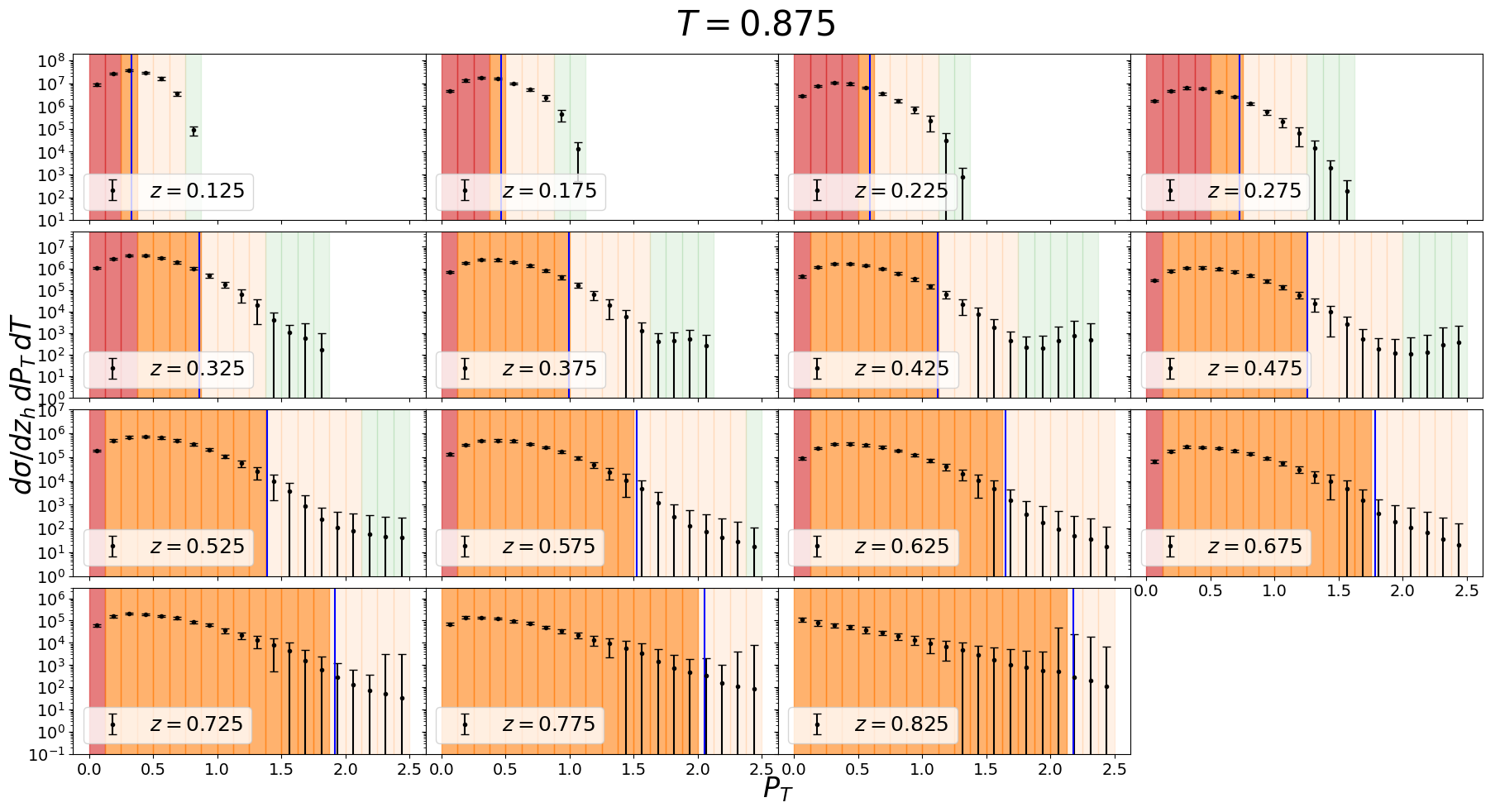}\\
\caption{
BELLE data for thrust $T=0.875$ selected according to the criteria of Eq.~\eqref{eq:makris_crit} (upper panel) and to the algorithm of  Fig.~\ref{fig:AS_algorithm} (lower panel). 
Here Region 3 (green bins) is not realized.
The criteria of Ref.~\cite{Makris:2020ltr} suggest that Region 1 and Region 2 are equally distributed on the left of the cut in $P_T$. Therefore, a phenomenological analysis performed according to this criteria necessarily requires a matching procedure in order to properly describe the bins at the boundaries between Region 1 and Region 2. Instead,  with the more refined algorithm proposed in this paper, we find that only the lower $z_h$ bins correspond to Region 1, while most of the BELLE data belong to Region 2. With this selection, the issues related to the matching problem are less severe.}
\label{fig:canvas_T_875_compare}
\end{figure}
%

\begin{figure}
\centering
\includegraphics[width=14cm]{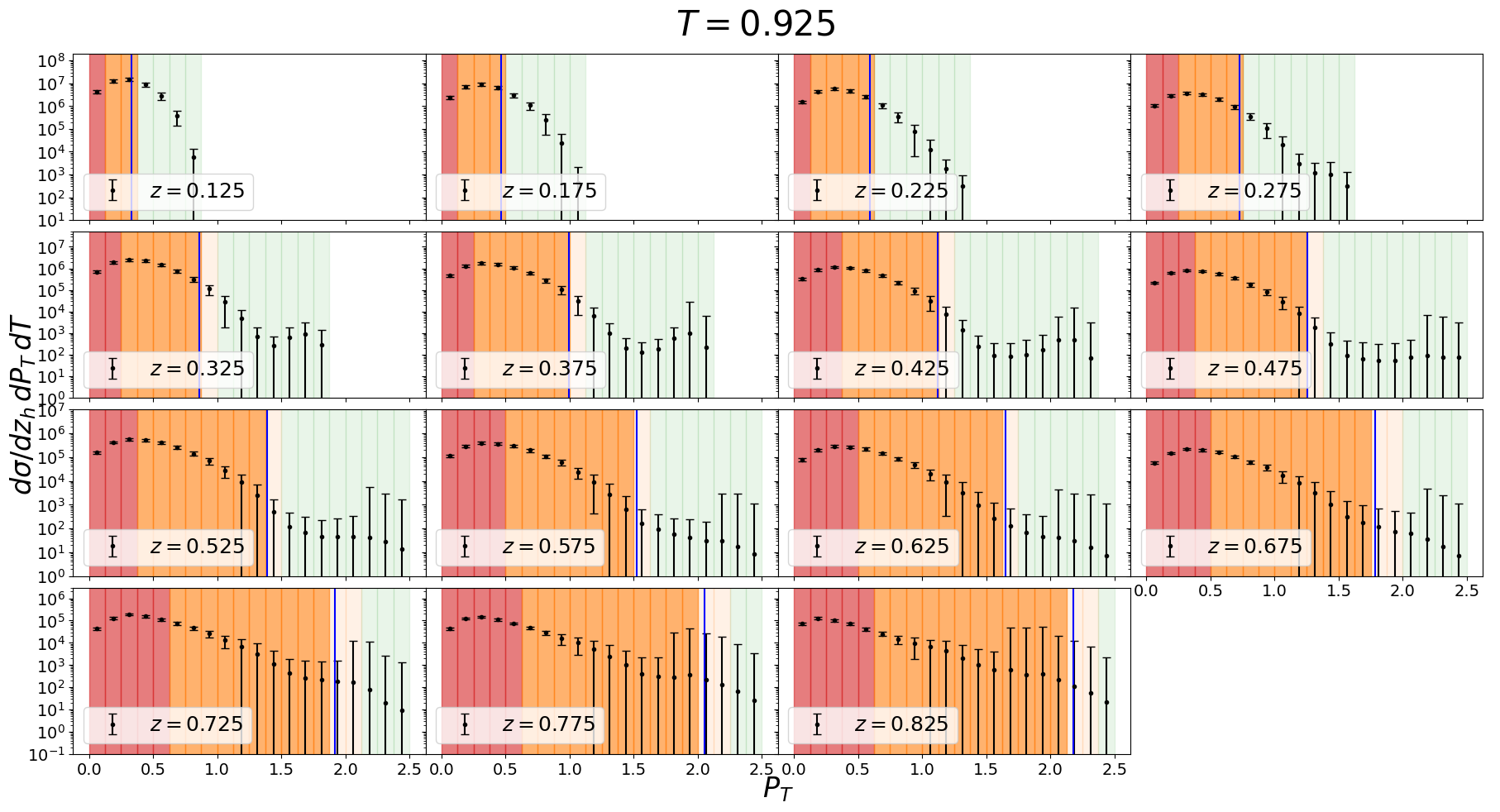}\\
\includegraphics[width=14cm]{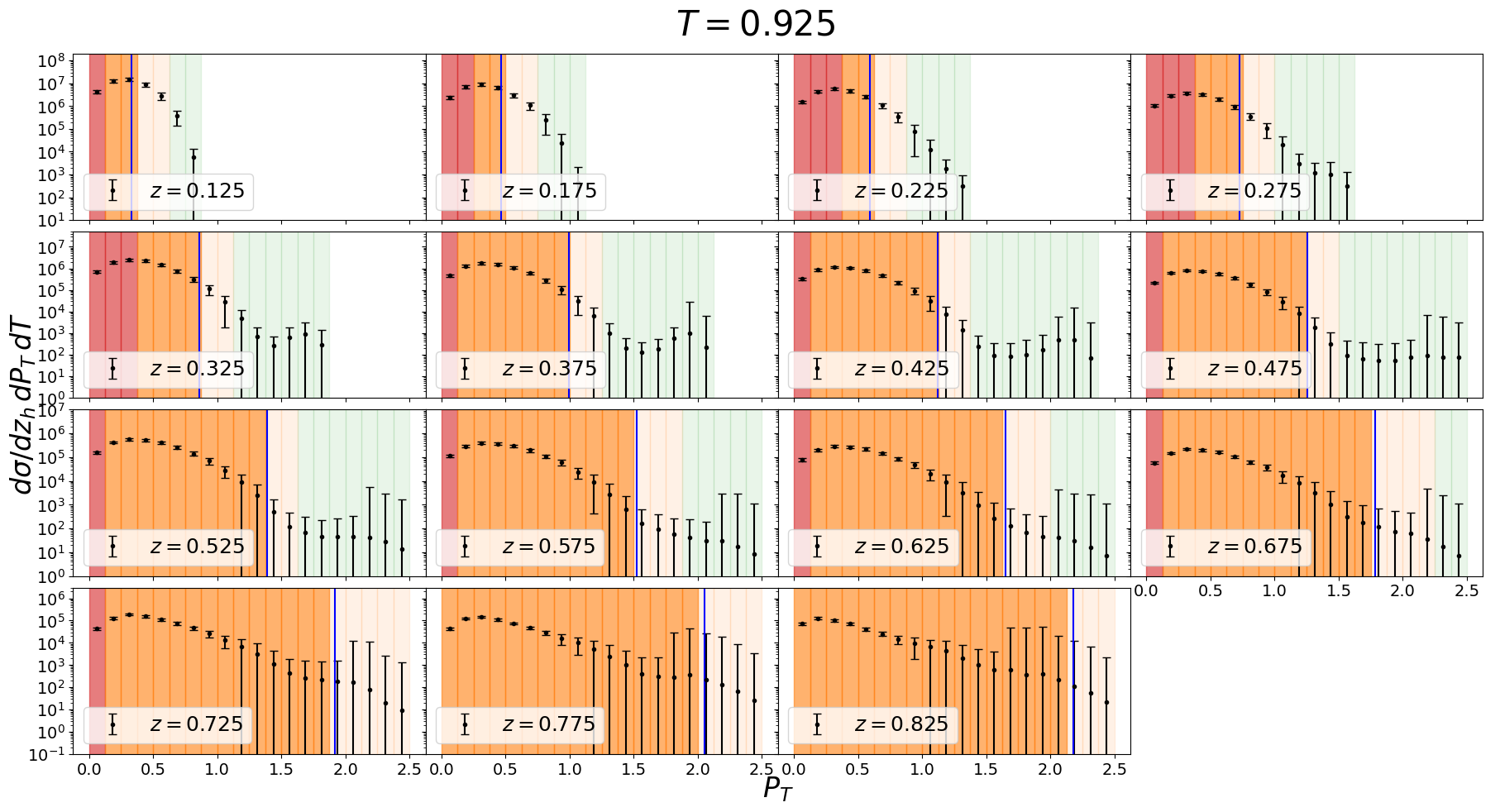}\\
\caption{
BELLE data for thrust $T=0.925$ selected according to the criteria of Eq.~\eqref{eq:makris_crit} (upper panel) and to the algorithm of  Fig.~\ref{fig:AS_algorithm} (lower panel). 
Here Region 3 (green bins) is not realized.
The criteria of Ref.~\cite{Makris:2020ltr} suggest that Region 1 and Region 2 are both present on the left of the cut in $P_T$. Therefore, a phenomenological analysis performed according to this criteria necessarily requires a matching procedure in order to properly describe the bins at the boundaries of Region 1 and Region 2. Instead,  with the more refined algorithm proposed in this paper, we find that only the lower $z_h$ bins correspond to Region 1, while most of the BELLE data belong to Region 2. With this selection, the issues related to the matching problem are less severe.}
\label{fig:canvas_T_925_compare}
\end{figure}

\begin{figure}
\centering
\includegraphics[width=14cm]{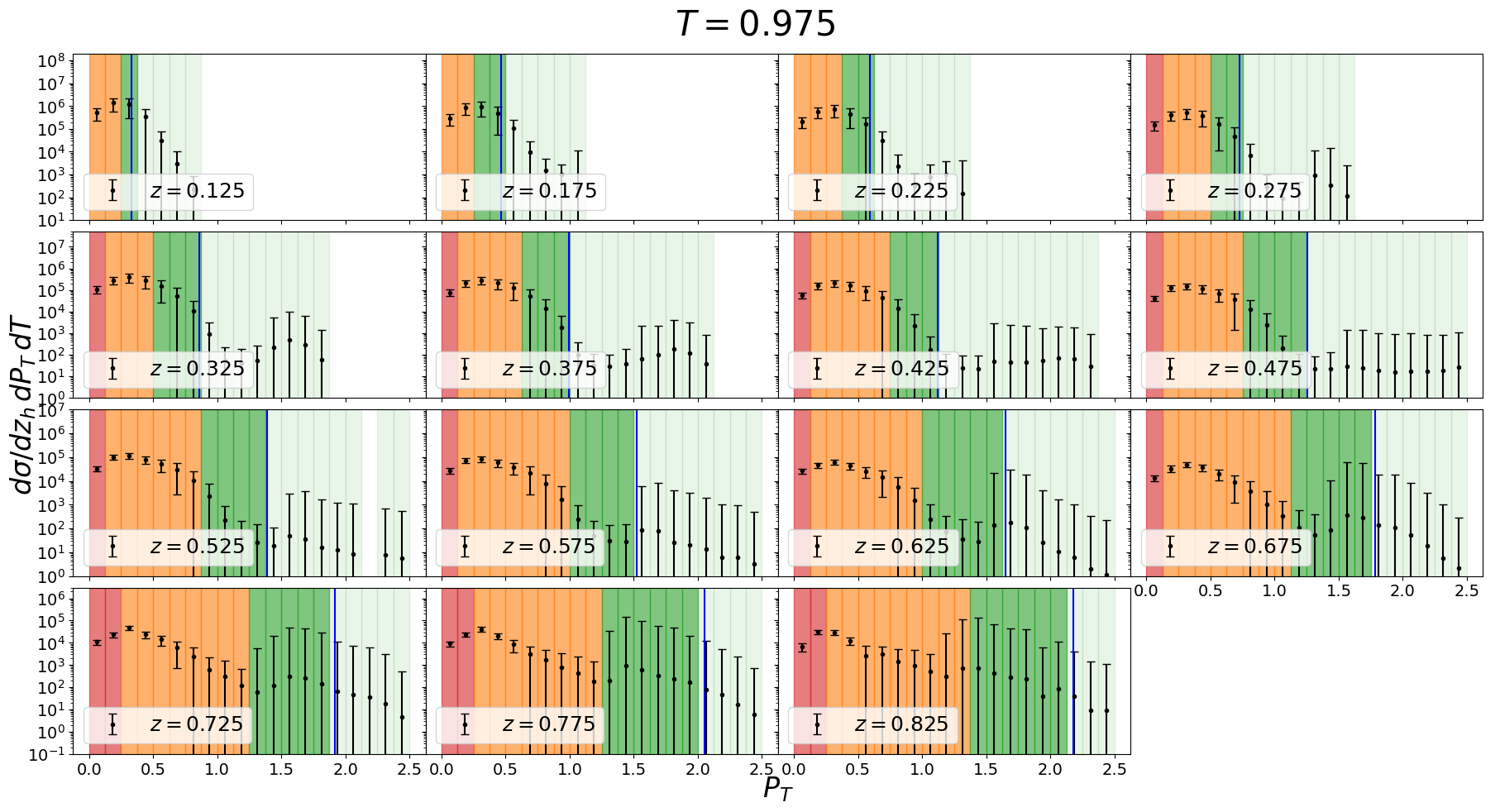}\\
\includegraphics[width=14cm]{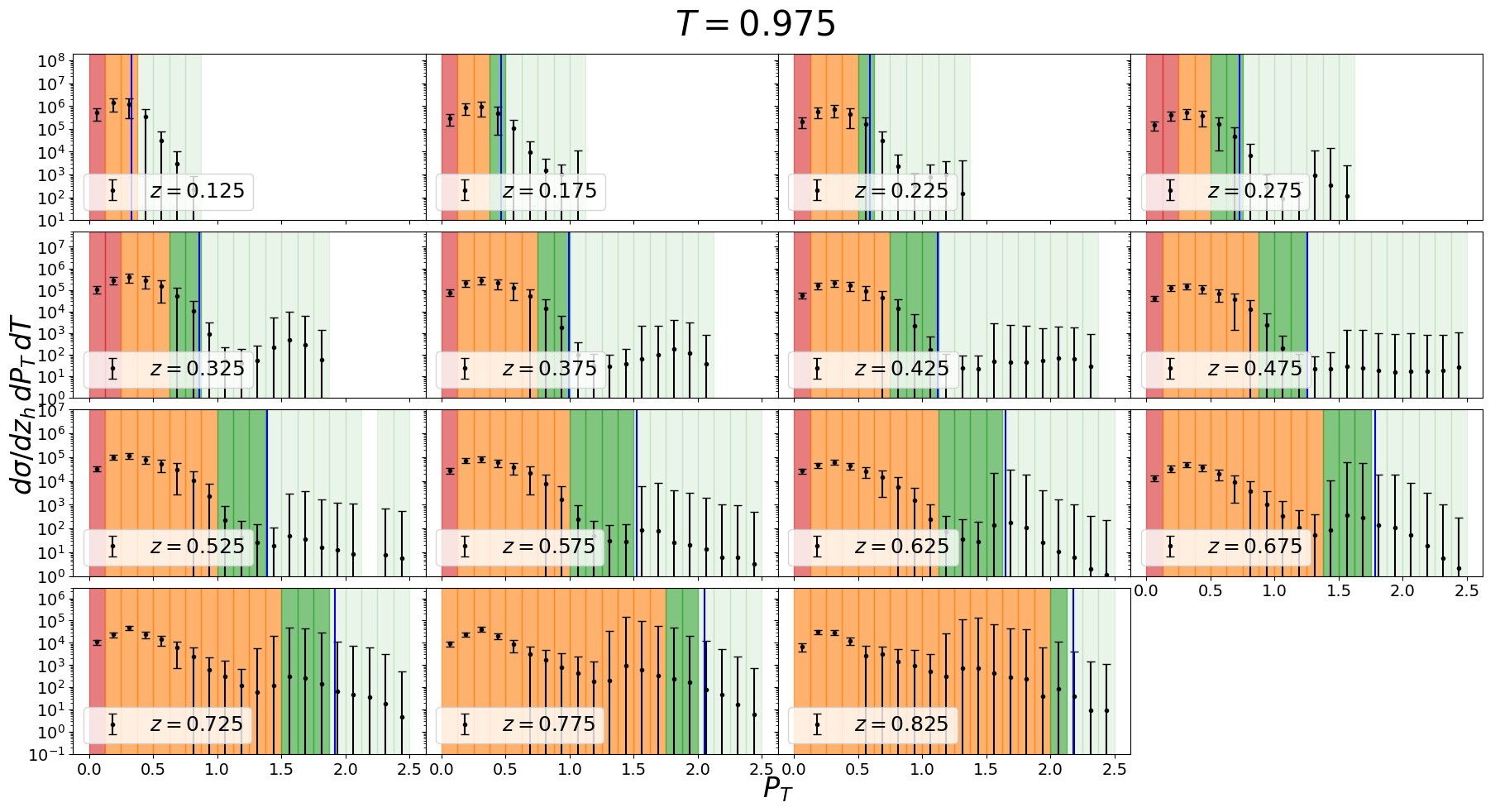}
\caption{BELLE data for thrust $T=0.925$ selected according to the criteria of Eq.~\eqref{eq:makris_crit} (upper panel) and to the algorithm of  Fig.~\ref{fig:AS_algorithm} (lower panel). 
Here Region 3 (green bins) appears together with the other two kinematic regions.
Both the criteria of Ref.~\cite{Makris:2020ltr} and the more refined algorithm proposed in this paper suggests that all three regions are relevant in all the $z_h$-bins. Therefore, in this case, a proper matching procedure would be necessary to appropriately describe the transitions from one region to the following one. Notice that there are two boundaries: one between Region 1 and Region 2, and one between Region 2 and Region 3.}
\label{fig:canvas_T_975_compare}
\end{figure}

\clearpage

\providecommand{\href}[2]{#2}\begingroup\raggedright\endgroup

\end{document}